\documentclass[aps,prm,showpacs,twocolumn,superscriptaddress,floatfix]{revtex4-2}

\pdfoutput=1
\newcommand{\figurescale}{1}

\usepackage{bm}
\usepackage{verbatim}
\usepackage{amsmath}
\usepackage{amssymb}
\usepackage[utf8]{inputenc}
\usepackage[pdftex]{graphicx}
\usepackage[separate-uncertainty=true]{siunitx}
\DeclareSIUnit{\rpm}{rpm}
\usepackage[colorlinks=true,urlcolor=blue,linkcolor=blue,citecolor=blue]{hyperref}
\usepackage{xcolor}
\usepackage[T1]{fontenc}
\usepackage{placeins}
\usepackage{nicefrac}
\usepackage{braket}
\usepackage{pdfcomment}
\usepackage{xcolor}
\usepackage[euler]{textgreek}
\usepackage[normalem]{ulem}

\usepackage{pbox}
\usepackage{makecell}
\usepackage{hhline}
\usepackage{multirow}
\usepackage{tabularx}
\usepackage{soul}

\definecolor{greentwo}{RGB}{0,180,0}

\begin{document}
%############################## TITLE #########################################
\title{Moir\'e straintronics: a universal platform for reconfigurable quantum materials}
%##############################################################################
%
%############################ AUTHORS #########################################
%
\author{M. K\"ogl}
\thanks{Contributed equally to this work.}
\email{mk220@hw.ac.uk}
\affiliation{Institute of Photonics and Quantum Sciences, School of Engineering and Physical Sciences, Heriot-Watt University, Edinburgh EH14 4AS, UK}
\author{P. Soubelet}
\thanks{Contributed equally to this work.}
\email{pedro.soubelet@wsi.tum.de}
\affiliation{Walter Schottky Institut and Physik Department, Technische Universit\"at M\"unchen, Am Coulombwall 4, 85748 Garching, Germany}
\author{M. Brotons-Gisbert}
\affiliation{Institute of Photonics and Quantum Sciences, School of Engineering and Physical Sciences, Heriot-Watt University, Edinburgh EH14 4AS, UK}
\author{A. V. Stier}
\affiliation{Walter Schottky Institut and Physik Department, Technische Universit\"at M\"unchen, Am Coulombwall 4, 85748 Garching, Germany}
\author{B. D. Gerardot}
\affiliation{Institute of Photonics and Quantum Sciences, School of Engineering and Physical Sciences, Heriot-Watt University, Edinburgh EH14 4AS, UK}
\author{J. J. Finley}
\affiliation{Walter Schottky Institut and Physik Department, Technische Universit\"at M\"unchen, Am Coulombwall 4, 85748 Garching, Germany}
%
%##############################################################################
%
\date{\today}
%
%##############################################################################
%									ABSTRACT
%##############################################################################
%

\begin{abstract}
Large scale two-dimensional (2D) moir\'e superlattices are driving a revolution in designer quantum materials. The electronic interactions in these superlattices, strongly dependent on the periodicity and symmetry of the moir\'e pattern, critically determine the emergent properties and phase diagrams. To date, the relative twist angle between two layers has been the primary tuning parameter for a given choice of constituent crystals. Here, we establish strain as a powerful mechanism to in-situ modify the moir\'e periodicity and symmetry. We develop an analytically exact mathematical description for the moir\'e lattice under arbitrary in-plane heterostrain acting on any bilayer structure. We demonstrate the ability to fine-tune the moir\'e lattice near critical points, such as the magic angle in bilayer graphene, or fully reconfigure the moir\'e lattice symmetry beyond that imposed by the unstrained constituent crystals. Due to this unprecedented simultaneous control over the strength of electronic interactions and lattice symmetry, 2D heterostrain provides a powerful platform to engineer, tune, and probe strongly correlated moir\'e materials.
\end{abstract}

%key words: Heterostain, Moir\'e lattice, Hubbard model, Strain-tunable On-chip quantum simulator.
%
%##############################################################################
%
\maketitle
%
%###############################################################################
%								MAIN TEXT
%###############################################################################
%

\section{Introduction}
Strong correlations among electrons arise when their mutual Coulomb interaction is similar or larger than their kinetic energy, and the delicate balance between these two energy scales determines the ground state of the system and its low-energy excitations. The rise of long period moir\'e patterns, formed by vertically stacking two atomically thin crystals with a lattice mismatch ($\delta$) and / or relative twist angle ($\theta$), provides an unique capability to tune the two critical energy scales (Coulomb interaction and kinetic energies) - and the electron density - over several orders of magnitude, creating a highly versatile solid-state quantum material platform \cite{Kennes2020}. The potential to engineer and probe strongly correlated states has been highlighted in several moir\'e material systems, including twisted bilayer graphene heterostructures near the magic angle \cite{Cao2018,cao2018correlated,Balents2020} and homo-\cite{Shimazaki2019,Wang2020b,Wang2022} or hetero- \cite{Tang2020,regan2020mott,Li2021,Campbell2022} bilayer transition metal dichalcogenide (TMD) moir\'e heterostructures. The physics of these systems is determined by the strong electronic interactions in the landscape defined by the underlying moiré potentials. Further examples of synthetic moir\'e materials in which new physical properties arise due to 
%the moiré dependent 
interlayer hybridization can be found in moir\'e trapped excitons \cite{seyler2019signatures,Baek2020a}, twisted 2D magnets \cite{Song2021,Xie2022,Xu2022}, moir\'e solitons \cite{edelberg2020tunable, Hu2020}, moir\'e polaritons \cite{Chen2020}, and ferroelectric moir\'e materials \cite{Vizner2021,Yasuda2021,Hao2021}. 

The simplest model to understand strongly interacting quantum systems is the Hubbard model, which consists of a kinetic term defined by the hopping parameter $t$ and the on-site Coulomb repulsion $U$. In the Hubbard picture, when $U/t > 1$ the kinetic energy is quenched and strong electronic correlations emerge. Although it is expected that the Hubbard model can provide valuable insight into phenomena such as high temperature superconductivity or exotic magnetic states \cite{Meda2009, jiang2019superconductivity}, only the 1D case can readily be solved. The solution of higher dimensional problems depends on a delicate balance between the different model parameters and the lattice geometry \cite{Kennes2020}. Hence, quantum simulators based on ultracold atomic optical lattices, where it is possible to preset the lattice geometry and tune $U/t$ by adjusting the laser power and $U$ using Feshbach resonance techniques \cite{Tarruell2018a}, have become an exciting avenue to probe the Hubbard model and explore emerging quantum materials \cite{Gross}. 

Although 2D moir\'e materials offer access to different energy scales and a broader range of density and temperature than what is achievable in cold atom optical lattices, the same wide-ranging in-situ tunability (of lattice symmetry, $U$, and $t$) possible in optical lattices has proven to be elusive for 2D quantum materials. While the interlayer coupling and band alignment can be tuned in-situ by displacement fields \cite{Shimazaki2019,Wang2020b,Hao2021}, $U$ and $t$ can only be tuned via the choice of constituent materials, the stacking angle $\theta$ or adjusting the dielectric environment of the active moir\'e region \cite{Wu2018a,Pan2020a}. Since $\theta$ and the permittivity of the surrounding layers is set during the fabrication of the moir\'e heterostructure, direct continuous modification of the Hubbard model parameters is not readily achievable, restricting the capability to fine-tune parameters near critical points and broadly explore phase diagrams. This limitation constrains the usefulness of moir\'e materials as quantum simulators.

Here, we explore the effect of heterostrain on the size and geometry of moir\'e lattices formed in vertically stacked bilayers of van der Waals materials. Using a generic analytical approach that is independent of the intrinsic crystal structure, we describe the effects of biaxial, uniaxial, and shear heterostrains for realistic experimental conditions. We find that depending on the type of strain, we can tune the moir\'e wavelength or even smoothly modify the moir\'e lattice from e.g. triangular into rectangular lattices. Furthermore, the moir\'e wavelength can be tuned in situ in different lattice directions to break symmetries in the system, satisfying lattice conditions required for exotic quantum models beyond the standard Hubbard model such as coupled Luttinger chains \cite{Georges2000}. Finally, we show that the use of heterostrain can be implemented to  fine tune and tailor structures to overcome fabrication variabilities that are typically present, for instance, in magic-angle graphene and other moir\'e materials \cite{lau2022reproducibility}. 
We thus show that strain control is a promising strategy to provide the critical degree of freedom required to realize reconfigurable quantum materials and achieve a fully tunable on chip quantum simulator.

%\cite{cao2018correlated,cao2018correlated,Yankowitz2019}. 

\section{Mathematical description}
\label{I}

Moir\'e lattices are large-scale interference patterns formed when the unit cells of adjacent layers  deviate either by a small twist angle $\theta$ or a lattice constant mismatch $\delta$ \cite{kuwabara1990anomalous}. In this work, we assume that the unit cell of both layers of the homo-/heterostructure are of similar geometry while the size of the unit cell one layer may deviate by a scaling factor $(1+ \delta)$. The following formulation describes a pure geometrical deformation of moir\'e lattices under a global 2D strain tensor in which each layer is considered as a rigid object, i.e. atomic reconstruction effects can be neglected. Possible experimental implementations are presented in section \ref{SM-Exp}). 

\begin{figure}[t!!]
\includegraphics*[keepaspectratio=true, clip=true, angle=0, width=1.\columnwidth, trim={60mm, 95mm, 60mm, 0mm}]{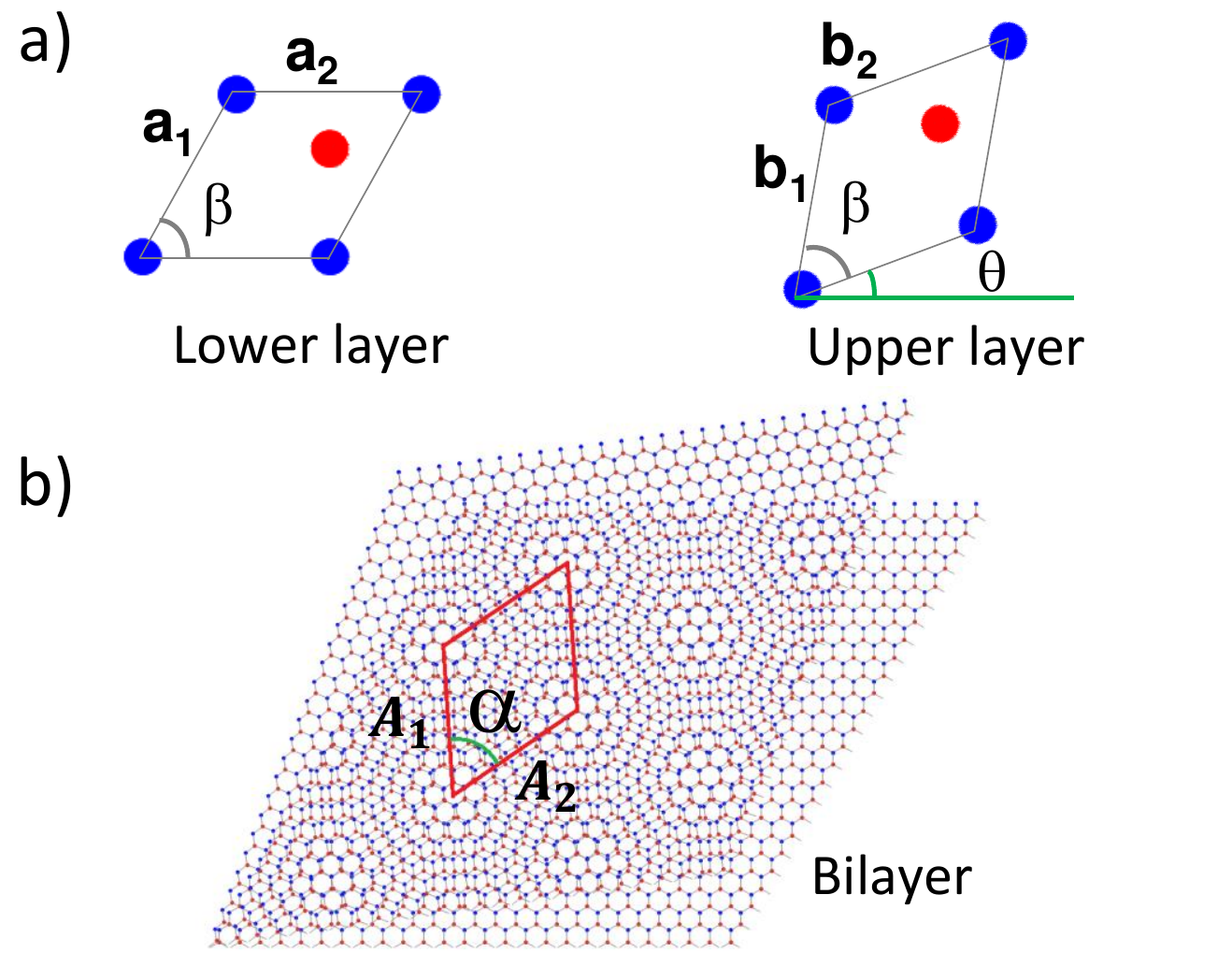}
\caption{\textbf{a)} Sketch of the different layers that compose the moiré lattice. The individual lattices chosen are hexagonal crystals, such as found in graphene and TMDs, and the overall rotation is $\theta_0 = 0$. The lattice parameters of the first (second) layer are $\mathbf{a}_1$ ($\mathbf{b}_1$) and $\mathbf{a}_2$ ($\mathbf{b}_2$), the angle between those vectors is $\beta$ for both layers and the relative angle between the lattices is $\theta$. \textbf{b)} Moiré lattice parameters $A_1$ and $A_2$ and the internal angle $\alpha$ between those vectors in a bilayer formed by the hexagonal monolayer crystals.}
\label{figura:-1}
\end{figure}

The most general moir\'e pattern is formed between two monoclinic lattices with Bravais lattice vectors $\mathbf{a}_i$ for the $lower$ layer and $\mathbf{b}_i$ for the $upper$ one, as sketched in figure \ref{figura:-1}a. These primitive lattice vectors can be written as
\begin{equation}
\begin{split}
\mathbf{a}_i&=a_i R_{\psi_i}\cdot \begin{pmatrix} 1\\ 0 \end{pmatrix} \\
\mathbf{b}_i&=a_i (1+\delta) R_{\psi_i+\theta}\cdot \begin{pmatrix} 1\\ 0 \end{pmatrix},
\end{split}
\end{equation}
where $a_i$ are the lattice constants of a general 2D monoclinic lattice and $R_{\psi_i}$ is the 2D rotation matrix with an angle $\psi_i$. The angle $\psi_i$ is
\begin{equation}
\psi_i=\left\{
\begin{aligned}
\theta_0 ~&;~ i=1 \\ \theta_0 + \beta ~&; ~i=2 
\end{aligned} \right.
\end{equation}
with $\theta_0$ being the overall rotation of the lattices compared to the $x-$axis and $\beta$ the angle between the primitive lattice vectors $\mathbf{a}_1$($\mathbf{b}_1$) and $\mathbf{a}_2$($\mathbf{b}_2$). The stacking of these TMDs results in a bilayer system with a moir\'e pattern as presented in Fig.\ref{figura:-1}b. Such a lattice is characterized by the moiré lattice vectors $\mathbf{A}_1$ and $\mathbf{A}_2$ whose magnitudes are $A_1$ and $A_2$, respectively. In addition, we define as $\alpha$ the angle between those vectors.

%==FIGURA======FIGURA======FIGURA======FIGURA======FIGURA======FIGURA=====
%==FIGURA======FIGURA======FIGURA======FIGURA======FIGURA======FIGURA=====

In the next step, both lattices are subject to a geometric deformation due to an applied strain on the bottom layer, expressed by a general 2D strain tensor $\epsilon$. We restrict our discussion to a global strain that is uniform across the whole layer. To account for any slippage between layers, we introduce the strain transfer parameter $\mu$. This strain transforms the Bravais lattice vectors $\mathbf{a}_i$ and $\mathbf{b}_i$ of the individual layers in the form
\begin{equation}
\begin{split}
\mathbf{a'}_i&=a_i \left(\mathbb{I}+\epsilon \right) \cdot R_{\psi_i}\cdot \begin{pmatrix} 1\\ 0 \end{pmatrix} \\
\mathbf{b'}_i&=a_i (1+\delta) \left(\mathbb{I}+\mu \epsilon \right) \cdot R_{\psi_i+\theta}\cdot \begin{pmatrix} 1\\ 0 \end{pmatrix}.
\end{split}
\end{equation}

The parameter $\mu$ allows us to model different experimental setups (see section \ref{SM-S6}), for instance, a system where the lower layer is clamped directly to the substrate while the upper one is only in contact with the first one. Since the interlayer friction is high for commensurate structures (small $\theta$ or $\delta$) and low for incommensurate structures (high $\theta$ or $\delta$), the transferred strain in the second layer is different for each case \cite{Ru2020, Song2018a,RN35}. In the incommensurate  case, the strain on the lower layer is not transferred to the upper one, i.e. $\mu=0$, which we refer to as $heterostrain$. For small $\theta$ or $\delta$ the layers are more commensurate such that $0<\mu<1$. A full transfer of strain $\mu=1$ is referred to as $homostrain$.

The strain tensor $\epsilon$ is commonly written as a symmetric 2x2 matrix with three independent parameters \cite{Bi2019}. Hence, we express the strain matrix as 
\begin{equation}
\label{eq:4}
\epsilon = \left(\begin{matrix} \epsilon_{xx} & \epsilon_{xy} \\ \epsilon_{xy} & \epsilon_{yy}
\end{matrix}\right)= \epsilon_c \mathbb{I}+ \epsilon_s S_{\phi_s},
\end{equation}
where we define
\begin{itemize}
    \item $\epsilon_c=\frac{\epsilon_{xx}+\epsilon_{yy}}{2}$ as \textit{biaxial strain},
    \item $\epsilon_s=\sqrt{\left(\frac{\epsilon_{xx}-\epsilon_{yy}}{2}\right)^2+\epsilon_{xy}^2}$ as \textit{shear strain},
    \item $S_{\phi_s}=\cos(\phi_s)\sigma_x + \sin(\phi_s)\sigma_z$ as \textit{shear matrix}, and
    \item $\phi_s=\arccos\left(\frac{\epsilon_{xy}}{\epsilon_s}\right)$ as \textit{shear strain angle}.
\end{itemize}
Our strain parameterization follows Ref. \onlinecite{halbertal2021moire}, which separates the strain into \textit{biaxial strain} $\epsilon_c$ (which changes the size of the unit cell) and \textit{shear strain} $\epsilon_s$ (which alters the shape of the unit cell). Details on the physical meaning of the shear angle and different experimental feasibilities can be found in section \ref{SM-Exp}. In this context, \textit{uniaxial strain} ($\epsilon_{xx}=\epsilon_u$, $\epsilon_{xy}=0$, $\epsilon_{yy}=-\nu \epsilon_u$) is a mixture of biaxial and shear strain with a shear angle set to $\phi_s=90^\circ$. Note that under uniaxial strain, the crystal is elongated along one direction while in the perpendicular direction the crystal deforms proportional to the Poisson ratio $-\nu$.

% An illustration of dilation and shear strain is shown in Fig.~\ref{figura:1} and Fig.~\ref{figura:5}.

Using the previous definitions it is possible to calculate the real space moir\'e lattice under general strain, as presented section \ref{SM-S1-1}. The new moir\'e lattice vectors $\mathbf{A'}_i$ take the form
\begin{equation}
\label{eq:final}
\begin{split}
\mathbf{A'}_i &=\frac{a_i(1+\delta)}{\Delta} \\
&\cdot [ (1+\delta)c_\mu \left((1+\epsilon_c)\mathbb{I}+ \epsilon_s S_{\phi_s}\right)\\
&-c_1\left((1+\mu \epsilon_c) R_{\theta}+\mu\epsilon_s S_{\phi_s+\theta}\right) ] \cdot R_{\psi_i} \cdot\begin{pmatrix} 1\\ 0 \end{pmatrix}
\end{split}
\end{equation}
where the denominator $\Delta$ is the expression
\begin{equation}
\label{eq:denominator}
\begin{split}
\Delta&=c_\mu (1+\delta)^2+c_1\\
&-2(1+\delta)\left[(1+\epsilon_c +\mu\epsilon_c)+\mu(\epsilon_c^2-\epsilon_s^2)\right]\cos(\theta),
\end{split}
\end{equation}
$c_1=(1+\epsilon_c)^2-\epsilon_s^2$, and $c_\mu=(1+\mu\epsilon_c)^2-\mu^2\epsilon_s^2$. Finally, we calculate the moir\'e unit cell area $M$ as
\begin{equation}
\label{eq:moiresize}
\begin{split}
M&=\|\mathbf{A'}_1 \times \mathbf{A'}_2\| =a_1a_2\|\sin(\beta)\|\frac{\left(1+\delta\right)^2 }{\|\Delta\|} c_1 c_{\mu}.
\end{split}
\end{equation}
The full calculation is provided in section \ref{SM-S1-2}.

We emphasize that all manifestations of 2D strain are covered by these calculations and no approximations are made, in contrast to previous descriptions that focus on hexagonal lattices only and use approximations that only apply to specific strain configurations \cite{Cosma2014,Tong2017a, Bi2019}. Note that expression~\ref{eq:final} defines the moir\'e lattice vectors and can be employed for any homo- and heterobilayer structure with a similarly shaped unit cell.

%Having found a description for the moir\'e lattice vector, we analyze the dependency of the moir\'e vectors on the different strain parameter $\epsilon_c$, $\epsilon_s$ and $\phi_s$. 
In the particular case of homostrain, expression \ref{eq:final} leads to $\mathbf{A'}_i= \left(\mathbb{I}+\epsilon \right)\cdot\mathbf{A}_i$ which means that the strain is applied to the moir\'e vectors as if the moir\'e lattice is strained itself. Even though 2D materials can withstand very high strain levels \cite{Bertolazzi2011}, the maximum applicable strain is $\epsilon_c,\epsilon_s<<1$. Hence, the homostrain effect on the moir\'e lattice is very small and can be neglected in comparison to the effect of tuning the twist angle $\theta$. However, in the case of imperfect strain transduction between the layers ($\mu<1$) we find that strain has a substantial effect on the moir\'e size and shape, since the magnitude of the moir\'e vectors is dominated by the denominator $\Delta$ in expression \ref{eq:final}. For example, in the homobilayer case, increasing biaxial strain $\epsilon_c$ decreases the moir\'e lattice parameters with the approximate dependence of ${A'}_i\propto \frac{1}{\|\epsilon_c\|}$. In the context of the Hubbard model, the hopping parameter $t$ decreases exponentially with $A'_i$ while $U$ scales inversely with $A'_i$ \cite{Wu2018a, Pan2020a} and, therefore, the ratio $U/t$ can be modified by tuning $\epsilon_c$. In contrast to $\theta$, in-situ strain tuning is experimentally viable which allows direct tuning of $U/t$ (see section \ref{SM-Exp}).

In contrast to biaxial strain, shear strain can make the denominator $\Delta$ identically zero, leading to the divergence of the moir\'e vectors. This condition, referred to as a one-dimensional (1D) moir\'e pattern \cite{Tong2017a}, has already been experimentally observed \cite{Bai2019}. This effect occurs when the deformation of one layer due to shear strain aligns the lattice sites in one direction while leaving a mismatch in the perpendicular direction. We emphasize that this effect is not dependent on the orientation of the lattice $\theta_0$, the shear angle $\phi_s$ or the shape of the layer lattices.

In addition to the size of the moir\'e pattern, shear strain is also able to change the moir\'e lattice geometry since $\epsilon_s$ changes the shape of the individual lattice depending on the orientation of $\phi_s$ and $\theta_0$. The main defining parameter for the geometry of the moir\'e pattern is the angle $\alpha$ between the moir\'e vectors (see Fig.\ref{figura:-1}b). This angle takes the form 
\begin{equation}
    \sin(\alpha)=\left\|\frac{\mathbf{A'}_1\times\mathbf{A'}_2}{{A'}_1 {A'}_2}\right\|.
\end{equation}
In the case of zero shear strain $(\epsilon_s=0)$, it can be shown that $\frac{a_1}{a_2}\stackrel{\epsilon_s=0}{=}\frac{{A'}_1}{{A'}_2}$ and that $\alpha\stackrel{\epsilon_s=0}{=}\beta$, proving that the shape of the moir\'e lattice is equal to the shape of the underlying lattice defined by $\mathbf{a_i}$, as presented in detail section \ref{SM-S1-4}.

In contrast to in situ tuning of $\theta$ \cite{Yao2021,Hu2022,RN33}, 2D strain has the potential to tune the size and shape of the moire pattern via the three independent strain parameters $\epsilon_c$, $\epsilon_s$ and $\phi_s$. In the following section, we will focus on the case of pure heterostrain ($\mu=0$) on homo- and heterobilayer structures and show how the different types of strain can affect the moir\'e lattice geometry.

\section{Heterostrain on hexagonal homo- and hetero-bilayers}
\label{III}

In this section, we center our analysis on moiré patterns generated by monolayers with hexagonal lattices, in particular, we take WSe$_2$ homobilayers and MoSe$_2$-WSe$_2$ heterobilayers as paradigmatic cases. The lattice parameters used to perform the calculations are $a_{MoSe_2} = 0.329$\,nm for MoSe$_2$ \cite{Kang2013a} and 0.4\% smaller for the WSe$_2$ \cite{Kang2013a, osti_1192989, SCHUTTE1987207}. As shown in section \ref{SM-S6}, strategies can be proposed to avoid strain transfer for most bilayer structures. In this work, we analyse the pure heterostrain case, in which the deformation is applied to the lower layer, which, in our specific heterobilayer example, corresponds to the WSe$_2$ monolayer. The Poisson ratios are $\nu_{MoSe_2} = 0.23$ and $\nu_{WSe_2} = 0.19$ for MoSe$_2$ and WSe$_2$, respectively \cite{Kang2013a}. For simplicity, in this section we focus on three different configurations: biaxial strain, uniaxial strain and shear strain.

\subsection{Biaxial strain}
\label{subsec:biaxial_strain}

%==FIGURA======FIGURA======FIGURA======FIGURA======FIGURA======FIGURA=====
\begin{figure}[t!!]
\includegraphics*[keepaspectratio=true, clip=true, angle=0, width=1.\columnwidth, trim={55mm, 72mm, 51mm, 5mm}]{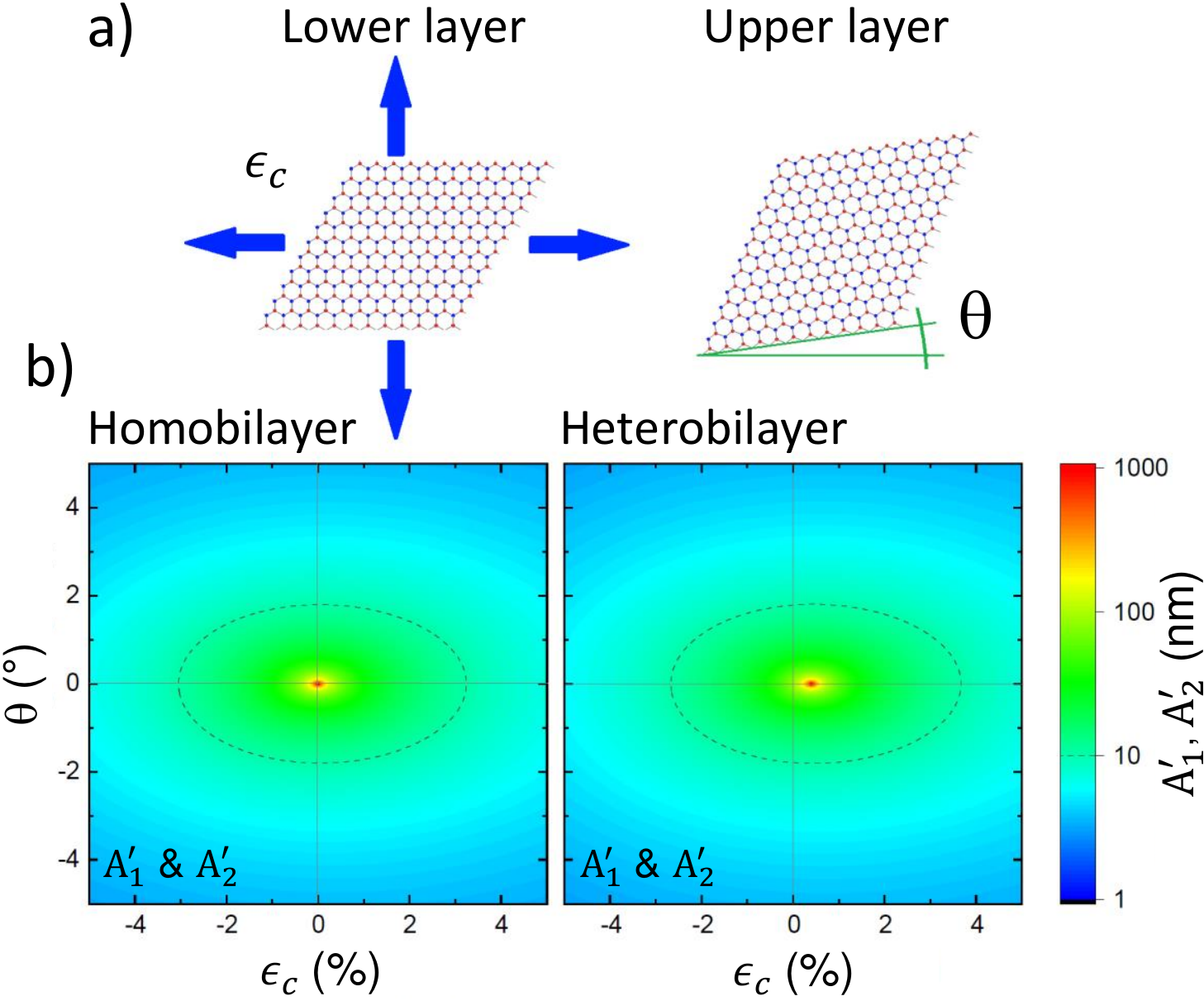}
\caption{\textbf{a)} Sketch of the moiré lattice formation. To perform the calculations, the biaxial strain $\epsilon_c$, indicated trough blue arrows, is applied to the lower layer. The upper layer is stacked with an angle $\theta$ respect to the lower one. \textbf{b)} Moiré lattice parameters $A_1'$ and $A_2'$ as function of biaxial strain and stacking angle for homo- (left) and hetero- (right) bilayer structures.
}
\label{figura:-2}
\end{figure}
%==FIGURA======FIGURA======FIGURA======FIGURA======FIGURA======FIGURA=====

Figure \ref{figura:-2}a shows a sketch of the bilayer formation in which $\epsilon_c$ produces an in-plane elongation/contraction of the lower layer unit cell and the upper layer is stacked at an angle $\theta$. The effect of $\epsilon_c$ and $\theta$ on the moiré pattern is depicted in figure \ref{figura:-2}b for homobilayer (left) and heterobilayer (right) structures. The vertical lines at $\epsilon_c = 0$ correspond to the well known behaviour of homo- and heterobilayers described in Ref.\onlinecite{Yankowitz2012}, i.e., in the homobilayer case, the structure presents a divergence of $A_1'$ and $A_2'$ at $\theta = 0$ that evidences the absence of moiré patterns in naturally stacked bilayers (2H and 3R stacking order). However, the stacking of layers with $\theta \neq 0$ and/or $\epsilon_c \neq 0$ gives rise to moiré patterns whose periodicity ($A_1'$ and $A_2'$) decreases by increasing $\epsilon_c$ and/or $\theta$. In the heterobilayer case, as the lattice parameter of WSe$_2$ is $\sim0.4\%$ smaller than MoSe$_2$, the stretching of the WSe$_2$ layer by $\sim0.4\%$ recovers the divergence observed at the point $\epsilon_c=0$, $\theta = 0$ for homobilayers. Therefore the use of biaxial heterostrain in heterobilayers enables the formation of arbitrarily large moiré lattices as observed in homobilayer structures. Another important point to note is that due to the hexagonal symmetry of each layer, neither $\epsilon_c$ nor $\theta$ can modify the moiré lattice geometry that is also hexagonal, i.e. $A_1' = A_2'$ and $\alpha = 60^{\circ}$, as demonstrated section \ref{SM-S1-4}. Consequently, different combinations of strain and stacking angle can lead to the same moiré lattice (but with a relative rotation). For instance, the dotted lines in both panels of figure \ref{figura:-2}b describe a unique moiré lattice with $A_1' = A_2' = 10$\,nm. Section \ref{SM-S1-3} provides details regarding the direction of the moir\'e lattice with respect to the underlying lattice in absence of shear strain.

\subsection{Uniaxial strain}
\label{US}

%==FIGURA======FIGURA======FIGURA======FIGURA======FIGURA======FIGURA=====
\begin{figure}[t!!]
\includegraphics*[keepaspectratio=true, clip=true, angle=0, width=1.\columnwidth, trim={58mm, 9mm, 39mm, 6mm}]{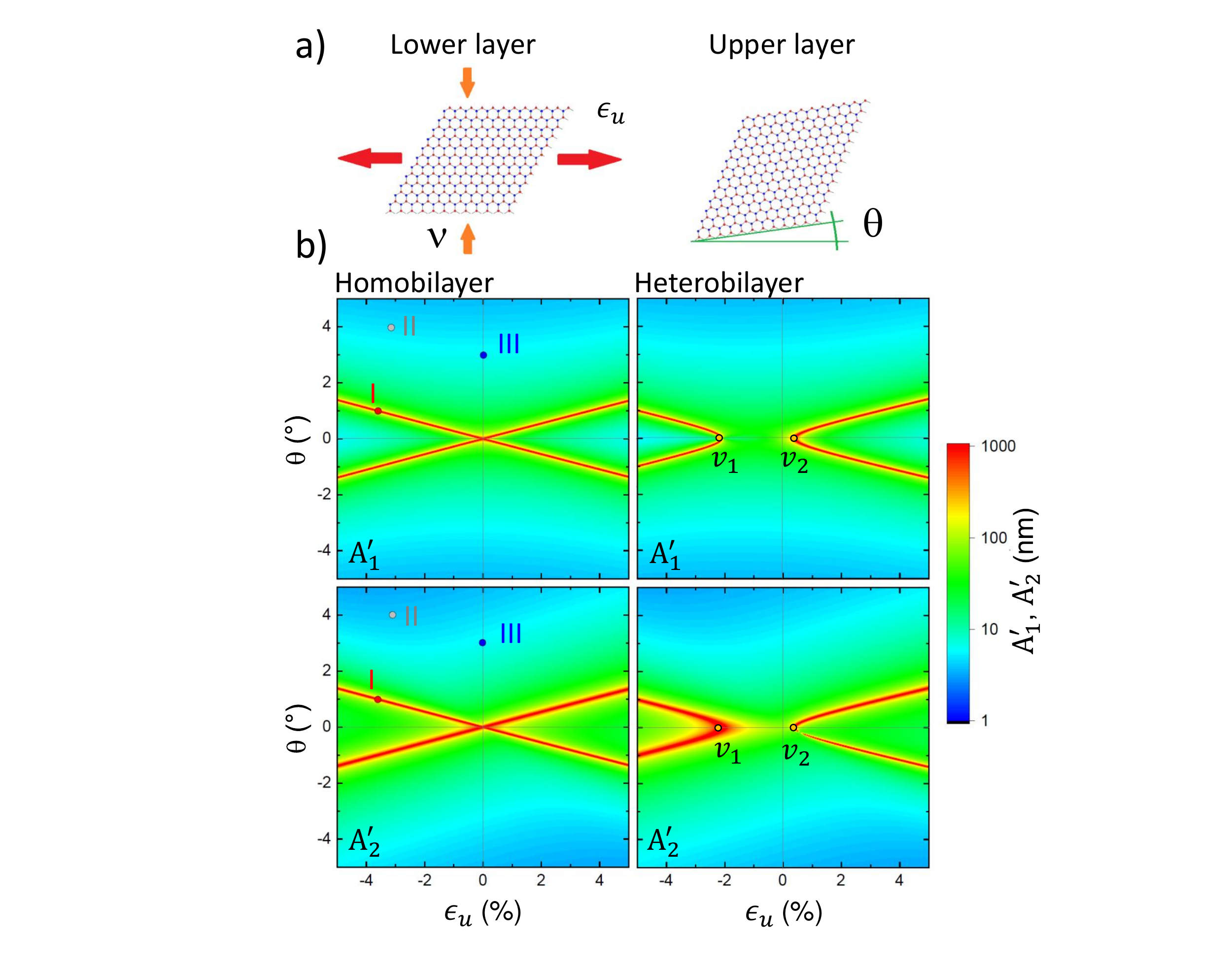}
\caption{\textbf{a)} Schematic of moiré lattice formation under uniaxial heterostrain $\epsilon_u$. The deformation of the lower layer due to strain along a zigzag direction and Poisson ratio $\nu$ are depicted by the arrows. The upper layer was stacked with an angle $\theta$ respect to the lower layer. \textbf{b)} Moiré lattice parameters $A_1'$ (upper panels) and $A_2'$ (lower panels) for the homobilayer- and heterobilayer structures (left and right panels, respectively) as function of uniaxial strain and stacking angle. Spots labeled as $I$, $II$, and $III$ correspond to the different moiré lattices presented in real space in Fig.\ref{figura:-4}b and those marked as $v_1$ and $v_2$ identify the vertexes of the divergence curves in the heterobilayer case.}
\label{figura:-3}
\end{figure}
%==FIGURA======FIGURA======FIGURA======FIGURA======FIGURA======FIGURA=====

We continue by describing the effect of uniaxial heterostrain on the moiré lattice generated in homo- and heterobilayers. Figure \ref{figura:-3}a shows a sketch of this situation in which the lower layer is stretched/contracted along the zigzag direction by applying a uniaxial strain $\epsilon_u$. Due to the Poisson effect, the direction perpendicular to $\epsilon_u$ is deformed by $-\nu \epsilon_u$, where $\nu$ is the Poisson ratio. In the case where the layer is stretched in the zigzag direction, the armchair direction is contracted (stretched) if $\nu>0 (\nu<0)$. The upper layer is stacked at an angle $\theta$ with respect to the lower layer. Figure \ref{figura:-3}b depicts the moiré lattice parameters $A_1'$ (upper panels) and $A_2'$ (lower panels) for homobilayer- and heterobilayer structures (left and right panels, respectively). Uniaxial strain has a stronger influence on the moiré lattice than biaxial strain: (i) For uniaxial strain, the moiré lattice parameters $A_1'$ and $A_2'$ diverge along curves in the $\theta-\epsilon_u$ space instead of a single point (see red curves in Fig.\ref{figura:-3}b). (ii) $\epsilon_u$ breaks the degeneracy between $A_1'$ and $A_2'$. (iii) In heterobilayers, uniaxial strain  cannot retrieve the moiré lattice geometry found in homobilayers. 

%==FIGURA======FIGURA======FIGURA======FIGURA======FIGURA======FIGURA=====
\begin{figure}[t!!]
\includegraphics*[keepaspectratio=true, clip=true, angle=0, width=1.\columnwidth, trim={50mm, 22mm, 42mm, 1mm}]{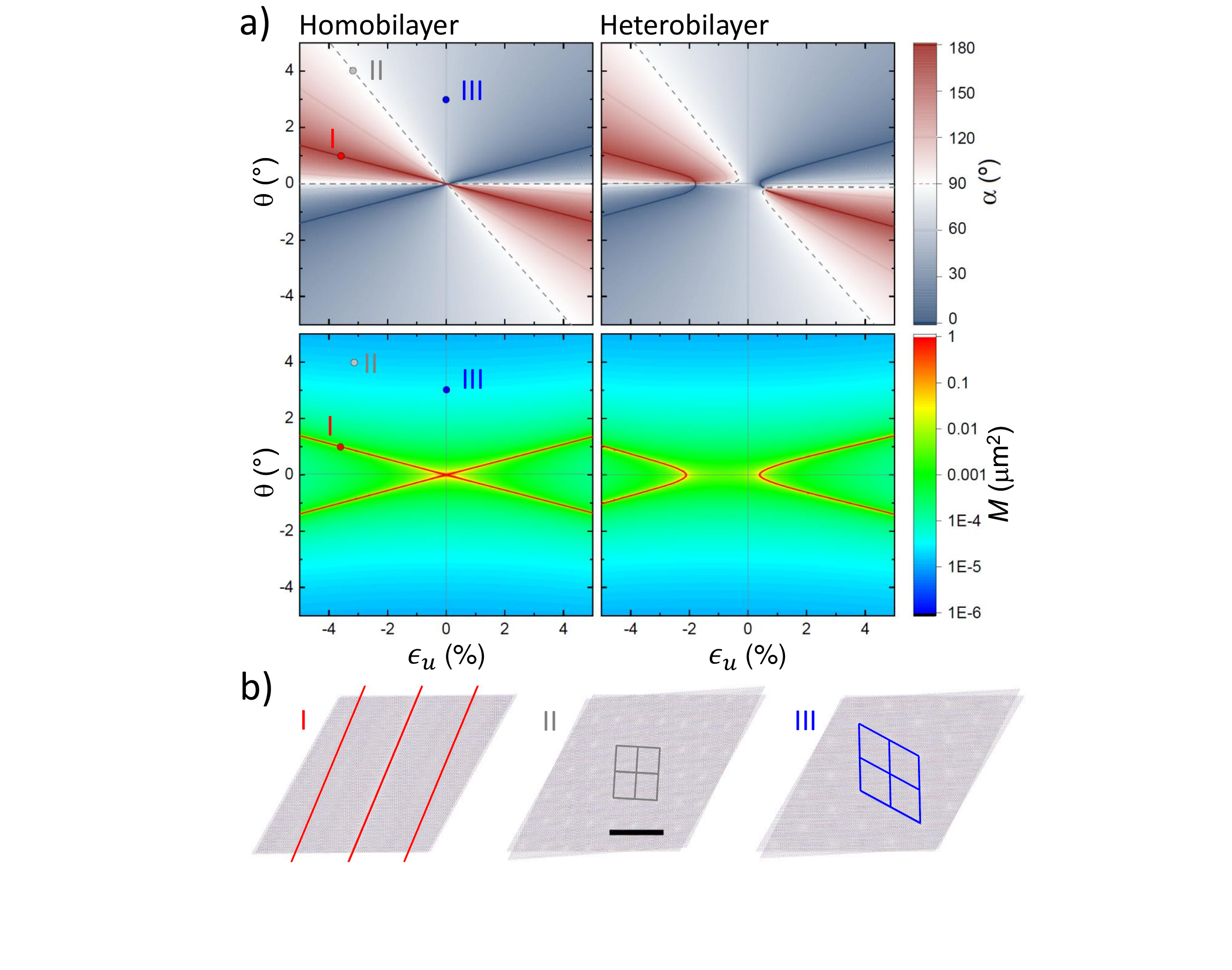}
\caption{\textbf{a)} Relative angle $\alpha$ (upper panels) and moiré lattice area $M$ (lower panels) as function of $\epsilon_u$ and $\theta$ for the homobilayer- (left panels) and heterobilayer structure (right panels). Grey dotted lines denote configurations in which the moiré lattice is rectangular ($\alpha = 90^{\circ}$). The spots labeled as $I$, $II$ and $III$ correspond to the moiré lattices presented in panel b. \textbf{b)} Illustration of moiré lattices corresponding to the points labelled as $I$, $II$, and $III$ in figures \ref{figura:-3}a and \ref{figura:-4}b. Case $I$ shows a 1D moiré lattice, $II$ a rectangular one, and $III$ a hexagonal one.}
\label{figura:-4}
\end{figure}
%==FIGURA======FIGURA======FIGURA======FIGURA======FIGURA======FIGURA=====

In homobilayers, $A_1'$ and $A_2'$ diverge along lines that intersect at the origin ($\epsilon_u=0$, $\theta = 0$). For heterobilayers, these curves do not intersect but show an avoided crossing behaviour instead with vertices located on the axis $\theta=0^{\circ}$ (points marked as $v_1$ and $v_2$ on Fig.\ref{figura:-3}b). The position of $v_2$ is given by the mismatch of the lattice parameters between the two layers, which in our case of a MoSe$_2$/WSe$_2$ heterostructure, is $\epsilon_u = 0.4$\%. The position of $v_1$ is weighted by $\nu_{WSe_2}$, as the deformation on the armchair axis is $19$\% of that of the zigzag axis. Now, the vertex is located at $-0.4/0.19 \simeq -2.1$\%. It can easily be seen that a vanishing Poisson ratio can eliminate the divergence of the lattice parameters and lattice area in the negative semi space. Further information about the effect of the Poisson ratio on the moiré lattice can be found section \ref{SM-C}.

A further insight in how uniaxial strain can affect the moiré lattice geometry can be observed in figure \ref{figura:-4}a and b. Fig.\ref{figura:-4}a shows $\alpha$ (upper panels) and $M$ (lower panels) as function of $\epsilon_u$ and $\theta$ for homo- and heterobilayers. $\alpha$ is a measure of the strong geometrical deformation of the moiré lattice. For instance, by combining $\theta$ and $\epsilon_u$, it is possible to generate a rectangular lattice ($\alpha = 90^{\circ}$), identified by grey dotted lines. On the other hand, for a 1D moiré lattice, indicated by diverging moiré lattice parameters (red lines in Fig. \ref{figura:-3}b), $\mathbf{A_1'}$ and $\mathbf{A_2'}$ tend to be collinear, i.e. $\alpha \to 0^{\circ}$ or $\alpha \to 180^{\circ}$. Note that the divergence of $A_1'$ and $A_2'$ also causes a divergence of the moiré unit cell area, even though these lattice vectors tend to be collinear. Figure \ref{figura:-4}b depicts real space images of some of the accessible moiré lattices by combining $\epsilon_u$ and $\theta$. The different geometries correspond to the three points marked as $I$, $II$ and $III$ in figures \ref{figura:-3}b and \ref{figura:-4}a. Point $I$ describes a 1D moiré lattice, point $II$ a rectangular geometry and $III$ a hexagonal arrangement.

\subsection{Shear strain}

%==FIGURA======FIGURA======FIGURA======FIGURA======FIGURA======FIGURA=====
\begin{figure}[t!!]
\includegraphics*[keepaspectratio=true, clip=true, angle=0, width=1.\columnwidth, trim={53mm, 13mm, 41mm, 8mm}]{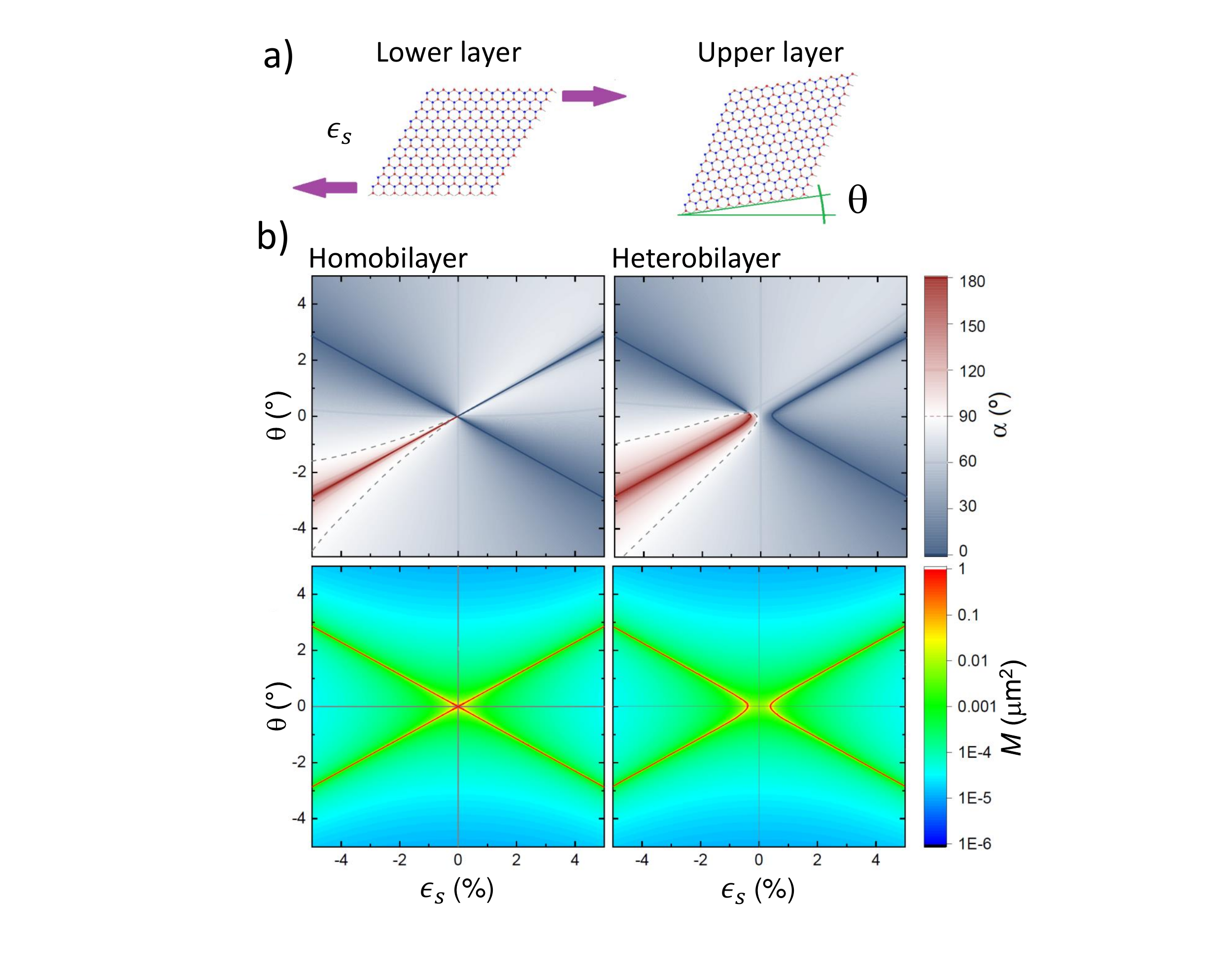}
\caption{\textbf{a)} Scheme of the moiré lattice formation under shear heterostrain $\epsilon_s$. The deformation of the lower layer due to shear strain along the zigzag direction is depicted through arrows. The upper layer was stacked with an angle $\theta$ respect to the lower one. \textbf{b)} Relative angle $\alpha$ (upper panels) and moiré lattice area $M$ (lower panels) as function of shear strain $\epsilon_s$ and stacking angle $\theta$ for homobilayer (left panels) and heterobilayer (right panels) systems. Grey dotted lines denote configurations in which the moiré lattice is rectangular ($\alpha = 90^{\circ}$).}
\label{figura:-6}
\end{figure}
%==FIGURA======FIGURA======FIGURA======FIGURA======FIGURA======FIGURA=====

%==FIGURA======FIGURA======FIGURA======FIGURA======FIGURA======FIGURA=====
\begin{figure*}[hht!!]
	\scalebox{\figurescale}{\includegraphics[width=1\linewidth, trim={42mm, 18mm, 42mm, 8mm}]{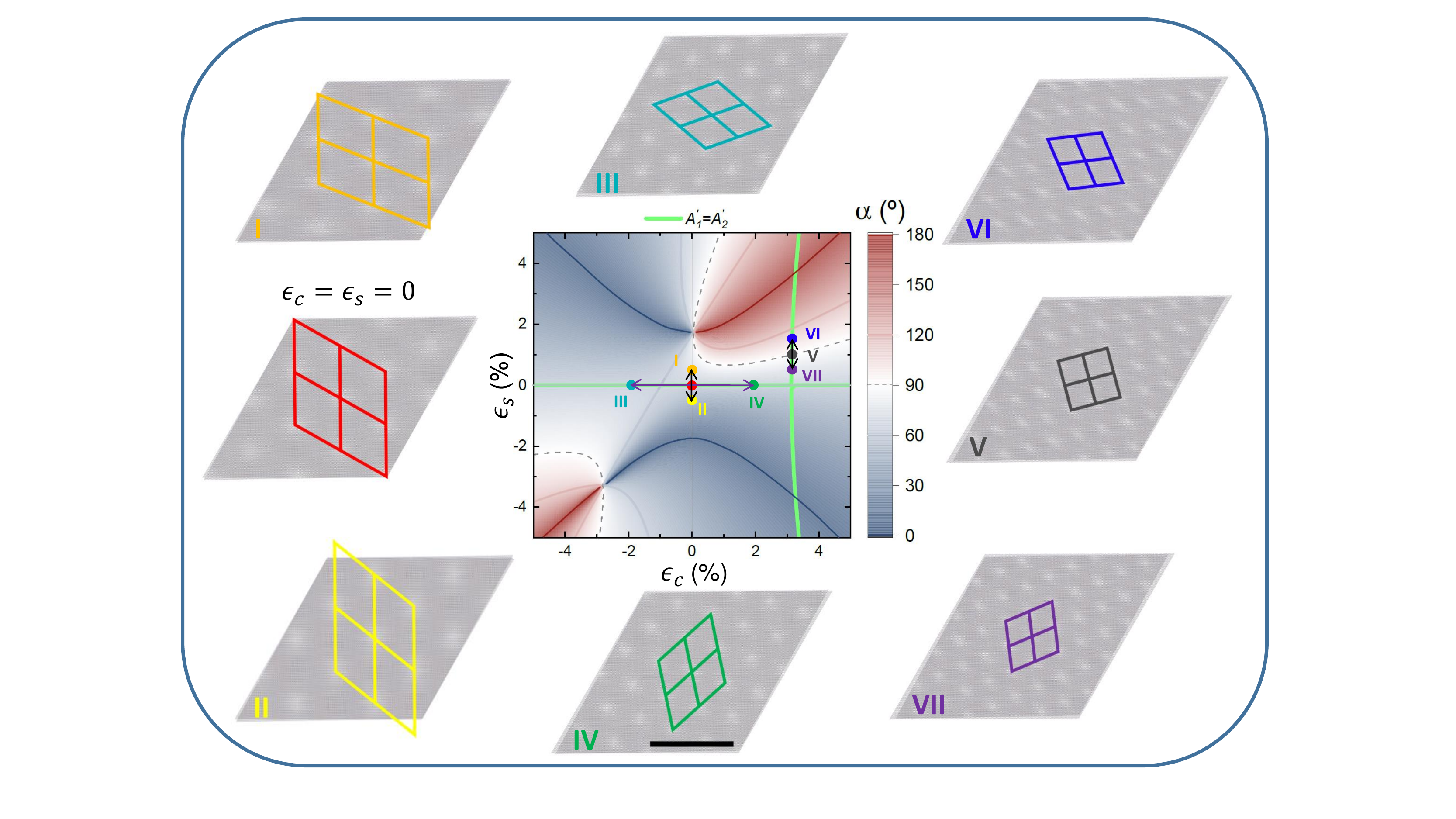}}
	\renewcommand{\figurename}{FIG.|}
	\caption{\label{figura:-7}
		\textbf{Geometrical deformation of a general homobilayer stacked at $\theta = 1^{\circ}$ under biaxial and shear heterostrain.} Light green lines denote moiré lattices in which $A_1' = A_2'$. Points labeled as $I$, $II$, $III$, $IV$, $V$, $VI$ and $VII$ corresponds to the different illustrations of the moir\'e pattern in real space. The scale bar in $IV$ has a length of $\SI{30}{\nano\meter}$ in case of a MoSe$_2$ bilayer and  is valid for all real space moir\'e patterns $I-VII$. An experimental setup for the simultaneous control of biaxial and shear strain is shown in section \ref{SM-subsec:expmoire-2}. The computational script to reproduce real space images and strain animations can be found in \cite{ESM}.
		}
\end{figure*}
%==FIGURA======FIGURA======FIGURA======FIGURA======FIGURA======FIGURA=====

We finish our discussion on the effect of heterostrain on moiré lattices by presenting the case of shear strain ($\epsilon_s$). In figure \ref{figura:-6}a, we sketch the different layers and their deformation/rotation under the influence of $\epsilon_s$. The case presented corresponds to a sideways load applied along the zigzag direction on the lower layer (violet arrows) while the upper layer is stacked at an angle $\theta$. For simplicity, we will focus our analysis on $\alpha$ and $M$, shown in figure \ref{figura:-6}b for homobilayer- (left) and heterobilayer structures (right). As in the case of uniaxial strain, the moiré lattice parameters and the moiré area diverge along curves in the $\theta -  \epsilon_s$ space (red lines). Once again, such divergence is accompanied by the formation of 1D moiré lattices, as the internal angle $\alpha$ tends to $0^{\circ}$ or $180^{\circ}$. Again, the main difference between homo- and heterobilayers is in the vicinity of zero strain and $\theta =0$, where for heterostructures the divergence curves do not cross  (see red lines in the moiré lattice area). The positions of the vertexes of the curves are located at $\pm 0.4\%$, since this strain is necessary to fulfill the mismatch between the lattice parameters of the TMDs.  As a confirmation of the universality of our mathematical description of moir\'e lattices, we show similar behavior for moir\'e patterns formed by rectangular unit cells, such as a homobilayer of WTe$_2$ in section \ref{SM-D}.

In summary, we have presented a general geometrical description of the effect of strain in homo- and heterobilayer systems. We show that heterostrain can be used to form a vast variety of moir\'e lattice geometries, independent of the underlying lattice size and shape. We demonstrate how an initial moir\'e lattice geometry can be tuned into a variety of particular moir\'e patterns, e.g. a hexagonal lattice can be made rectangular or even 1D. 

\section{Outlook}
\label{iv}

Our results are experimentally feasible and can be realized with state of the art strain tuning setups \cite{Iff2019, Wang2020, Hicks2014}. Although heterostrain is beginning to be explored experimentally \cite{edelberg2020tunable, Zhang2022}, most strain tuning experiments focus on homostrain. However, most existing homostrain setups can also achieve heterostrain by clamping one layer to the substrate while leaving the other one mechanically decoupled from the substrate \cite{RN125, Liu2018}. A thorough presentation of the different experimental setups and clamping configurations is given in section \ref{subsec:expmoire}, where we also display how to apply pure biaxial, uniaxial, or shear strain as well as their possible combinations . 

The combination of shear and biaxial strains can become powerful knobs to modify, on demand, the moir\'e pattern size and shape. For example, figure \ref{figura:-7} presents the combined effect of biaxial and shear strain in a generic hexagonal homobilayer stacked with $\theta = 1^{\circ}$. From the condition $\epsilon_c = \epsilon_s =0$, it is possible to change $U/t$, in-situ, by tuning the size of the moir\'e pattern through biaxial strain (point $\epsilon_c = \epsilon_s =0$ to point III or to point IV). On the other hand, light green lines in the plot depict the condition $A'_1 = A'_2$. Varying the shear strain along the line that connects point I with II modifies the ratio $A'_1/A'_2$ which realizes a triangular Hubbard model with two tunable hopping parameters $t_1$ and $t_2$. Finally, the point labeled as $V$ in Fig. \ref{figura:-7} presents a lattice in which the effect described in $\epsilon_c = \epsilon_s = 0$ can be equivalently applied to a perfect square lattice. The fact that all these moir\'e patterns are accessible within one experiment highlights the potential of this novel approach.

Moir\'e straintronics offers a powerful new avenue to explore highly correlated quantum systems. As commonly known, once the sample is fabricated, it is usually not possible to tune the stacking angle. Furthermore, achieving precise rotational alignment within the necessary accuracy required near critical points, such as in magic-angle graphene, is very challenging with state of the art fabrication procedures \cite{lau2022reproducibility}. In contrast to twist angle tuning \cite{RN35}, strain can be applied much more precise with an accuracy corresponding to $0.0001^\circ$ in the twist angle (as shown in section \ref{SM-subsec:expmoire-2}). Hence, the twist angle set during the fabrication can provide a rough alignment knob while heterostrain can be used to fine tune the moir\'e lattice. Further, heterostrain offers an additional mechansim to generate flat bands in graphene \cite{Zhang2022,Huder2018} or tune highly correlated moir\'e quantum materials around critical points in the phase diagram \cite{Pan2020a}. On the other hand, our mathematical framework presents an important starting point for exploring reconstructed moir\'e lattices \cite{edelberg2020tunable,Hu2020, Chen2020, Vizner2021,Yasuda2021,Hao2021}. Finally, the capability to in-situ tune the geometry \textit{and} the interaction strength in highly correlated moir\'e quantum systems is, to the best of our knowledge, unprecedented. We therefore expect strain tuning of moir\'e materials will have a major impact on the exploration of highly interacting quantum systems, from fine-tuning magic-angle graphene to the realization of moir\'e quantum simulators for Luttinger liquids, the Hubbard model, and beyond.

%\cite{Jaoui2022}
%
%##############################################################################
%    Acknowledgements & Contributions
%##############################################################################
%
\section{Acknowledgements}
P.S., A.V.S. and J.J.F. gratefully acknowledge the German Science Foundation (DFG) for financial support via grants FI 947/8-1, DI 2013/5-1 and SPP-2244, as well as the clusters of excellence MCQST (EXS-2111) and e-conversion (EXS-2089). M.K., and B.D.G. are supported by the ERC (grant no. 725920). M.B.-G. acknowledges the Royal Society for support via a University Research Fellowship. B.D.G. acknowledges the Royal Society for a Wolfson Merit Award and the Royal Academy of Engineering for a Chair in Emerging Technology.

\section{Competing financial interests}
The Authors declare no Competing Financial or Non-Financial Interests.
\section{Author contributions}
A.S., M. B.-G., B.G., and J.F. conceived the project. P.S. developed an algorithm for specific strain tunings. M.K. developed the mathematical algorithm for the exact description of a universal Moire strain tuning. P.S. and M.K. interpreted the calculations and wrote the manuscript equally, with contributions from all authors.

\section{Data availability}
The data presented were generated from the mathematical algorithm outlined in the main text. Code for geometrical illustrations of (hetero)strained moir\'e patterns with the calculated moir\'e lattice as those presented Fig.~\ref{figura:-7} can be found in \cite{ESM}.

\clearpage
\onecolumngrid

\section{Additional information}

\subsection{Calculation of the Moir\'e lattice vectors}
\label{SM-S1-1}

The moir\'e lattice vector is calculated via the reciprocal space. First, we start by calculating the reciprocal lattice vectors $\mathbf{g'}_i$ of the lower layer and $\mathbf{h'}_i$ of the upper layer which are
\begin{equation}
\begin{split}
\mathbf{g'}_i &=\frac{2\pi}{\|\mathbf{a'}_1^T \cdot R_{\pi/2}\cdot \mathbf{a'}_2\|} R_{\pi/2} \cdot \mathbf{a'}_j\\
\mathbf{h'}_i &=\frac{2\pi}{\|\mathbf{b'}_1^T \cdot R_{\pi/2}\cdot \mathbf{b'}_2\|} R_{\pi/2} \cdot \mathbf{b}'_j,
\end{split}
\end{equation}
where $j\neq i$. These expressions can be simplified by using equation~\ref{eq:norm1} (See sub-calculation \ref{SUB3}) such that they take the form
\begin{equation}
\begin{split}
\mathbf{g'}_i&=\frac{2\pi}{a_i \|\left((1+\epsilon_c)^2-\epsilon_s^2\right)\cos(\beta+\frac{\pi}{2})\|} R_{\pi/2} \cdot \left(\mathbb{I}+\epsilon \right) \cdot R_{\psi_j}\cdot \begin{pmatrix} 1\\ 0 \end{pmatrix}\\
\mathbf{h'}_i&=\frac{2\pi}{a_i(1+\delta) \|\left((1+\mu\epsilon_c)^2-\mu^2\epsilon_s^2\right)\cos(\theta+\beta+\frac{\pi}{2}-\theta)\|} R_{\pi/2} \cdot \left(\mathbb{I}+\mu\epsilon \right) \cdot R_{\psi_j}\cdot \begin{pmatrix} 1\\ 0 \end{pmatrix}.
\end{split}
\end{equation}
With the definitions $c_1=(1+\epsilon_c)^2-\epsilon_s^2$ and  $c_\mu=(1+\mu\epsilon_c)^2-\mu^2\epsilon_s^2$, the reciprocal lattice vectors can be written as
\begin{equation}
\begin{split}
\mathbf{g'}_i&=\frac{2\pi}{a_i c_1 \cos(\beta+\frac{\pi}{2})} R_{\pi/2} \cdot \left(\mathbb{I}+\epsilon \right) \cdot R_{\psi_j}\cdot \begin{pmatrix} 1\\ 0 \end{pmatrix}\\
\mathbf{h'}_i&=\frac{2\pi}{a_i(1+\delta) c_\mu \cos(\beta+\frac{\pi}{2})} R_{\pi/2} \cdot \left(\mathbb{I}+\mu\epsilon \right) \cdot R_{\psi_j}\cdot \begin{pmatrix} 1\\ 0 \end{pmatrix}.
\end{split}
\end{equation}
Knowing $\mathbf{g'}_i$ and $\mathbf{h'}_i$, it is possible to calculate the reciprocal moir\'e lattice vectors $\mathbf{G'}_i$ by subtracting the individual lattice vectors from one another such that
\begin{equation}
\begin{split}
\mathbf{G'}_i &=\mathbf{g'}_i-\mathbf{h'}_i \\
\mathbf{G'}_i&=\frac{2\pi}{a_i(1+\delta) c_1 \cos(\beta+\frac{\pi}{2})} R_{\pi/2} \cdot \left[\left(\mathbb{I}+\epsilon \right)(1+\delta) \cdot R_{\psi_j}-\left(\mathbb{I}+\mu\epsilon \right)\frac{c_1}{c_\mu} \cdot R_{\psi_j+\theta}\right] \cdot  \begin{pmatrix} 1\\ 0 \end{pmatrix}.
\end{split}
\end{equation}
Finally the moir\'e lattice vectors $\mathbf{A'}_i$ is calculated by transforming $\mathbf{G'}_i$ into real space. Hence $\mathbf{A'}_i$ becomes
\begin{equation}
\label{eq:moirevectorcal}
\begin{split}
\mathbf{A'}_i &= \frac{2\pi}{\|\mathbf{G'}_1^T R_{\pi/2}\mathbf{G'}_2\|} R_{\pi/2}\cdot\mathbf{G'}_i \\
&=\frac{(2\pi)^2 R_{\pi} \cdot \left[\left(\mathbb{I}+\epsilon \right)(1+\delta) \cdot R_{\psi_i}-\left(\mathbb{I}+\mu\epsilon \right)\frac{c_1}{c_\mu} \cdot R_{\psi_i+\theta}\right]}{a_i(1+\delta) c_1 \cos(\beta+\frac{\pi}{2}) \|\mathbf{G'}_1^T R_{\pi/2}\mathbf{G'}_2\|}  \cdot  \begin{pmatrix} 1\\ 0 \end{pmatrix}.
\end{split}
\end{equation}
By applying equation~\ref{eq:cal4} (See sub-calculation \ref{SUB4}) and reordering, the real space moir\'e lattice vectors take the form
\begin{equation}
\label{eq:realmoirevector}
\begin{split}
\mathbf{A'}_i &=\frac{a_i(1+\delta)}{\Delta} \\
&\cdot\left[(1+\delta)c_\mu \left((1+\epsilon_c)R_{\psi_i}+ \epsilon_s S_{\phi_s+\psi_i}\right)-c_1\left((1+\mu \epsilon_c) R_{\psi_i+\theta}+\mu\epsilon_s S_{\phi_s+\psi_i+\theta}\right) \right]\cdot \begin{pmatrix} 1\\ 0 \end{pmatrix}.
\end{split}
\end{equation}
where the denominator $\Delta$ is 
\begin{equation}
\Delta=c_\mu (1+\delta)^2+c_1-2(1+\delta)\left[(1+\epsilon_c +\mu\epsilon_c)+\mu(\epsilon_c^2-\epsilon_s^2)\right]\cos(\theta).
\end{equation}
The equation for the real space Moir\'e vector can be simplified to:
\begin{equation}
\begin{split}
\mathbf{A'}_i &=\frac{a_j(1+\delta)}{\Delta} \\
&\cdot\left[(1+\delta)c_\mu \left((1+\epsilon_c)\mathbb{I}+ \epsilon_s S_{\phi_s}\right)-c_1\left((1+\mu \epsilon_c) R_{\theta}+\mu\epsilon_s S_{\phi_s+\theta}\right) \right]\cdot R_{\psi_i} \cdot\begin{pmatrix} 1\\ 0 \end{pmatrix}.
\end{split}
\end{equation}

\paragraph{Sub-Calculation 1}\mbox{}\\
In this paragraph, the expression $\left(\mathbb{I}+\alpha \epsilon)\mathbb{R}_{\pi/2}(\mathbb{I}+\beta \epsilon\right)$ is calculated. Here 
\begin{equation}
\epsilon = \epsilon_c \mathbb{I}+ \epsilon_s \left( \cos(\phi_s)\sigma_x + \sin(\phi_s)\sigma_z \right),
\end{equation}
where $\sigma_{x}$($\sigma_{z}$) is the $x$($z$) Pauli matrix and $\epsilon_c$, $\epsilon_s$ and $\phi_s$ are the parameters defined in the main text. The parameters $\alpha \in \mathbb{R}$ and $\beta \in \mathbb{R}$ are introduced to generalize the calculation. Therefore, the matrix 
\begin{equation}
\begin{split}
A :&=\left(\mathbb{I}+\alpha \epsilon)R_{\pi/2}(\mathbb{I}+\beta \epsilon\right) \\
&=R_{\pi/2}+\alpha \epsilon R_{\pi/2}+\beta R_{\pi/2} \epsilon +\alpha \beta \epsilon R_{\pi/2} \epsilon 
\end{split}
\end{equation}
calculates the desired value. In the next step, $R_{\pi/2}$ is rewritten in terms of the Pauli matrices $\sigma_y$ by 
\begin{equation}
R_{\pi/2}=\begin{pmatrix}
\cos(\pi/2) & -\sin(\pi/2) \\ \sin(\pi/2) & \cos(\pi/2) 
\end{pmatrix}=\begin{pmatrix}
0 & -1 \\ 1 & 0
\end{pmatrix}=-\mathrm{i}\begin{pmatrix}
0 & -\mathrm{i} \\ \mathrm{i} & 0
\end{pmatrix}=-\mathrm{i} \sigma_y
\end{equation}
where $\mathrm{i}$ is the imaginary unit. Introducing $\mathrm{i}$ into the calculation of $A$ gives
\begin{equation}
\begin{split}
A=(-\mathrm{i}\sigma_y)&+\alpha \epsilon (-\mathrm{i}\sigma_y)+\beta (-\mathrm{i}\sigma_y) \epsilon +\alpha \beta \epsilon (-\mathrm{i}\sigma_y) \epsilon \\
=(-\mathrm{i}\sigma_y)&+\alpha \left[\epsilon_c \mathbb{I}+\epsilon_s (\cos (\phi_s)\sigma_x+\sin(\phi_s)\sigma_z)\right]\cdot(-\mathrm{i}\sigma_y)\\
&+\beta (-\mathrm{i}\sigma_y)\cdot \left[\epsilon_c \mathbb{I}+\epsilon_s (\cos (\phi_s)\sigma_x+\sin(\phi_s)\sigma_z)\right]\\
&+\alpha\beta \left[\epsilon_c \mathbb{I}+\epsilon_s (\cos (\phi_s)\sigma_x+\sin(\phi_s)\sigma_z)\right]\cdot(-\mathrm{i}\sigma_y) \\
&\cdot \left[\epsilon_c \mathbb{I}+\epsilon_s (\cos (\phi_s)\sigma_x+\sin(\phi_s)\sigma_z)\right]\\
=(-\mathrm{i}\sigma_y)&+\epsilon_c(\alpha+\beta)(-\mathrm{i}\sigma_y)\\
&+\alpha \epsilon_s (\cos(\phi_s)\sigma_z-\sin(\phi_s)\sigma_x)+\beta \epsilon_s (-\cos(\phi_s)\sigma_z+\sin(\phi_s)\sigma_x)\\
&+\alpha\beta \left[\epsilon_c \mathbb{I}+\epsilon_s (\cos (\phi_s)\sigma_x+\sin(\phi_s)\sigma_z)\right]\\
&\cdot \left[\epsilon_c (-\mathrm{i}\sigma_y)+\epsilon_s (\cos (\phi_s)(-\sigma_z)+\sin(\phi_s)\sigma_x\right]\\
=(-\mathrm{i}\sigma_y)&+\epsilon_c(\alpha+\beta)(-\mathrm{i}\sigma_y)+\epsilon_s(\alpha-\beta) \left[\cos (\phi_s)\sigma_z-\sin(\phi_s)\sigma_x\right ]\\
&+\alpha\beta(-\mathrm{i}\sigma_y)\epsilon_c^2 \\
&+\alpha\beta\epsilon_s\epsilon_c\left[\cos (\phi_s)(-\sigma_z)+\sin(\phi_s)\sigma_x +\cos (\phi_s)\sigma_z-\sin(\phi_s)\sigma_x\right]\\
&+\alpha\beta\epsilon_s^2\left[ \mathrm{i}\sigma_y \cos^2(\phi_s)+\cos(\phi_s)\sin(\phi_s)(\sigma_x^2-\sigma_z^2)+\mathrm{i}\sigma_y\sin^2(\phi_s)\right]\\
&=(1+\epsilon_c(\alpha+\beta))(-\mathrm{i}\sigma_y)+\epsilon_s(\alpha-\beta) \left[\cos (\phi_s)\sigma_z-\sin(\phi_s)\sigma_x\right]+\alpha\beta(-\mathrm{i}\sigma_y)(\epsilon_c^2-\epsilon_s^2),
\end{split}
\end{equation}
where the identity $\sigma_i\sigma_j=\delta_{ij}\mathbb{I}+i\epsilon_{ijk}\sigma_k$ of the Pauli matrices is used. The shear angle matrix 
\begin{equation}
\label{eq:shearanglematrix}
S_{\phi_s}=\cos(\phi_s)\sigma_z-\sin(\phi_s)\sigma_x
\end{equation}
is introduced to further simplify the expression. Hence $A$ takes the form
\begin{equation}
\label{eq:cal1}
\begin{split}
A &=\left(\mathbb{I}+\alpha \epsilon)+R_{\pi/2}(\mathbb{I}+\beta \epsilon\right)\\
&=(1+\epsilon_c(\alpha+\beta)+\alpha\beta(\epsilon_c^2-\epsilon_s^2))R_{\pi/2}+\epsilon_s(\alpha-\beta)S_{\phi_s}.
\end{split}
\end{equation}

\paragraph{Sub-calculation 2}\mbox{}\\
In this paragraph the properties of the shear matrix $S_{\phi_s}$, as defined in equation~\ref{eq:shearanglematrix}, are calculated when multiplied with a rotation matrix. This leads to 
\begin{equation}
\begin{split}
S_{\phi_s}\cdot R_\alpha&=\begin{pmatrix}
\cos(\phi_s) & -\sin(\phi_s) \\ -\sin(\phi_s) & -\cos(\phi_s) 
\end{pmatrix}\cdot\begin{pmatrix}
\cos(\alpha) & -\sin(\alpha) \\ \sin(\alpha) & \cos(\alpha) 
\end{pmatrix}\\
&=\begin{pmatrix}
\cos(\phi_s+\alpha) & -\sin(\phi_s+\alpha) \\ -\sin(\phi_s+\alpha) & -\cos(\phi_s+\alpha) 
\end{pmatrix}=S_{\phi_s+\alpha}
\end{split}
\end{equation} 
and
\begin{equation}
\begin{split}
R_\alpha\cdot S_{\phi_s} &=\begin{pmatrix}
\cos(\alpha) & -\sin(\alpha) \\ \sin(\alpha) & \cos(\alpha) 
\end{pmatrix}\cdot\begin{pmatrix}
\cos(\phi_s) & -\sin(\phi_s) \\ -\sin(\phi_s) & -\cos(\phi_s) 
\end{pmatrix}\\
&=\begin{pmatrix}
\cos(\phi_s-\alpha) & -\sin(\phi_s-\alpha) \\ -\sin(\phi_s-\alpha) & -\cos(\phi_s-\alpha) 
\end{pmatrix}=S_{\phi_s-\alpha}
\end{split}
\end{equation}
where we use $\sin(\alpha\pm\beta)=\sin(\alpha)\cos(\beta)\pm\sin(\beta)\cos(\alpha)$ and $\cos(\alpha\pm\beta)=\cos(\alpha)\cos(\beta)\mp\sin(\alpha)\sin(\beta)$. Altogether this leads to the properties of the shear angle matrix
\begin{equation}
\label{eq:shearanglematrixprop}
\begin{split}
S_{\phi_s} \cdot R_\alpha &=S_{\phi_s+\alpha}\\
R_\alpha \cdot S_{\phi_s} &=S_{\phi_s-\alpha}.
\end{split}
\end{equation}
\paragraph{Sub-calculation 3}\mbox{}\label{SUB3}\\
This paragraph calculates the expression $B$ which is the inner product between 2 different hexagonal lattice vectors 
\begin{equation}
B:=\begin{pmatrix} 1 & 0\end{pmatrix}\cdot R_{\gamma}^T\cdot\left(\mathbb{I}+\alpha \epsilon)\mathbb{R}_{\pi/2}(\mathbb{I}+\beta \epsilon\right)\cdot R_{\delta} \cdot \begin{pmatrix} 1 \\ 0\end{pmatrix}.
\end{equation}
Using equation~\ref{eq:cal1}, $B$ can be simplified to 
\begin{equation}
\begin{split}
B= (1+\epsilon_c(\alpha+\beta)+\alpha\beta(\epsilon_c^2-\epsilon_s^2)) & \cdot\left(\begin{pmatrix} 1 & 0\end{pmatrix}\cdot R_{\gamma}^T \cdot R_{\pi/2}\cdot R_{\delta} \cdot \begin{pmatrix} 1 \\ 0\end{pmatrix}\right) \\ 
+\epsilon_s(\alpha-\beta) & \cdot\left(\begin{pmatrix} 1 & 0\end{pmatrix}\cdot R_{\gamma}^T \cdot S_{\phi_s}\cdot R_{\delta} \cdot \begin{pmatrix} 1 \\ 0\end{pmatrix}\right)
\end{split}
\end{equation}
which can be calculated, using equation~\ref{eq:shearanglematrixprop} and using $R_\gamma\cdot R_\delta=R_{\gamma+\delta}$, to the form
\begin{equation}
\label{eq:norm1}
\begin{split}
B= \left(1+\epsilon_c(\alpha+\beta)+\alpha\beta(\epsilon_c^2-\epsilon_s^2)\right) \cos(\delta+\frac{\pi}{2}-\gamma) 
+\epsilon_s(\alpha-\beta) \cos(\delta+\frac{\pi}{2}+\gamma).
\end{split}
\end{equation}
\paragraph{Sub-calculation 4}\mbox{}\label{SUB4}\\
In this part, the inner product between the two Moir\'e pattern reciprocal lattice vectors $\mathbf{G}_1$ and $\mathbf{G}_2$ while $\mathbf{G}_2$ is turned by $90^\circ$. The calculated expression is $C=\mathbf{G'}_1^T R_{\pi/2}\mathbf{G'}_2$. In a first step, $C$ is rewritten in the form
\begin{equation}
\begin{split}
C=\mathbf{G'}_1^T R_{\pi/2}&\mathbf{G'}_2\\
=\left(\frac{2\pi}{a(1+\delta) c_1 \cos(\beta+\frac{\pi}{2})}\right)^2 &\cdot \begin{pmatrix} 1& 0 \end{pmatrix}   \cdot \left[R_{\psi_1}^T\cdot\left(\mathbb{I}+\epsilon \right)(1+\delta) +R_{\psi_1+\theta}^T\cdot\left(\mathbb{I}+\mu\epsilon \right)\frac{c_1}{c_\mu} \right] \\
&\cdot R_{\pi/2} \cdot \left[\left(\mathbb{I}+\epsilon \right)(1+\delta) \cdot R_{\psi_2}+\left(\mathbb{I}+\mu\epsilon \right)\frac{c_1}{c_\mu} \cdot R_{\psi_2+\theta}\right] \cdot  \begin{pmatrix} 1\\ 0 \end{pmatrix} 
\end{split}
\end{equation}
such that it can be calculated. To simplify the calculation, the temporary constant
\begin{equation}
\kappa=\left(\frac{2\pi}{a(1+\delta) c_1 \cos(\beta+\frac{\pi}{2})}\right)
\end{equation}
is introduced such that 
\begin{equation}
\begin{split}
C&=\underbrace{\kappa^2\begin{pmatrix} 1& 0 \end{pmatrix}   \cdot \left[R_{\psi_1}^T\cdot\left(\mathbb{I}+\epsilon \right)(1+\delta) \right] \cdot R_{\pi/2} \cdot \left[\left(\mathbb{I}+\epsilon \right)(1+\delta) \cdot R_{\psi_2}\right]\cdot  \begin{pmatrix} 1\\ 0 \end{pmatrix}}_{C_1}\\
&+\underbrace{\kappa^2\begin{pmatrix} 1& 0 \end{pmatrix}   \cdot \left[R_{\psi_1+\theta}^T\cdot\left(\mathbb{I}+\mu\epsilon \right)\frac{c_1}{c_\mu} \right] \cdot R_{\pi/2} \cdot \left[\left(\mathbb{I}+\mu\epsilon \right)\frac{c_1}{c_\mu} \cdot R_{\psi_2+\theta}\right]\cdot  \begin{pmatrix} 1\\ 0 \end{pmatrix}}_{C_2}\\
&-\underbrace{\kappa^2\begin{pmatrix} 1& 0 \end{pmatrix}   \cdot \left[R_{\psi_1+\theta}^T\cdot\left(\mathbb{I}+\mu\epsilon \right)\frac{c_1}{c_\mu} \right] \cdot R_{\pi/2} \cdot \left[\left(\mathbb{I}+\epsilon \right)(1+\delta) \cdot R_{\psi_2}\right]\cdot  \begin{pmatrix} 1\\ 0 \end{pmatrix}}_{C_3}\\
&-\underbrace{\kappa^2\begin{pmatrix} 1& 0 \end{pmatrix}   \cdot \left[R_{\psi_1}^T\cdot\left(\mathbb{I}+\epsilon \right)(1+\delta) \right] \cdot R_{\pi/2} \cdot \left[\left(\mathbb{I}+\mu\epsilon \right)\frac{c_1}{c_\mu} \cdot R_{\psi_2+\theta}\right]\cdot  \begin{pmatrix} 1\\ 0 \end{pmatrix}}_{C_4}.
\end{split}
\end{equation}
The different parts of the equation can then be evaluated. The first part $C_1$ can be calculated using Eq.~\ref{eq:norm1} such that 
\begin{equation}
C_1=\kappa^2 (1+\delta)^2 \left((\epsilon_c+1)^2-\epsilon_s^2\right)\cos(\frac{\pi}{2}-\beta)=\kappa^2 c_1 \cos(\frac{\pi}{2}-\beta)(1+\delta)^2.
\end{equation}
The next part of the equation $C_2$ can be calculated similar to $C_1$ which makes
\begin{equation}
C_2=\kappa^2 \frac{c_1^2}{c_\mu} \cos(\frac{\pi}{2}-\beta). 
\end{equation} 
The terms $C_3$ and $C_4$ are a little bit different and it is actually useful to calculate their sum, which gives
\begin{equation}
\begin{split}
C_3+C_4 &= \kappa^2\frac{c_1}{c_\mu}\cdot\left((1+\epsilon_c+\mu \epsilon_c) +\mu (\epsilon_c^2-\epsilon_s^2)\right)\left[\cos(\frac{\pi}{2}-\beta+\theta)+\cos(\frac{\pi}{2}-\beta-\theta)\right]\\
&+\kappa^2\frac{c_1}{c_\mu}\cdot\left[\epsilon_s (\mu-1)\cos(\psi_1+\psi_2+\phi_s)+\epsilon_s (1-\mu)\cos(\psi_1+\psi_2+\phi_s)\right]\\
&=\kappa^2\frac{c_1}{c_\mu}\left((1+\epsilon_c+\mu \epsilon_c) +\mu (\epsilon_c^2-\epsilon_s^2)\right)2\cos(\frac{\pi}{2}-\beta)\cos(\theta)
\end{split}
\end{equation}
Unifying all the terms of the equation the expression $C=\mathbf{G'}_1^T R_{\pi/2}\mathbf{G'}_2$ can be simplified 
\begin{equation}
\label{eq:cal4}
\begin{split}
C&=\mathbf{G'}_1^T R_{\pi/2}\mathbf{G'}_2\\
&=\kappa^2 \cos(\frac{\pi}{2}-\beta)\frac{c_1}{c_\mu}\left(c_\mu (1+\delta)^2 +c_1-2 (1+\delta)\left((1+\epsilon_c+\mu \epsilon_c) +\mu (\epsilon_c^2-\epsilon_s^2)\right)\cos(\theta)\right)
\end{split}
\end{equation}

\subsubsection{Size of the Moir\'e lattice}
\label{SM-S1-2}

The size of the Moir\'e lattice, $M$, is calculated in the following way:
\begin{equation}
M=\|\mathbf{A'}_1 \times \mathbf{A'}_2\|=\det\begin{pmatrix} \mathbf{A'}_1 & \mathbf{A'}_2 \end{pmatrix}  = \begin{vmatrix} A'_{11} & A'_{12}  \\
A'_{21}  & A'_{22} 
\end{vmatrix}= \det A.
\end{equation}
Using equation~5 the matrix $A$ can be rewritten to
\begin{equation}
\begin{split}
A&=\frac{(1+\delta)}{\Delta}\cdot\underbrace{\left[(1+\delta)c_\mu \left((1+\epsilon_c)\mathbb{I}+ \epsilon_s S_{\phi_s}\right)-c_1\left((1+\mu \epsilon_c) R_{\theta}+\mu\epsilon_s S_{\phi_s+\theta}\right) \right]}_S\\
&\cdot \begin{pmatrix}
R_{\psi_2} \cdot\begin{pmatrix} a_1\\ 0 \end{pmatrix} & R_{\psi_i} \cdot\begin{pmatrix} a_2\\ 0 \end{pmatrix}
\end{pmatrix}.
\end{split}
\end{equation}
Hence the Moir\'e size can be rewritten in the form 
\begin{equation}
M=\det A = a_1a_2\left(\frac{(1+\delta)}{\Delta}\right)^2 \cdot \det S \cdot \begin{vmatrix}
\cos(\theta_0)& \cos(\theta_0+\beta)  \\
\sin(\theta_0)  & \sin(\theta_0+\beta) 
\end{vmatrix}
\end{equation}
and 
\begin{equation}
\begin{split}
\begin{vmatrix}
\cos(\theta_0)& \cos(\theta_0+\beta)  \\
\sin(\theta_0)  & \sin(\theta_0+\beta) 
\end{vmatrix}&=\cos(\theta_0)\sin(\theta_0+\beta)-\cos(\theta_0+\beta)\sin(\theta_0)\\
&=\sin(\theta_0+\beta-\theta_0)=\sin(\beta)
\end{split}
\end{equation}
which simplifies $M$ to 
\begin{equation}
M=a_1a_2\sin(\beta)\left(\frac{(1+\delta)}{\Delta}\right)^2 \cdot \det S.
\end{equation}
$\det S$ is calculated using some linear algebra such that 
\begin{equation}
\det S=c_\mu c_1 \Delta,
\end{equation}
which leads to the final form of $M$
\begin{equation}
\label{eq:calMoiresize}
M=a_1a_2\sin(\beta)\frac{(1+\delta)^2 }{\Delta}\cdot c_\mu c_1
\end{equation}

\subsubsection{Moir\'e lattice shape without applied shear strain}
\label{SM-S1-4}

In this section we prove that the moir\'e lattice $\mathbf{A_i'}$ has the same geometry as the underlying lattice $\mathbf{a_i}$ when $\epsilon_s=0$. To prove that the moir\'e unit cell has the same geometry, we start by proving that $\frac{a_1}{a_2}\stackrel{\epsilon_s=0}{=}\frac{\|\mathbf{A'}_1\|}{\|\mathbf{A'}_2\|}$. First, the length of the moir\'e vector is calculated 
\begin{equation}
\label{eq:noshearstrainmoire}
\begin{split}
\mathbf{A'}_i &\stackrel{\epsilon_s=0}{=}\frac{a_i(1+\delta)}{\Delta} \cdot[(1+\delta)c_\mu (1+\epsilon_c)\mathbb{I}-c_1(1+\mu \epsilon_c) R_{\theta} ]\cdot R_{\psi_i} \cdot\begin{pmatrix} 1\\ 0 \end{pmatrix}\\
&=\frac{a_i(1+\delta)}{\Delta}\cdot R_{\psi_i}\cdot \underbrace{\left[(1+\delta)c_\mu (1+\epsilon_c)\mathbb{I}-c_1(1+\mu \epsilon_c) R_{\theta}\right] \begin{pmatrix} 1\\ 0 \end{pmatrix}}_{\mathbf{v}}.
\end{split}
\end{equation}
where we use the property that the rotation matrix and the unit matrix commute: $[R_{\theta},R_{\psi_i}]=[\mathbf{1},R_{\psi_i}]=0$. With this result we now calculate the size of the Moir\'e lattice vector
\begin{equation}
\|\mathbf{A'}_i\|=\sqrt{\mathbf{A'}_i^T\cdot\mathbf{A'}_i}=\frac{a_i(1+\delta)}{\Delta}\sqrt{\mathbf{v}^T \cdot R_{\psi_i}^T R_{\psi_i}\cdot\mathbf{v}}=\frac{a_i(1+\delta)}{\Delta}\sqrt{\mathbf{v}^T \cdot \mathbf{v}}.
\end{equation}
Using this expression we are now able to prove that
\begin{equation}
\frac{\|\mathbf{A'}_1\|}{\|\mathbf{A'}_2\|}\stackrel{\epsilon_s=0}{=}\frac{\frac{a_1(1+\delta)}{\Delta}\sqrt{\mathbf{v}^T \cdot \mathbf{v}}}{\frac{a_2(1+\delta)}{\Delta}\sqrt{\mathbf{v}^T \cdot \mathbf{v}}}=\frac{a_1}{a_2}.
\end{equation}
To show that the moir\'e pattern has the same geometry as the underlying lattice we need to prove that the angle between the moir\'e vectors $\alpha$ is equal to the angle between the underlying  lattice vectors $\beta$. The angle $\alpha$ is calculated 
\begin{equation}
\sin{\alpha}=\frac{M}{\|\mathbf{A'}_1\|\|\mathbf{A'}_1\|}\stackrel{\epsilon_s=0}{=}\frac{a_1a_2\sin(\beta)\frac{(1+\delta)^2 }{\Delta}\cdot c_\mu c_1}{\frac{a_1a_2(1+\delta)^2}{\Delta^2}\mathbf{v}^T \cdot \mathbf{v}}=\sin{\beta}\frac{\Delta \cdot c_\mu c_1}{\mathbf{v}^T \cdot \mathbf{v}}.
\end{equation}
With some algebra the product $\mathbf{v}^T \cdot \mathbf{v}$ can be simplified to
\begin{equation}
\mathbf{v}^T \cdot \mathbf{v} \stackrel{\epsilon_s=0}{=} c_\mu c_1 \left((1+\delta)^2 (1+\epsilon_c)^2\frac{c_\mu}{c_1}+\frac{c_1}{c_\mu}(1+\mu \epsilon_c)^2-2(1+\delta)(1+\epsilon_c)(1+\mu \epsilon_c)\cos{\theta} \right). 
\end{equation}
Finally, the angle $\alpha$ can be calculated to be
\begin{equation}
\begin{split}
\sin{\alpha}&\stackrel{\epsilon_s=0}{=}\sin{\beta}\frac{\Delta}{(1+\delta)^2 (1+\epsilon_c)^2\frac{c_\mu}{c_1}+\frac{c_1}{c_\mu}(1+\mu \epsilon_c)^2-2(1+\delta)(1+\epsilon_c)(1+\mu \epsilon_c)\cos{\theta}}\\
&\stackrel{\epsilon_s=0}{=}\sin{\beta}\frac{(1+\delta)^2 (1+\mu\epsilon_c)^2+(1+\epsilon_c)^2-2(1+\delta)(1+\epsilon_c+\mu \epsilon_c+\mu\epsilon_c^2)\cos{\theta}}{(1+\delta)^2 (1+\mu\epsilon_c)^2+(1+\epsilon_c)^2-2(1+\delta)(1+\epsilon_c+\mu \epsilon_c+\mu\epsilon_c^2)\cos{\theta}}\\
&\stackrel{\epsilon_s=0}{=}\sin{\beta}.
\end{split}
\end{equation}
Altogether, this shows that the geommetry of the moir\'e lattice is the same as the underlying lattice in case there is no shear strain $\epsilon_s=0$.

\subsubsection{Moir\'e lattice orientation in absence of shear strain}
\label{SM-S1-3}

In this section we calculate the orientation of the moir\'e lattice as function of stacking angle and biaxial strain. As in the previous section, the angle $\kappa$ between the underlying lattice $\mathbf{a_i}$ and the moir\'e vector $\mathbf{A_i'}$ can be calculated via
\begin{equation}
\tan(\kappa)\stackrel{\epsilon_s=0}{=}\frac{\sin(\gamma}{\cos(\gamma)}=\frac{\frac{\mathbf{a_i}\times \mathbf{A_i}}{a_i A_i}}{\frac{\mathbf{a_i}\cdot \mathbf{A_i}}{a_i A_i}}=\frac{\mathbf{a_i}\times \mathbf{A_i}}{\mathbf{a_i}\cdot \mathbf{A_i}}.
\end{equation}
The equation can be simplified to 
\begin{equation}
\tan(\kappa)\stackrel{\epsilon_s=0}{=}\frac{a_i^2 \frac{1+\delta}{\Delta}\sin(\theta)(1+\mu\epsilon_c) c_1}{a_i^2 \frac{1+\delta}{\Delta}\left( (1+\delta)(1+\mu\epsilon_c)^2(1+\epsilon_c)-(1+\epsilon_c)^2\cdot \cos(\theta)\right)}=\frac{\sin(\theta)}{(1+\delta)\frac{1+\mu\epsilon_c}{1+\epsilon_c}-\cos(\theta)}
\end{equation}
which is the final form for the direction of the moir\'e lattice with respect to the underlying lattice.

\subsection{Experimental feasibility of moir\'e strain tuning setup}
\label{SM-Exp}

%==FIGURA======FIGURA======FIGURA======FIGURA======FIGURA======FIGURA=====
\begin{figure}[t!!]
\includegraphics*[keepaspectratio=true, clip=true, angle=0, width=1.\columnwidth, trim={15mm, 1mm, 20mm, 1mm}]{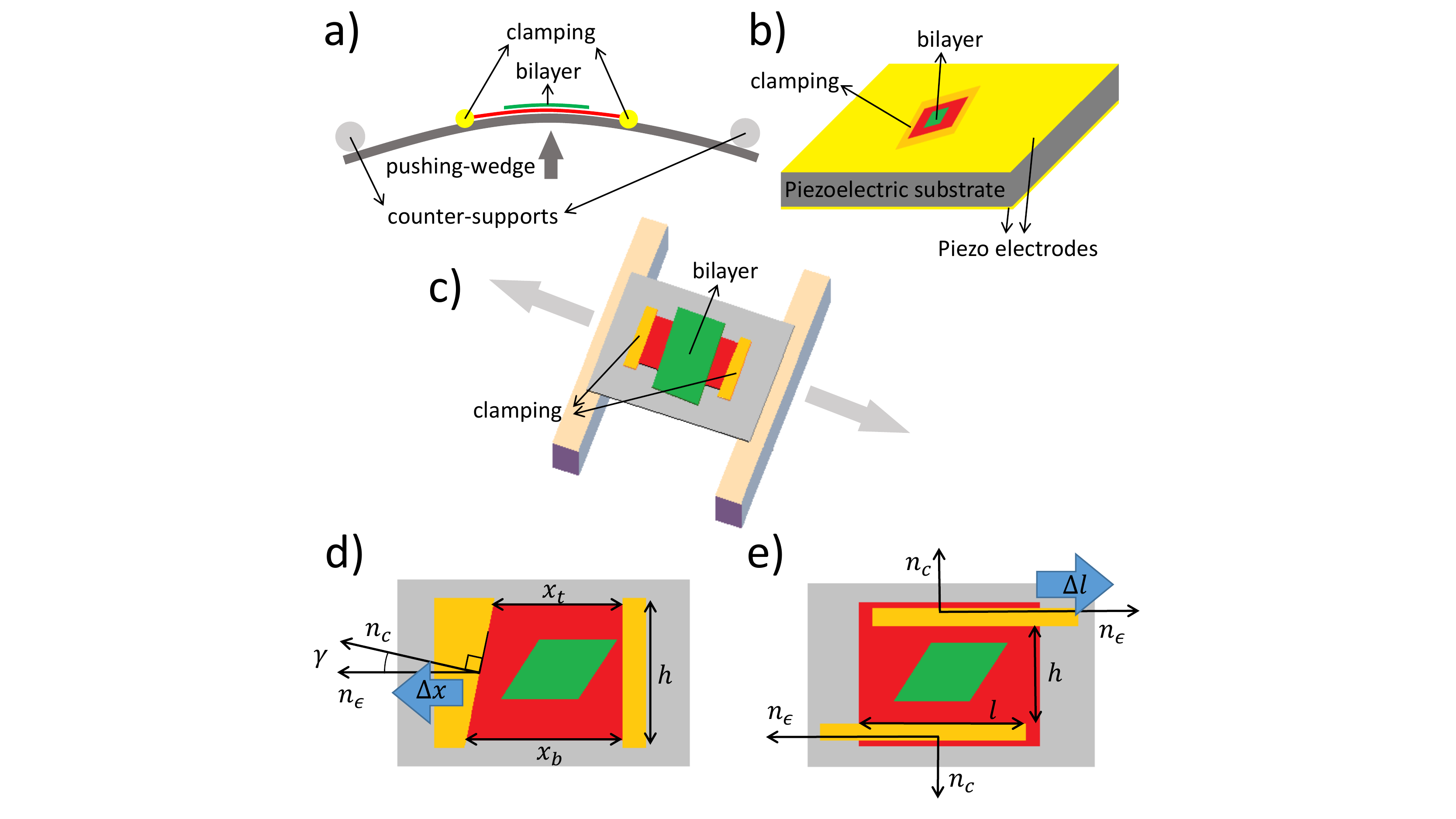}
\caption{Sketch of possible heterostrain devices and clamping techniques to apply strain with different shear strain to biaxial strain ratios. Metal evaporation (yellow) clamps the lower layer (red) to ensure the strain transmission. The upper layer (green) is not adhered to the substrate. \textbf{a)} Counter supports and a pushing-wedge setup to bend the substrate and transmit the uniaxial elongation-contraction to the sample \cite{RN125, Guan2015, Wang2020, McRae2019, Wang2020f, Guan2017}.  \textbf{b)}Biaxial strain tuning device provided by the in-plane expansion/contraction of the substrate \cite{RN83, Aslan2018, Iff2019, Hui2013}.  \textbf{c)} Uniaxial strain cell setups using ductile substrates \cite{McRae2019, edelberg2020tunable, Cenker2022, Kuklewicz2012, Seidl2006}. \textbf{d)} Uniaxial strain with angled clamp to tune $\phi_s$. The lower layer is clamped perpendicularly to the uniaxial strain direction ($n_{\epsilon}$) while the left side is clamped with an angle $\gamma$ respect to $n_{\epsilon}$. \textbf{e)} Clamping to apply pure shear strain via perpendicular clamps. The clamps normal ($n_c$) is perpendicular to the applied strain direction ($n_{\epsilon}$).}
\label{figura:00}
\end{figure}
%==FIGURA======FIGURA======FIGURA======FIGURA======FIGURA======FIGURA=====

In contrast to in situ tuning of $\theta$ \cite{Yao2021,Hu2022,RN33}, strain tuning allows a more diverse variation of the moir\'e size and shape. In this section, we discuss the topic from the experimental point of view.

\subsubsection{How to avoid atomic reconstruction in strain tuning experiments}

Along our discussion, we have assumed the individual layers of the 2D moir\'e sample as rigid objects, however, to reach an experimental realization, it is important to take atomic reconstruction into account. It is known that, for samples with small stacking angles, atoms within each layer try to adjust the stacking landscape forming commensurate regions \cite{Woods2014, RN88, edelberg2020tunable, lin2021large}. This effect increases interlayer adhesion and leads to higher friction between layers that hinders the possibility to modify the atomic registry \cite{Ru2020, Song2018a,RN35}. Conversely, higher stacking angles reduce the reconstruction effects and avoid strain transfer between layers ($\mu \neq 0$). For MoSe$_2$-WSe$_2$ heterostructures, these critical angles are $2.5^\circ$ and $\sim 1.0^\circ$ for 2H and 3R stacking, respectively \cite{RN88, Andersen2021, Enaldiev2020}, and increases to $\sim 3.0^\circ$ in the case of 2H homobilayers \cite{lin2021large}. Hence, it is possible to avoid atomic reconstruction by using a stacking angle above the critical angle of the bilayer.

On the other hand, while atomic reconstruction is always present, heterostrain can also reduce interlayer interaction and prevents atomic reconstruction \cite{Wang2019b}, as it has been shown that $\epsilon_u \sim 3\%$ can transform a totally commensurate moir\'e lattice ($\theta=0^\circ$) into a partially incommensurate structure with reduced atomic reconstruction \cite{edelberg2020tunable}.

\subsubsection{Experimental moir\'e strain tuning setups}
\label{subsec:expmoire}
So far, 2D crystal strain tuning experiments focus on homostrain and, to the best of our knowledge, there is only one experimental heterostrain realization \cite{edelberg2020tunable}. However, most existing setups can achieve the heterostrain condition in samples with low friction between adjacent layers by clamping one layer to the substrate via metal evaporation while leaving the other one mechanically decoupled from the substrate\cite{RN125, Liu2018}. For instance, biaxial strain on a 2D heterostructure can be achieved by mounting it on top of a biaxially strained substrate, as presented in figure \ref{figura:00}b. The substrate is either a polymer which elongates by thermal expansion \cite{RN83} or a piezoelectric crystal which expands according to an applied voltage\cite{Aslan2018,Iff2019,Hui2013}. In this configuration, the lower TMD layer has to be clamped in both directions. Using this devices, a biaxial strain of up to $\epsilon_c=0.9\%$ can be achieved \cite{RN83}.

Uniaxial and shear strain can be realized with a similar setup, by mounting the 2D structure on top of a stretchable substrate, for example, a three-point pushing wedge setup as presented in figure \ref{figura:00}a \cite{RN125, Guan2015,Wang2020, McRae2019, Wang2020f, Guan2017, dadgar2018strain, RN24} or in a uniaxial strain cell \cite{McRae2019, edelberg2020tunable, Cenker2022}, as illustrated in figure \ref{figura:00}c. These experimental setups apply a uniaxial strain $\epsilon_0$ to the substrate in which is transferred to the 2D heterostructure and are able to reach deformations up to $3\%$ \cite{edelberg2020tunable}.

Figure \ref{figura:00}d and e show how the tailoring of the clamps, defined by lithography, gives full control over the mixture between uniaxial and shear strain. Pure uniaxial strain is reached when the clamps face normal and $\mathbf{n}_c$ is parallel to the direction of the applied strain $\mathbf{n}_\epsilon$. For a mixture of uniaxial and shear strain, one of the clamps can be evaporated at an angle $\gamma$, which is the angle between the clamp face normal $\mathbf{n}_c$ and the direction of the strain $\mathbf{n}_e$. In this way, the proper design of the metal clamps gives full control over the mixture of uniaxial and shear strain by changing the angle $\gamma$ of the clamps.

\begin{figure}[t!!]
\includegraphics*[keepaspectratio=true, clip=true, angle=0, width=1.\columnwidth, trim={0mm, 0mm, 0mm, 0mm}]{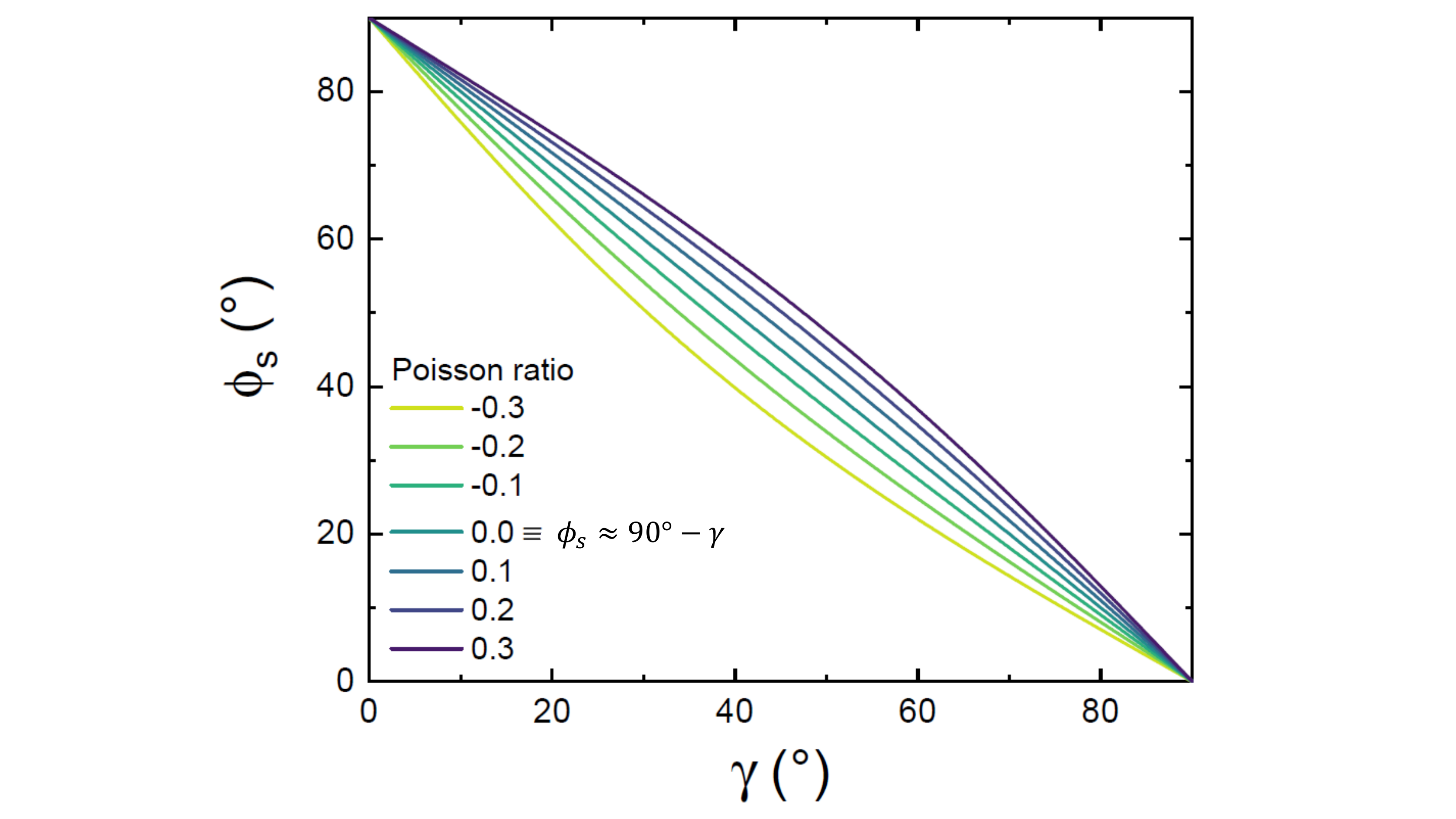}
\caption{Shear strain angle as function of $\gamma$ for different values of $\nu$. In case $\nu = 0$, $\phi_s(\gamma) \approx 90^\circ -\gamma$}
\label{figura:SI-4}
\end{figure}

The setup in Fig.~\ref{figura:00}d has the following dependencies regarding the strain
\begin{equation}
\begin{split}    
\epsilon_t&=\frac{\Delta x}{x_t} \\
\epsilon_b&=\frac{\Delta x}{x_b}.
\end{split}
\end{equation}
In case of applying an angled clamp of the left side with $0^\circ<\gamma<90^\circ$, the top strain $\epsilon_b$ will be higher than at the bottom $\epsilon_b$ since $x_b>x_t$. Hence, the overall layer is subject to a shear strain. The overall uniaxial strain on the layer is $\epsilon_b=\epsilon_u$. If we subtract the overall uniaxial strain from the strain applied at the top layer we can calculate the residual displacement $\Delta x'$ of the top layer
\begin{equation}
\Delta x'= \Delta x - \epsilon_b *x_t= \Delta x \left(\frac{x_b-x_t}{x_b}\right)= \frac{\Delta x}{x_b}\tan(\gamma) h.
\end{equation}
The shear strain is defined as $\epsilon_{xy} = \frac{\Delta x'}{2 h}$ such that the final strain matrix of the system in Fig.~\ref{figura:00}d becomes
\begin{equation}
\epsilon=\epsilon_u \begin{pmatrix} 1 &\frac{\tan(\gamma)}{2} \\ \frac{\tan(\gamma)}{2}& -\nu \end{pmatrix}
\end{equation}
where we have used that the layer deforms by $-\nu \epsilon_u$ in the opposite direction of the applied strain via the Poisson effect. With this result we can calculate the parameters 
\begin{equation}
    \begin{split}
        \epsilon_c &= \epsilon_u \frac{1-\nu}{2} \\
        \epsilon_s &= \epsilon_u \sqrt{\left(\frac{1+\nu}{2}\right)^2 +\tan(\gamma)^2} \\
        \phi_s&= \arccos\left(\frac{\tan(\gamma)}{\sqrt{\left(\frac{1+\nu}{2}\right)^2 +\tan(\gamma)^2}}\right).
    \end{split}
\end{equation}
For a vanishing Poisson ratio $\nu\approx 0$ the shear strain angle can be simplified to $\phi_s\stackrel{\nu\approx 0}{\approx} 90^\circ-\gamma$ as can be seen in Fig~\ref{figura:SI-4}.

%==FIGURA======FIGURA======FIGURA======FIGURA======FIGURA======FIGURA=====
%==FIGURA======FIGURA======FIGURA======FIGURA======FIGURA======FIGURA=====

Depending on $\gamma$, the different strain parameters take the form 
 \begin{itemize}
     \item $\epsilon_c=\frac{\epsilon_0}{2}(1-\nu)$
     \item $\epsilon_s=\frac{\epsilon_0}{2} \sqrt{(1+\nu)^2+\tan^2(\gamma)}$
     \item $\phi_s= \arccos\left(\frac{\tan(\gamma)}{\sqrt{(1+\nu)^2+\tan^2(\gamma)}})\right)\stackrel{\nu=0}{=} 90^\circ-\gamma$,
 \end{itemize}
which shows that the amount of shear strain can continuously increase by simply increasing the angle $\gamma$. Even more, the latter expression shows that the shear angle turns out to be the angle between the faces of the clamp when $\nu = 0$, and $\gamma$ is therefore the ideal parameter to tune the ratio between $\epsilon_c$ and $\epsilon_s$.

Finally, the pure shear strain case is achieved by setting $\mathbf{n}_c$ perpendicular to $\mathbf{n}_\epsilon$, as presented in Fig~\ref{figura:00}e. In this configuration, the clamp length $l$ and the distance $h$ between the clamps determine the amount of shear heterostrain applied to one of the layers, resulting $\epsilon_c=0$, $\phi_s=0^\circ$ and 
\begin{center}
$\epsilon_s=\frac{\epsilon_0}{2} \frac{l}{h}$.
\end{center}
Note that only the clamps geometry define the amount of shear strain and reducing $h$ by half has the same effect as doubling the size of $l$. For graphene or TMDs it's easily possible to find flakes with a length $l$ of more than $\SI{50}{\micro\meter}$. If the flake is clamped with a height $h$ of only $\SI{5}{\micro\meter}$ it is possible to magnify the shear strain by a factor of $10$ compared to an equal height and length clamping. This experimental configuration, therefore, allows reaching unprecedented values of strain compared to the configurations presented in the literature.

After having established experimental setups for in-situ tuning either shear or biaxial strain we continue by presenting a setup that would allow tuning of biaxial and shear strain individually for a single sample. The experimental setup involves a combination of the biaxial and shear strain devices mentioned previously. Although the construction of such a setup is challenging compared to that presented for pure biaxial or shear strain, it is experimentally viable and, if realized, would provide unprecedented control over the moir\'e pattern. In this proposed scheme, the bottom layer is mounted on top of a piezoelectric substrate and clamped such that it can be biaxially strained. The top layer is transferred on top of the bottom layer and clamped to apply shear strain as illustrated in Fig.~\ref{figura:00}e, with the only difference that the clamps are mounted on the top layer, mechanically decoupled from the bottom layer. The whole structure is mounted on top of a flexible substrate and put into a strain cell as depicted in Fig.~\ref{figura:00}c. Thus, the bottom layer is biaxially strain tuned by applying a voltage to the piezoelectric substrate while the top layer can be shear strain tuned using the strain cell.

Since the biaxial strain is applied relative to the top layer, the system can be modelled as if the top layer is biaxially strained opposite to the strain applied by the piezoelectric substrate. Hence, it is possible to simultaneously and individually apply biaxial and shear strain to the top layer with this setup. Even when complex, such  device would have a great potential, as can be seen in Fig.~6. in the main manuscript. We want to emphasize that the proposed setup is quite challenging to realize compared to the setups applying pure biaxial or shear strain. However, our proposed or similar setups can be experimentally viable and the unprecedented control over the moir\'e pattern would be worth the necessary engineering tasks.

\subsubsection{Precision of strain tuning compared to twist angle tuning}
\label{SM-subsec:expmoire-2}

The fabrication of magic angle graphene is challenging due to the precise rotational alignment at the angle $1.1^\circ$ \cite{lau2022reproducibility, Cao2018, cao2018correlated, Yankowitz2019}. Recently, there have been efforts of allowing in situ tuning the stacking angle using an atomic force microscope tips \cite{Yao2021, Hu2022, RN33}. However, the experimental setup involves using bulk hexagonal boron nitride rotors which lowers the flexibility of the moir\'e sample structure.

Heterostrain can also be used to generate flat bands in twisted bilayer graphene at non magic angles \cite{Zhang2022,Huder2018}. Since strain is usually set by applying a voltage to a piezoelectric material strain can be tuned very precisely. Using commercial strain cells it is easily possible to achieve strain tuning precision of $<0.05\%$ \cite{Schindler2020}. For a magic angle graphene bilayer stacked at exactly $1.1^\circ$, a heterostrain of $0.05\%$ would correspond to a twist angle increase of $0.0001^\circ$, three orders of magnitude more precise than necessary to achieve strong correlations \cite{Cao2018,cao2018correlated,Yankowitz2019}. Therefore, heterostrain tuning is a powerful tool for high precision tuning 2D materials around critical point and show unprecedented accuracy controlling the moir\'e pattern. For future experiments it is hence possible to use the twist angle roughly at the desired angle for the moir\'e and subsequently use strain to fine tune the moir\'e to the desired size. For example, fabricating a $1^\circ$ twisted bilayer graphene and then using a heterostrain of about $0.8\%$(experimentally feasible  sec:~\ref{subsec:expmoire})  to achieve correlated states in twisted bilayer graphene. 

\subsubsection{Strain transfer in moir\'e bilayer structures}
\label{SM-S6}

%==FIGURA======FIGURA======FIGURA======FIGURA======FIGURA======FIGURA=====
\begin{figure}[ht!!]
\includegraphics*[keepaspectratio=true, clip=true, angle=0, width=1.\columnwidth, trim={0mm, 2mm, 0mm, 0mm}]{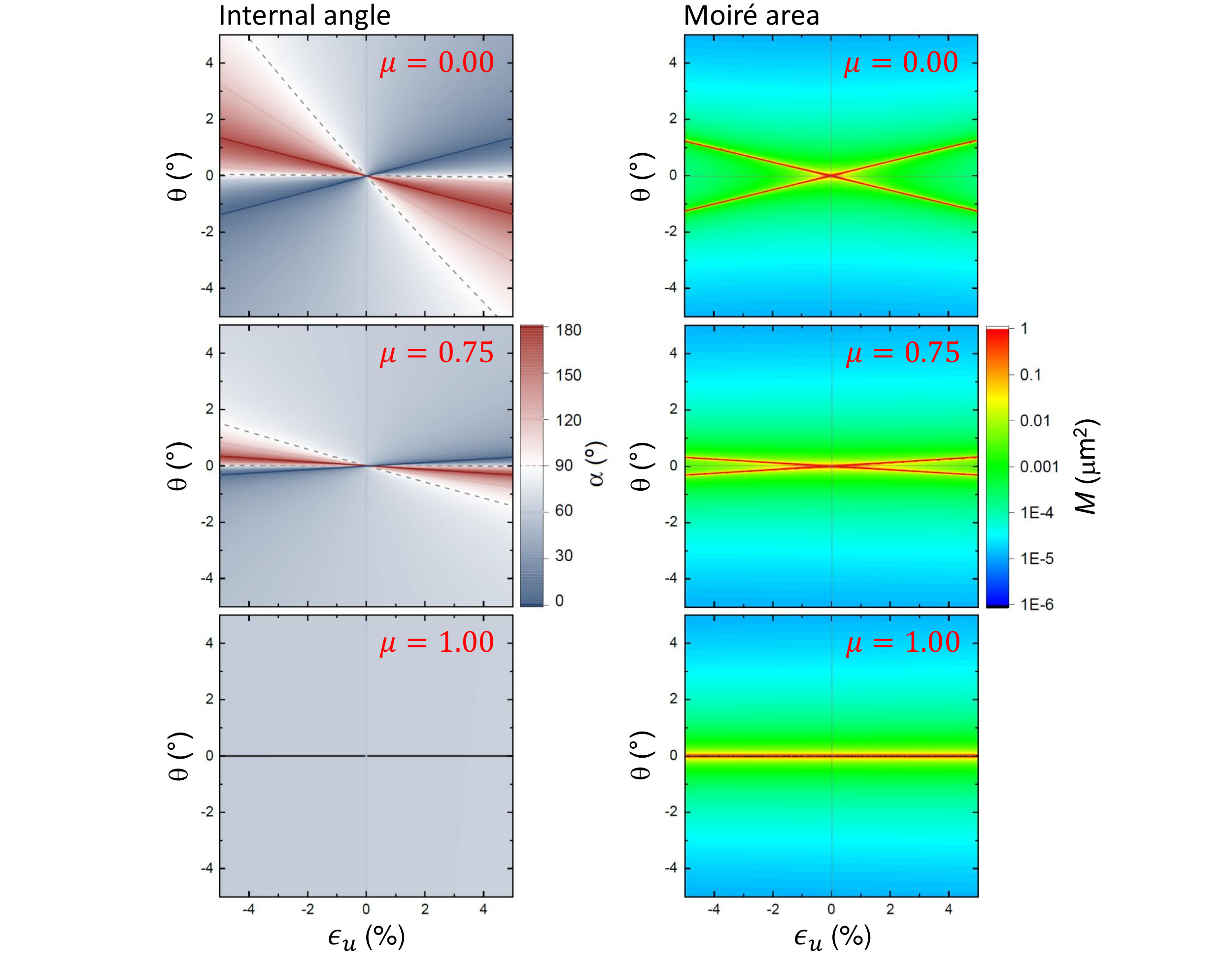}
\caption{Internal angle $\alpha$ and moiré area of a homobilayer as function of $\epsilon_u$ and $\theta$ for three different conditions. From top to bottom, $\mu=0.00$, $\mu=0.75$ and $\mu=1.00$.}
\label{figura:5}
\end{figure}
%==FIGURA======FIGURA======FIGURA======FIGURA======FIGURA======FIGURA=====

In our mathematical algorithm we introduce the strain transfer parameter $\mu$ which we set to $\mu=0$ assuming no strain transfer from the strained bottom layer to the top layer. This assumption is valid in realistic experiments due to the low interlayer friction between 2D materials \cite{RN30, Kobayashi2017}. Due to the low friction, the strain is only transferred to the top layer up to a critical strain above which the top layer relaxes to be unstrained. The critical strain for 2H stacked bilayer graphene is around $1\%$ \cite{Georgoulea2022}. Hence, above this critical strain, no strain is transferred. The relatively high critical strain does, however, reduce a lot when the layers are twisted to a moire bilayer since the interlayer friction decreases when the twist angle is away from the commensurate angles $0^\circ$ and $60^\circ$ and is generally lower for TMDs \cite{Ru2020}. In the main text we are discussing TMD moir\'e bilayers with a twist angle above $1^\circ$ such that almost all strain values are above the critical strain and $\mu=0$.

Even if $\mu$ is finite, the qualitative effects of pure heterostrain are still present. Figure \ref{figura:5} shows, as example, the effect of the adhesion parameter $\mu$ on $\alpha$ and $M$ in the particular case of uniaxial strain in a homobilayers. The figure depicts, from top to bottom panels, the evolution of the system from perfect heterostrain ($\mu=0.00$) to homostrain ($\mu = 1.00$). 

The perfect homostrain case shows almost no dependence on the applied deformation, where the main variation is observed in the moiré area as function of $\theta$. Note that as there is no dependence of $M$ on $\epsilon_u$, the divergence line at $\theta = 0$ describes the 2H(3R) stacked bilayer that lacks of moiré lattice. By decreasing $\mu$, such divergence line tears in the double divergence lines that give rise to 1D moiré lattices. For $\mu < 1.00$, the system shows qualitatively the same behaviour as for $\mu = 0$ but at higher strain.

\subsection{Effect of Poisson ratio on the moiré lattice}
\label{SM-C}

%==FIGURA======FIGURA======FIGURA======FIGURA======FIGURA======FIGURA=====
\begin{figure}[ht!!]
\includegraphics*[keepaspectratio=true, clip=true, angle=0, width=1.\columnwidth, trim={15mm, 0mm, 15mm, 0mm}]{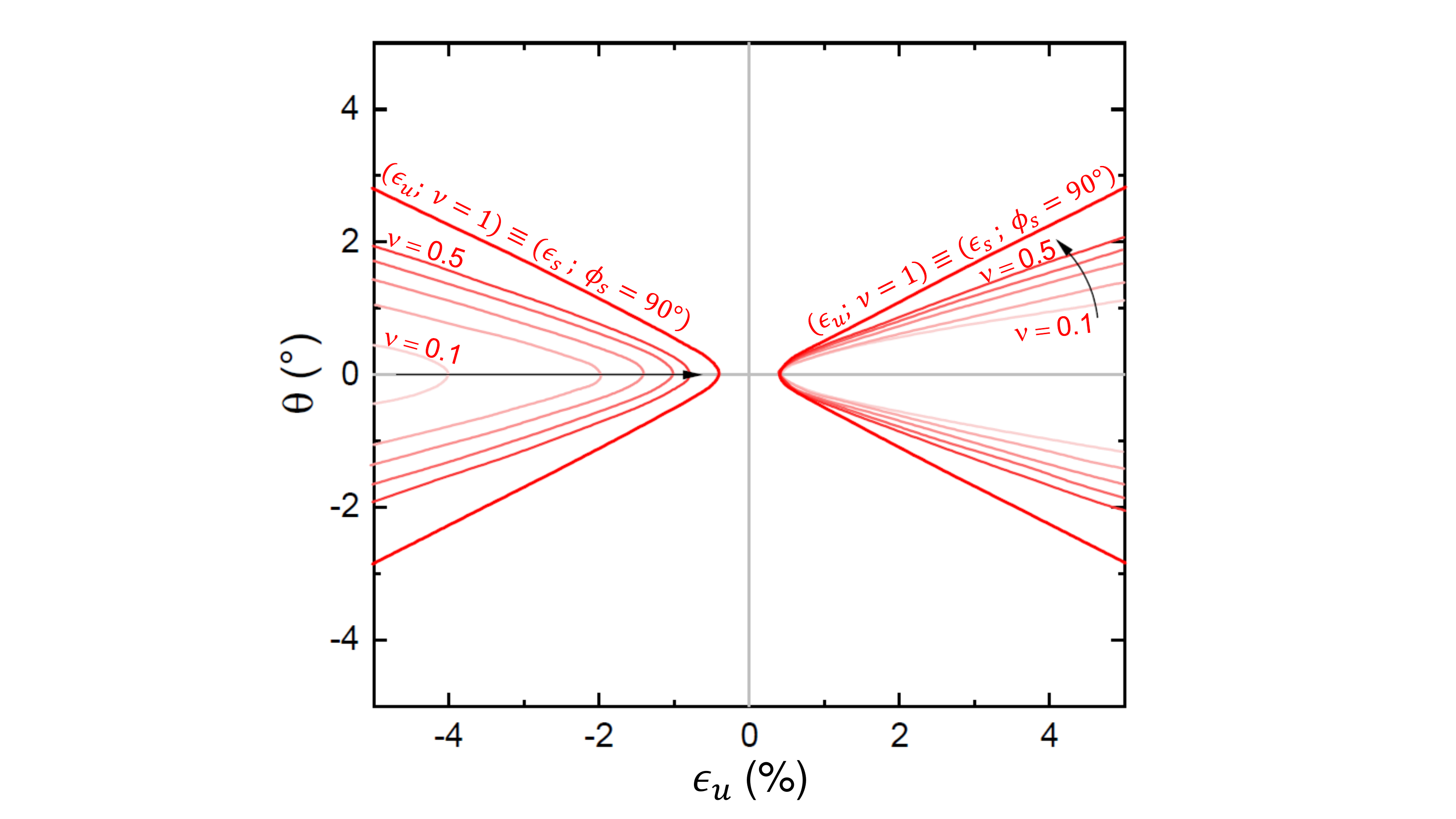}
\caption{Divergence curve as function of stacking angle and uniaxial strain for different Poisson ratios from $\nu=0.1$ to $\nu=0.5$. The limit case $\nu=1$ is equivalent to the divergence curve for shear strain with $\phi_s=90^\circ$.}
\label{figura:3}
\end{figure}
%==FIGURA======FIGURA======FIGURA======FIGURA======FIGURA======FIGURA=====

Figure \ref{figura:3} presents the numerical calculation of the divergence curve in the moir\'e lattice of a heterobilayer with $0.4\%$ lattice mismatch as function of stacking angle and uniaxial heterostrain for different Poisson ratios ranging from $\nu_1 = 0.1$ to $\nu_5 = 0.5$. The limit case $\epsilon_u$; $\nu=1$ is equivalent the shear strain case with a shear angle $\phi_s=90^\circ$. The Poisson ratio has two major effects: I) In homo and heterobilayer cases, it affects the slope of the divergence curves. II) As described in the main text for heterobilayers, the Poisson ratio defines the vertex position of the divergence curve in the negative semi space. As the mismatch between the lattices of each TMD is $\sim 0.4$\%, those points, are located on $\sim-0.4/\nu$.

\subsection{Heterostrain on rectangular homobilayers: WTe$_2$ case}
\label{SM-D}

%==FIGURA======FIGURA======FIGURA======FIGURA======FIGURA======FIGURA=====
\begin{figure}[hh!]
\includegraphics*[keepaspectratio=true, clip=true, angle=0, width=1.\columnwidth, trim={22mm, 2mm, 19mm, 2mm}]{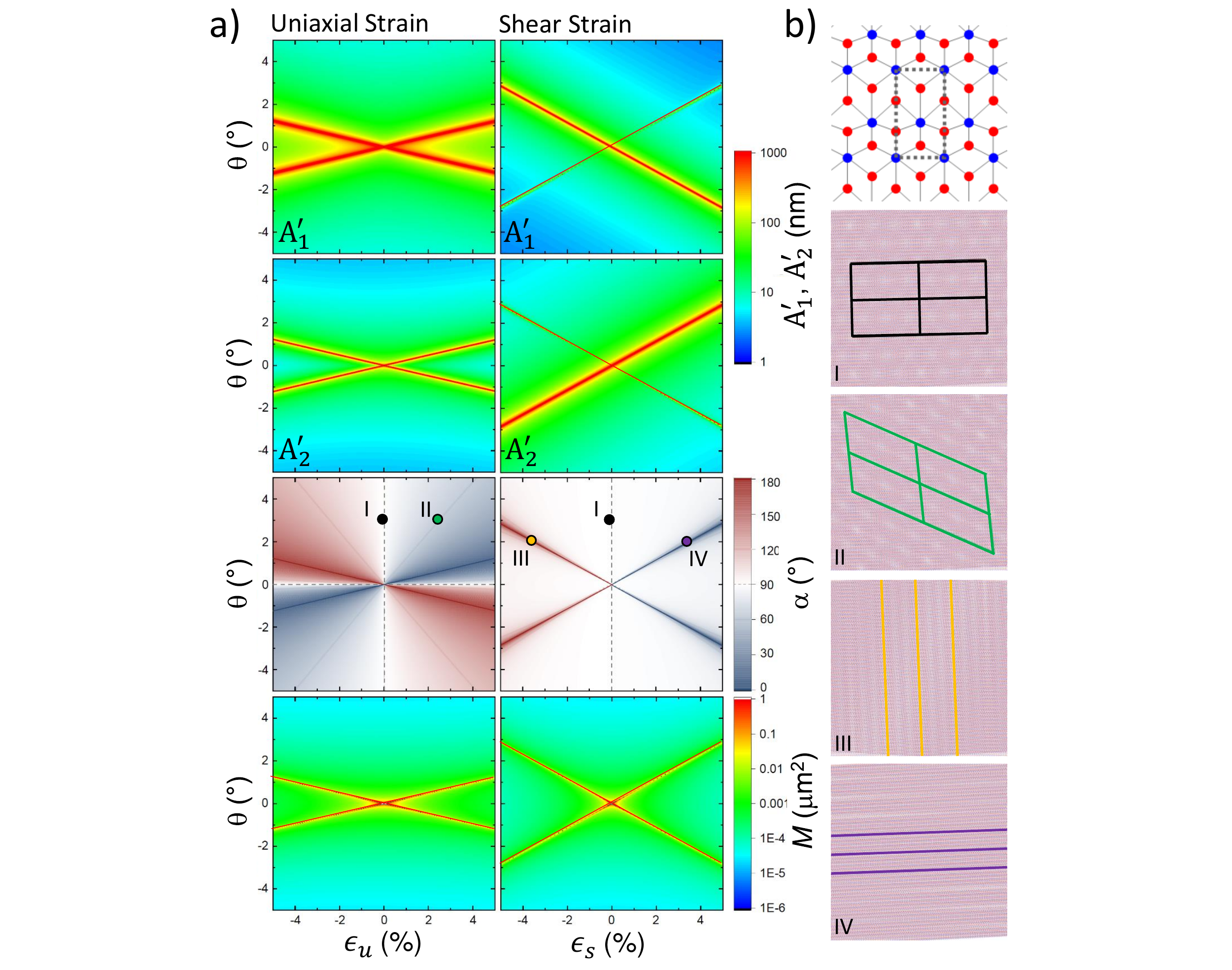}
\caption{\textbf{a)} Effect of uniaxial (left) and shear (right) strain on a moir\'e lattice generated by a WTe$_2$ homobilayer. From top to bottom: $A_1'$, $A_2'$, $\alpha$ and $M$. Spots labeled as $I$, $II$, $III$ and $IV$ correspond to the moiré lattices presented in real space in panel b. \textbf{b)} From top to bottom: WTe$_2$ unit cell, $I$ moir\'e lattice of a bilayer stacked at $\theta = 3^{\circ}$ in absence of strain, $II$ moir\'e lattice of bilayer stacked at $\theta = 3^{\circ}$ and $\epsilon_u \simeq 2.5\%$ (lattice with $\alpha = 60^{\circ}$) and $III$ and $IV$ showing 1D lattices for a bilayer stacked at $\theta \simeq 2.5^{\circ}$ and $\epsilon_s \simeq -3.8\%$ and $\epsilon_s \simeq 3.8\%$, respectively.}
\label{figura:4}
\end{figure}
%==FIGURA======FIGURA======FIGURA======FIGURA======FIGURA======FIGURA=====

As proof for the universality of our mathematical description, we calculate the moir\'e size and geometry for a twisted homobilayer of WTe$_2$ which has a rectangular unit cell. The results are depicted in Fig.~\ref{figura:4}. Similar to the hexagonal lattice case, uniaxial strain effectively changes the relative angle $\alpha$ which can be continuously tuned from $0^\circ$ to $180^\circ$. On the other hand, shear strain keeps $\alpha$ relatively constant around $90^\circ$ for different twist angles and strain levels, except for the regions of the 1D moir\'e. This is in agreement with our findings that the 1D moir\'e patterns are independent of the underlying lattice and only depend on the stacking angle, lattice mismatch and strain.

\twocolumngrid
%
%###############################################################################
%								BIBLIOGRAPHY
%##############################################################################
%

\bibliographystyle{apsrev4-1}
\bibliography{bibliography}

\begin{thebibliography}{78}%
\makeatletter
\providecommand \@ifxundefined [1]{%
 \@ifx{#1\undefined}
}%
\providecommand \@ifnum [1]{%
 \ifnum #1\expandafter \@firstoftwo
 \else \expandafter \@secondoftwo
 \fi
}%
\providecommand \@ifx [1]{%
 \ifx #1\expandafter \@firstoftwo
 \else \expandafter \@secondoftwo
 \fi
}%
\providecommand \natexlab [1]{#1}%
\providecommand \enquote  [1]{``#1''}%
\providecommand \bibnamefont  [1]{#1}%
\providecommand \bibfnamefont [1]{#1}%
\providecommand \citenamefont [1]{#1}%
\providecommand \href@noop [0]{\@secondoftwo}%
\providecommand \href [0]{\begingroup \@sanitize@url \@href}%
\providecommand \@href[1]{\@@startlink{#1}\@@href}%
\providecommand \@@href[1]{\endgroup#1\@@endlink}%
\providecommand \@sanitize@url [0]{\catcode `\\12\catcode `\$12\catcode
  `\&12\catcode `\#12\catcode `\^12\catcode `\_12\catcode `\%12\relax}%
\providecommand \@@startlink[1]{}%
\providecommand \@@endlink[0]{}%
\providecommand \url  [0]{\begingroup\@sanitize@url \@url }%
\providecommand \@url [1]{\endgroup\@href {#1}{\urlprefix }}%
\providecommand \urlprefix  [0]{URL }%
\providecommand \Eprint [0]{\href }%
\providecommand \doibase [0]{http://dx.doi.org/}%
\providecommand \selectlanguage [0]{\@gobble}%
\providecommand \bibinfo  [0]{\@secondoftwo}%
\providecommand \bibfield  [0]{\@secondoftwo}%
\providecommand \translation [1]{[#1]}%
\providecommand \BibitemOpen [0]{}%
\providecommand \bibitemStop [0]{}%
\providecommand \bibitemNoStop [0]{.\EOS\space}%
\providecommand \EOS [0]{\spacefactor3000\relax}%
\providecommand \BibitemShut  [1]{\csname bibitem#1\endcsname}%
\let\auto@bib@innerbib\@empty
%</preamble>
\bibitem [{\citenamefont {Kennes}\ \emph {et~al.}(2021)\citenamefont {Kennes},
  \citenamefont {Claassen}, \citenamefont {Xian}, \citenamefont {Georges},
  \citenamefont {Millis}, \citenamefont {Hone}, \citenamefont {Dean},
  \citenamefont {Basov}, \citenamefont {Pasupathy},\ and\ \citenamefont
  {Rubio}}]{Kennes2020}%
  \BibitemOpen
  \bibfield  {author} {\bibinfo {author} {\bibfnamefont {D.~M.}\ \bibnamefont
  {Kennes}}, \bibinfo {author} {\bibfnamefont {M.}~\bibnamefont {Claassen}},
  \bibinfo {author} {\bibfnamefont {L.}~\bibnamefont {Xian}}, \bibinfo {author}
  {\bibfnamefont {A.}~\bibnamefont {Georges}}, \bibinfo {author} {\bibfnamefont
  {A.~J.}\ \bibnamefont {Millis}}, \bibinfo {author} {\bibfnamefont
  {J.}~\bibnamefont {Hone}}, \bibinfo {author} {\bibfnamefont {C.~R.}\
  \bibnamefont {Dean}}, \bibinfo {author} {\bibfnamefont {D.~N.}\ \bibnamefont
  {Basov}}, \bibinfo {author} {\bibfnamefont {A.~N.}\ \bibnamefont
  {Pasupathy}}, \ and\ \bibinfo {author} {\bibfnamefont {A.}~\bibnamefont
  {Rubio}},\ }\href {\doibase 10.1038/s41567-020-01154-3} {\bibfield  {journal}
  {\bibinfo  {journal} {Nature Physics}\ }\textbf {\bibinfo {volume} {17}},\
  \bibinfo {pages} {155} (\bibinfo {year} {2021})},\ \Eprint
  {http://arxiv.org/abs/2011.12638} {2011.12638} \BibitemShut {NoStop}%
\bibitem [{\citenamefont {Cao}\ \emph {et~al.}(2018{\natexlab{a}})\citenamefont
  {Cao}, \citenamefont {Fatemi}, \citenamefont {Fang}, \citenamefont
  {Watanabe}, \citenamefont {Taniguchi}, \citenamefont {Kaxiras},\ and\
  \citenamefont {Jarillo-Herrero}}]{Cao2018}%
  \BibitemOpen
  \bibfield  {author} {\bibinfo {author} {\bibfnamefont {Y.}~\bibnamefont
  {Cao}}, \bibinfo {author} {\bibfnamefont {V.}~\bibnamefont {Fatemi}},
  \bibinfo {author} {\bibfnamefont {S.}~\bibnamefont {Fang}}, \bibinfo {author}
  {\bibfnamefont {K.}~\bibnamefont {Watanabe}}, \bibinfo {author}
  {\bibfnamefont {T.}~\bibnamefont {Taniguchi}}, \bibinfo {author}
  {\bibfnamefont {E.}~\bibnamefont {Kaxiras}}, \ and\ \bibinfo {author}
  {\bibfnamefont {P.}~\bibnamefont {Jarillo-Herrero}},\ }\href {\doibase
  10.1038/nature26160} {\bibfield  {journal} {\bibinfo  {journal} {Nature}\
  }\textbf {\bibinfo {volume} {556}},\ \bibinfo {pages} {43} (\bibinfo {year}
  {2018}{\natexlab{a}})}\BibitemShut {NoStop}%
\bibitem [{\citenamefont {Cao}\ \emph {et~al.}(2018{\natexlab{b}})\citenamefont
  {Cao}, \citenamefont {Fatemi}, \citenamefont {Demir}, \citenamefont {Fang},
  \citenamefont {Tomarken}, \citenamefont {Luo}, \citenamefont
  {Sanchez-Yamagishi}, \citenamefont {Watanabe}, \citenamefont {Taniguchi},
  \citenamefont {Kaxiras} \emph {et~al.}}]{cao2018correlated}%
  \BibitemOpen
  \bibfield  {author} {\bibinfo {author} {\bibfnamefont {Y.}~\bibnamefont
  {Cao}}, \bibinfo {author} {\bibfnamefont {V.}~\bibnamefont {Fatemi}},
  \bibinfo {author} {\bibfnamefont {A.}~\bibnamefont {Demir}}, \bibinfo
  {author} {\bibfnamefont {S.}~\bibnamefont {Fang}}, \bibinfo {author}
  {\bibfnamefont {S.~L.}\ \bibnamefont {Tomarken}}, \bibinfo {author}
  {\bibfnamefont {J.~Y.}\ \bibnamefont {Luo}}, \bibinfo {author} {\bibfnamefont
  {J.~D.}\ \bibnamefont {Sanchez-Yamagishi}}, \bibinfo {author} {\bibfnamefont
  {K.}~\bibnamefont {Watanabe}}, \bibinfo {author} {\bibfnamefont
  {T.}~\bibnamefont {Taniguchi}}, \bibinfo {author} {\bibfnamefont
  {E.}~\bibnamefont {Kaxiras}},  \emph {et~al.},\ }\href@noop {} {\bibfield
  {journal} {\bibinfo  {journal} {Nature}\ }\textbf {\bibinfo {volume} {556}},\
  \bibinfo {pages} {80} (\bibinfo {year} {2018}{\natexlab{b}})}\BibitemShut
  {NoStop}%
\bibitem [{\citenamefont {Balents}\ \emph {et~al.}(2020)\citenamefont
  {Balents}, \citenamefont {Dean}, \citenamefont {Efetov},\ and\ \citenamefont
  {Young}}]{Balents2020}%
  \BibitemOpen
  \bibfield  {author} {\bibinfo {author} {\bibfnamefont {L.}~\bibnamefont
  {Balents}}, \bibinfo {author} {\bibfnamefont {C.~R.}\ \bibnamefont {Dean}},
  \bibinfo {author} {\bibfnamefont {D.~K.}\ \bibnamefont {Efetov}}, \ and\
  \bibinfo {author} {\bibfnamefont {A.~F.}\ \bibnamefont {Young}},\ }\href
  {\doibase 10.1038/s41567-020-0906-9} {\bibfield  {journal} {\bibinfo
  {journal} {Nature Physics}\ }\textbf {\bibinfo {volume} {16}},\ \bibinfo
  {pages} {725} (\bibinfo {year} {2020})}\BibitemShut {NoStop}%
\bibitem [{\citenamefont {Shimazaki}\ \emph {et~al.}(2020)\citenamefont
  {Shimazaki}, \citenamefont {Schwartz}, \citenamefont {Watanabe},
  \citenamefont {Taniguchi}, \citenamefont {Kroner},\ and\ \citenamefont
  {Imamoğlu}}]{Shimazaki2019}%
  \BibitemOpen
  \bibfield  {author} {\bibinfo {author} {\bibfnamefont {Y.}~\bibnamefont
  {Shimazaki}}, \bibinfo {author} {\bibfnamefont {I.}~\bibnamefont {Schwartz}},
  \bibinfo {author} {\bibfnamefont {K.}~\bibnamefont {Watanabe}}, \bibinfo
  {author} {\bibfnamefont {T.}~\bibnamefont {Taniguchi}}, \bibinfo {author}
  {\bibfnamefont {M.}~\bibnamefont {Kroner}}, \ and\ \bibinfo {author}
  {\bibfnamefont {A.}~\bibnamefont {Imamoğlu}},\ }\href {\doibase
  10.1038/s41586-020-2191-2} {\bibfield  {journal} {\bibinfo  {journal}
  {Nature}\ }\textbf {\bibinfo {volume} {580}},\ \bibinfo {pages} {472}
  (\bibinfo {year} {2020})}\BibitemShut {NoStop}%
\bibitem [{\citenamefont {Wang}\ \emph
  {et~al.}(2020{\natexlab{a}})\citenamefont {Wang}, \citenamefont {Shih},
  \citenamefont {Ghiotto}, \citenamefont {Xian}, \citenamefont {Rhodes},
  \citenamefont {Tan}, \citenamefont {Claassen}, \citenamefont {Kennes},
  \citenamefont {Bai}, \citenamefont {Kim}, \citenamefont {Watanabe},
  \citenamefont {Taniguchi}, \citenamefont {Zhu}, \citenamefont {Hone},
  \citenamefont {Rubio}, \citenamefont {Pasupathy},\ and\ \citenamefont
  {Dean}}]{Wang2020b}%
  \BibitemOpen
  \bibfield  {author} {\bibinfo {author} {\bibfnamefont {L.}~\bibnamefont
  {Wang}}, \bibinfo {author} {\bibfnamefont {E.~M.}\ \bibnamefont {Shih}},
  \bibinfo {author} {\bibfnamefont {A.}~\bibnamefont {Ghiotto}}, \bibinfo
  {author} {\bibfnamefont {L.}~\bibnamefont {Xian}}, \bibinfo {author}
  {\bibfnamefont {D.~A.}\ \bibnamefont {Rhodes}}, \bibinfo {author}
  {\bibfnamefont {C.}~\bibnamefont {Tan}}, \bibinfo {author} {\bibfnamefont
  {M.}~\bibnamefont {Claassen}}, \bibinfo {author} {\bibfnamefont {D.~M.}\
  \bibnamefont {Kennes}}, \bibinfo {author} {\bibfnamefont {Y.}~\bibnamefont
  {Bai}}, \bibinfo {author} {\bibfnamefont {B.}~\bibnamefont {Kim}}, \bibinfo
  {author} {\bibfnamefont {K.}~\bibnamefont {Watanabe}}, \bibinfo {author}
  {\bibfnamefont {T.}~\bibnamefont {Taniguchi}}, \bibinfo {author}
  {\bibfnamefont {X.}~\bibnamefont {Zhu}}, \bibinfo {author} {\bibfnamefont
  {J.}~\bibnamefont {Hone}}, \bibinfo {author} {\bibfnamefont {A.}~\bibnamefont
  {Rubio}}, \bibinfo {author} {\bibfnamefont {A.~N.}\ \bibnamefont
  {Pasupathy}}, \ and\ \bibinfo {author} {\bibfnamefont {C.~R.}\ \bibnamefont
  {Dean}},\ }\href {\doibase 10.1038/s41563-020-0708-6} {\bibfield  {journal}
  {\bibinfo  {journal} {Nature Materials}\ } (\bibinfo {year}
  {2020}{\natexlab{a}}),\ 10.1038/s41563-020-0708-6}\BibitemShut {NoStop}%
\bibitem [{\citenamefont {Wang}\ \emph {et~al.}(2022)\citenamefont {Wang},
  \citenamefont {Yu}, \citenamefont {Kwan}, \citenamefont {Jia}, \citenamefont
  {Lei}, \citenamefont {Klemenz}, \citenamefont {Cevallos}, \citenamefont
  {Singha}, \citenamefont {Devakul}, \citenamefont {Watanabe}, \citenamefont
  {Taniguchi}, \citenamefont {Sondhi}, \citenamefont {Cava}, \citenamefont
  {Schoop}, \citenamefont {Parameswaran},\ and\ \citenamefont {Wu}}]{Wang2022}%
  \BibitemOpen
  \bibfield  {author} {\bibinfo {author} {\bibfnamefont {P.}~\bibnamefont
  {Wang}}, \bibinfo {author} {\bibfnamefont {G.}~\bibnamefont {Yu}}, \bibinfo
  {author} {\bibfnamefont {Y.~H.}\ \bibnamefont {Kwan}}, \bibinfo {author}
  {\bibfnamefont {Y.}~\bibnamefont {Jia}}, \bibinfo {author} {\bibfnamefont
  {S.}~\bibnamefont {Lei}}, \bibinfo {author} {\bibfnamefont {S.}~\bibnamefont
  {Klemenz}}, \bibinfo {author} {\bibfnamefont {F.~A.}\ \bibnamefont
  {Cevallos}}, \bibinfo {author} {\bibfnamefont {R.}~\bibnamefont {Singha}},
  \bibinfo {author} {\bibfnamefont {T.}~\bibnamefont {Devakul}}, \bibinfo
  {author} {\bibfnamefont {K.}~\bibnamefont {Watanabe}}, \bibinfo {author}
  {\bibfnamefont {T.}~\bibnamefont {Taniguchi}}, \bibinfo {author}
  {\bibfnamefont {S.~L.}\ \bibnamefont {Sondhi}}, \bibinfo {author}
  {\bibfnamefont {R.~J.}\ \bibnamefont {Cava}}, \bibinfo {author}
  {\bibfnamefont {L.~M.}\ \bibnamefont {Schoop}}, \bibinfo {author}
  {\bibfnamefont {S.~A.}\ \bibnamefont {Parameswaran}}, \ and\ \bibinfo
  {author} {\bibfnamefont {S.}~\bibnamefont {Wu}},\ }\href {\doibase
  10.1038/s41586-022-04514-6} {\bibfield  {journal} {\bibinfo  {journal}
  {Nature}\ }\textbf {\bibinfo {volume} {605}},\ \bibinfo {pages} {57}
  (\bibinfo {year} {2022})}\BibitemShut {NoStop}%
\bibitem [{\citenamefont {Tang}\ \emph {et~al.}(2020)\citenamefont {Tang},
  \citenamefont {Li}, \citenamefont {Li}, \citenamefont {Xu}, \citenamefont
  {Liu}, \citenamefont {Barmak}, \citenamefont {Watanabe}, \citenamefont
  {Taniguchi}, \citenamefont {MacDonald}, \citenamefont {Shan},\ and\
  \citenamefont {Mak}}]{Tang2020}%
  \BibitemOpen
  \bibfield  {author} {\bibinfo {author} {\bibfnamefont {Y.}~\bibnamefont
  {Tang}}, \bibinfo {author} {\bibfnamefont {L.}~\bibnamefont {Li}}, \bibinfo
  {author} {\bibfnamefont {T.}~\bibnamefont {Li}}, \bibinfo {author}
  {\bibfnamefont {Y.}~\bibnamefont {Xu}}, \bibinfo {author} {\bibfnamefont
  {S.}~\bibnamefont {Liu}}, \bibinfo {author} {\bibfnamefont {K.}~\bibnamefont
  {Barmak}}, \bibinfo {author} {\bibfnamefont {K.}~\bibnamefont {Watanabe}},
  \bibinfo {author} {\bibfnamefont {T.}~\bibnamefont {Taniguchi}}, \bibinfo
  {author} {\bibfnamefont {A.~H.}\ \bibnamefont {MacDonald}}, \bibinfo {author}
  {\bibfnamefont {J.}~\bibnamefont {Shan}}, \ and\ \bibinfo {author}
  {\bibfnamefont {K.~F.}\ \bibnamefont {Mak}},\ }\href {\doibase
  10.1038/s41586-020-2085-3} {\bibfield  {journal} {\bibinfo  {journal}
  {Nature}\ }\textbf {\bibinfo {volume} {579}},\ \bibinfo {pages} {353}
  (\bibinfo {year} {2020})}\BibitemShut {NoStop}%
\bibitem [{\citenamefont {Regan}\ \emph {et~al.}(2020)\citenamefont {Regan},
  \citenamefont {Wang}, \citenamefont {Jin}, \citenamefont {Bakti~Utama},
  \citenamefont {Gao}, \citenamefont {Wei}, \citenamefont {Zhao}, \citenamefont
  {Zhao}, \citenamefont {Zhang}, \citenamefont {Yumigeta} \emph
  {et~al.}}]{regan2020mott}%
  \BibitemOpen
  \bibfield  {author} {\bibinfo {author} {\bibfnamefont {E.~C.}\ \bibnamefont
  {Regan}}, \bibinfo {author} {\bibfnamefont {D.}~\bibnamefont {Wang}},
  \bibinfo {author} {\bibfnamefont {C.}~\bibnamefont {Jin}}, \bibinfo {author}
  {\bibfnamefont {M.~I.}\ \bibnamefont {Bakti~Utama}}, \bibinfo {author}
  {\bibfnamefont {B.}~\bibnamefont {Gao}}, \bibinfo {author} {\bibfnamefont
  {X.}~\bibnamefont {Wei}}, \bibinfo {author} {\bibfnamefont {S.}~\bibnamefont
  {Zhao}}, \bibinfo {author} {\bibfnamefont {W.}~\bibnamefont {Zhao}}, \bibinfo
  {author} {\bibfnamefont {Z.}~\bibnamefont {Zhang}}, \bibinfo {author}
  {\bibfnamefont {K.}~\bibnamefont {Yumigeta}},  \emph {et~al.},\ }\href
  {\doibase 10.1038/s41586-020-2092-4} {\bibfield  {journal} {\bibinfo
  {journal} {Nature}\ }\textbf {\bibinfo {volume} {579}},\ \bibinfo {pages}
  {359} (\bibinfo {year} {2020})}\BibitemShut {NoStop}%
\bibitem [{\citenamefont {Li}\ \emph {et~al.}(2021)\citenamefont {Li},
  \citenamefont {Jiang}, \citenamefont {Shen}, \citenamefont {Zhang},
  \citenamefont {Li}, \citenamefont {Devakul}, \citenamefont {Watanabe},
  \citenamefont {Taniguchi}, \citenamefont {Fu}, \citenamefont {Shan},\ and\
  \citenamefont {Mak}}]{Li2021}%
  \BibitemOpen
  \bibfield  {author} {\bibinfo {author} {\bibfnamefont {T.}~\bibnamefont
  {Li}}, \bibinfo {author} {\bibfnamefont {S.}~\bibnamefont {Jiang}}, \bibinfo
  {author} {\bibfnamefont {B.}~\bibnamefont {Shen}}, \bibinfo {author}
  {\bibfnamefont {Y.}~\bibnamefont {Zhang}}, \bibinfo {author} {\bibfnamefont
  {L.}~\bibnamefont {Li}}, \bibinfo {author} {\bibfnamefont {T.}~\bibnamefont
  {Devakul}}, \bibinfo {author} {\bibfnamefont {K.}~\bibnamefont {Watanabe}},
  \bibinfo {author} {\bibfnamefont {T.}~\bibnamefont {Taniguchi}}, \bibinfo
  {author} {\bibfnamefont {L.}~\bibnamefont {Fu}}, \bibinfo {author}
  {\bibfnamefont {J.}~\bibnamefont {Shan}}, \ and\ \bibinfo {author}
  {\bibfnamefont {K.~F.}\ \bibnamefont {Mak}},\ }\href {\doibase
  10.1038/s41586-021-04171-1} {\bibfield  {journal} {\bibinfo  {journal}
  {Nature}\ }\textbf {\bibinfo {volume} {600}},\ \bibinfo {pages} {641}
  (\bibinfo {year} {2021})}\BibitemShut {NoStop}%
\bibitem [{\citenamefont {Campbell}\ \emph {et~al.}(2022)\citenamefont
  {Campbell}, \citenamefont {Brotons-Gisbert}, \citenamefont {Baek},
  \citenamefont {Vitale}, \citenamefont {Taniguchi}, \citenamefont {Watanabe},
  \citenamefont {Lischner},\ and\ \citenamefont {Gerardot}}]{Campbell2022}%
  \BibitemOpen
  \bibfield  {author} {\bibinfo {author} {\bibfnamefont {A.~J.}\ \bibnamefont
  {Campbell}}, \bibinfo {author} {\bibfnamefont {M.}~\bibnamefont
  {Brotons-Gisbert}}, \bibinfo {author} {\bibfnamefont {H.}~\bibnamefont
  {Baek}}, \bibinfo {author} {\bibfnamefont {V.}~\bibnamefont {Vitale}},
  \bibinfo {author} {\bibfnamefont {T.}~\bibnamefont {Taniguchi}}, \bibinfo
  {author} {\bibfnamefont {K.}~\bibnamefont {Watanabe}}, \bibinfo {author}
  {\bibfnamefont {J.}~\bibnamefont {Lischner}}, \ and\ \bibinfo {author}
  {\bibfnamefont {B.~D.}\ \bibnamefont {Gerardot}},\ }\href {\doibase
  10.48550/arXiv.2202.08879} {\bibfield  {journal} {\bibinfo  {journal} {arXiv
  preprint arXiv:2202.08879}\ } (\bibinfo {year} {2022}),\
  10.48550/arXiv.2202.08879}\BibitemShut {NoStop}%
\bibitem [{\citenamefont {Seyler}\ \emph {et~al.}(2019)\citenamefont {Seyler},
  \citenamefont {Rivera}, \citenamefont {Yu}, \citenamefont {Wilson},
  \citenamefont {Ray}, \citenamefont {Mandrus}, \citenamefont {Yan},
  \citenamefont {Yao},\ and\ \citenamefont {Xu}}]{seyler2019signatures}%
  \BibitemOpen
  \bibfield  {author} {\bibinfo {author} {\bibfnamefont {K.~L.}\ \bibnamefont
  {Seyler}}, \bibinfo {author} {\bibfnamefont {P.}~\bibnamefont {Rivera}},
  \bibinfo {author} {\bibfnamefont {H.}~\bibnamefont {Yu}}, \bibinfo {author}
  {\bibfnamefont {N.~P.}\ \bibnamefont {Wilson}}, \bibinfo {author}
  {\bibfnamefont {E.~L.}\ \bibnamefont {Ray}}, \bibinfo {author} {\bibfnamefont
  {D.~G.}\ \bibnamefont {Mandrus}}, \bibinfo {author} {\bibfnamefont
  {J.}~\bibnamefont {Yan}}, \bibinfo {author} {\bibfnamefont {W.}~\bibnamefont
  {Yao}}, \ and\ \bibinfo {author} {\bibfnamefont {X.}~\bibnamefont {Xu}},\
  }\href {\doibase 10.1038/s41586-019-0957-1} {\bibfield  {journal} {\bibinfo
  {journal} {Nature}\ }\textbf {\bibinfo {volume} {567}},\ \bibinfo {pages}
  {66} (\bibinfo {year} {2019})}\BibitemShut {NoStop}%
\bibitem [{\citenamefont {Baek}\ \emph {et~al.}(2020)\citenamefont {Baek},
  \citenamefont {Brotons-Gisbert}, \citenamefont {Koong}, \citenamefont
  {Campbell}, \citenamefont {Rambach}, \citenamefont {Watanabe}, \citenamefont
  {Taniguchi},\ and\ \citenamefont {Gerardot}}]{Baek2020a}%
  \BibitemOpen
  \bibfield  {author} {\bibinfo {author} {\bibfnamefont {H.}~\bibnamefont
  {Baek}}, \bibinfo {author} {\bibfnamefont {M.}~\bibnamefont
  {Brotons-Gisbert}}, \bibinfo {author} {\bibfnamefont {Z.~X.}\ \bibnamefont
  {Koong}}, \bibinfo {author} {\bibfnamefont {A.}~\bibnamefont {Campbell}},
  \bibinfo {author} {\bibfnamefont {M.}~\bibnamefont {Rambach}}, \bibinfo
  {author} {\bibfnamefont {K.}~\bibnamefont {Watanabe}}, \bibinfo {author}
  {\bibfnamefont {T.}~\bibnamefont {Taniguchi}}, \ and\ \bibinfo {author}
  {\bibfnamefont {B.~D.}\ \bibnamefont {Gerardot}},\ }\href {\doibase
  10.1126/sciadv.aba8526} {\bibfield  {journal} {\bibinfo  {journal} {Science
  Advances}\ }\textbf {\bibinfo {volume} {6}},\ \bibinfo {pages} {1} (\bibinfo
  {year} {2020})}\BibitemShut {NoStop}%
\bibitem [{\citenamefont {Song}\ \emph {et~al.}(2021)\citenamefont {Song},
  \citenamefont {Sun}, \citenamefont {Anderson}, \citenamefont {Wang},
  \citenamefont {Qian}, \citenamefont {Taniguchi}, \citenamefont {Watanabe},
  \citenamefont {McGuire}, \citenamefont {Stöhr}, \citenamefont {Xiao},
  \citenamefont {Cao}, \citenamefont {Wrachtrup},\ and\ \citenamefont
  {Xu}}]{Song2021}%
  \BibitemOpen
  \bibfield  {author} {\bibinfo {author} {\bibfnamefont {T.}~\bibnamefont
  {Song}}, \bibinfo {author} {\bibfnamefont {Q.-C.}\ \bibnamefont {Sun}},
  \bibinfo {author} {\bibfnamefont {E.}~\bibnamefont {Anderson}}, \bibinfo
  {author} {\bibfnamefont {C.}~\bibnamefont {Wang}}, \bibinfo {author}
  {\bibfnamefont {J.}~\bibnamefont {Qian}}, \bibinfo {author} {\bibfnamefont
  {T.}~\bibnamefont {Taniguchi}}, \bibinfo {author} {\bibfnamefont
  {K.}~\bibnamefont {Watanabe}}, \bibinfo {author} {\bibfnamefont {M.~A.}\
  \bibnamefont {McGuire}}, \bibinfo {author} {\bibfnamefont {R.}~\bibnamefont
  {Stöhr}}, \bibinfo {author} {\bibfnamefont {D.}~\bibnamefont {Xiao}},
  \bibinfo {author} {\bibfnamefont {T.}~\bibnamefont {Cao}}, \bibinfo {author}
  {\bibfnamefont {J.}~\bibnamefont {Wrachtrup}}, \ and\ \bibinfo {author}
  {\bibfnamefont {X.}~\bibnamefont {Xu}},\ }\href {\doibase
  10.1126/science.abj7478} {\bibfield  {journal} {\bibinfo  {journal}
  {Science}\ }\textbf {\bibinfo {volume} {374}},\ \bibinfo {pages} {1140}
  (\bibinfo {year} {2021})}\BibitemShut {NoStop}%
\bibitem [{\citenamefont {Xie}\ \emph {et~al.}(2022)\citenamefont {Xie},
  \citenamefont {Luo}, \citenamefont {Ye}, \citenamefont {Ye}, \citenamefont
  {Ge}, \citenamefont {Sung}, \citenamefont {Rennich}, \citenamefont {Yan},
  \citenamefont {Fu}, \citenamefont {Tian}, \citenamefont {Lei}, \citenamefont
  {Hovden}, \citenamefont {Sun}, \citenamefont {He},\ and\ \citenamefont
  {Zhao}}]{Xie2022}%
  \BibitemOpen
  \bibfield  {author} {\bibinfo {author} {\bibfnamefont {H.}~\bibnamefont
  {Xie}}, \bibinfo {author} {\bibfnamefont {X.}~\bibnamefont {Luo}}, \bibinfo
  {author} {\bibfnamefont {G.}~\bibnamefont {Ye}}, \bibinfo {author}
  {\bibfnamefont {Z.}~\bibnamefont {Ye}}, \bibinfo {author} {\bibfnamefont
  {H.}~\bibnamefont {Ge}}, \bibinfo {author} {\bibfnamefont {S.~H.}\
  \bibnamefont {Sung}}, \bibinfo {author} {\bibfnamefont {E.}~\bibnamefont
  {Rennich}}, \bibinfo {author} {\bibfnamefont {S.}~\bibnamefont {Yan}},
  \bibinfo {author} {\bibfnamefont {Y.}~\bibnamefont {Fu}}, \bibinfo {author}
  {\bibfnamefont {S.}~\bibnamefont {Tian}}, \bibinfo {author} {\bibfnamefont
  {H.}~\bibnamefont {Lei}}, \bibinfo {author} {\bibfnamefont {R.}~\bibnamefont
  {Hovden}}, \bibinfo {author} {\bibfnamefont {K.}~\bibnamefont {Sun}},
  \bibinfo {author} {\bibfnamefont {R.}~\bibnamefont {He}}, \ and\ \bibinfo
  {author} {\bibfnamefont {L.}~\bibnamefont {Zhao}},\ }\href {\doibase
  10.1038/s41567-021-01408-8} {\bibfield  {journal} {\bibinfo  {journal}
  {Nature Physics}\ }\textbf {\bibinfo {volume} {18}},\ \bibinfo {pages} {30}
  (\bibinfo {year} {2022})}\BibitemShut {NoStop}%
\bibitem [{\citenamefont {Xu}\ \emph {et~al.}(2022)\citenamefont {Xu},
  \citenamefont {Ray}, \citenamefont {Shao}, \citenamefont {Jiang},
  \citenamefont {Lee}, \citenamefont {Weber}, \citenamefont {Goldberger},
  \citenamefont {Watanabe}, \citenamefont {Taniguchi}, \citenamefont {Muller},
  \citenamefont {Mak},\ and\ \citenamefont {Shan}}]{Xu2022}%
  \BibitemOpen
  \bibfield  {author} {\bibinfo {author} {\bibfnamefont {Y.}~\bibnamefont
  {Xu}}, \bibinfo {author} {\bibfnamefont {A.}~\bibnamefont {Ray}}, \bibinfo
  {author} {\bibfnamefont {Y.-T.}\ \bibnamefont {Shao}}, \bibinfo {author}
  {\bibfnamefont {S.}~\bibnamefont {Jiang}}, \bibinfo {author} {\bibfnamefont
  {K.}~\bibnamefont {Lee}}, \bibinfo {author} {\bibfnamefont {D.}~\bibnamefont
  {Weber}}, \bibinfo {author} {\bibfnamefont {J.~E.}\ \bibnamefont
  {Goldberger}}, \bibinfo {author} {\bibfnamefont {K.}~\bibnamefont
  {Watanabe}}, \bibinfo {author} {\bibfnamefont {T.}~\bibnamefont {Taniguchi}},
  \bibinfo {author} {\bibfnamefont {D.~A.}\ \bibnamefont {Muller}}, \bibinfo
  {author} {\bibfnamefont {K.~F.}\ \bibnamefont {Mak}}, \ and\ \bibinfo
  {author} {\bibfnamefont {J.}~\bibnamefont {Shan}},\ }\href {\doibase
  10.1038/s41565-021-01014-y} {\bibfield  {journal} {\bibinfo  {journal}
  {Nature Nanotechnology}\ }\textbf {\bibinfo {volume} {17}},\ \bibinfo {pages}
  {143} (\bibinfo {year} {2022})}\BibitemShut {NoStop}%
\bibitem [{\citenamefont {Edelberg}\ \emph {et~al.}(2020)\citenamefont
  {Edelberg}, \citenamefont {Kumar}, \citenamefont {Shenoy}, \citenamefont
  {Ochoa},\ and\ \citenamefont {Pasupathy}}]{edelberg2020tunable}%
  \BibitemOpen
  \bibfield  {author} {\bibinfo {author} {\bibfnamefont {D.}~\bibnamefont
  {Edelberg}}, \bibinfo {author} {\bibfnamefont {H.}~\bibnamefont {Kumar}},
  \bibinfo {author} {\bibfnamefont {V.}~\bibnamefont {Shenoy}}, \bibinfo
  {author} {\bibfnamefont {H.}~\bibnamefont {Ochoa}}, \ and\ \bibinfo {author}
  {\bibfnamefont {A.~N.}\ \bibnamefont {Pasupathy}},\ }\href@noop {} {\bibfield
   {journal} {\bibinfo  {journal} {Nature Physics}\ }\textbf {\bibinfo {volume}
  {16}},\ \bibinfo {pages} {1097} (\bibinfo {year} {2020})}\BibitemShut
  {NoStop}%
\bibitem [{\citenamefont {Hu}\ \emph {et~al.}(2020)\citenamefont {Hu},
  \citenamefont {Ou}, \citenamefont {Si}, \citenamefont {Wu}, \citenamefont
  {Wu}, \citenamefont {Dai}, \citenamefont {Krasnok}, \citenamefont {Mazor},
  \citenamefont {Zhang}, \citenamefont {Bao}, \citenamefont {Qiu},\ and\
  \citenamefont {Alù}}]{Hu2020}%
  \BibitemOpen
  \bibfield  {author} {\bibinfo {author} {\bibfnamefont {G.}~\bibnamefont
  {Hu}}, \bibinfo {author} {\bibfnamefont {Q.}~\bibnamefont {Ou}}, \bibinfo
  {author} {\bibfnamefont {G.}~\bibnamefont {Si}}, \bibinfo {author}
  {\bibfnamefont {Y.}~\bibnamefont {Wu}}, \bibinfo {author} {\bibfnamefont
  {J.}~\bibnamefont {Wu}}, \bibinfo {author} {\bibfnamefont {Z.}~\bibnamefont
  {Dai}}, \bibinfo {author} {\bibfnamefont {A.}~\bibnamefont {Krasnok}},
  \bibinfo {author} {\bibfnamefont {Y.}~\bibnamefont {Mazor}}, \bibinfo
  {author} {\bibfnamefont {Q.}~\bibnamefont {Zhang}}, \bibinfo {author}
  {\bibfnamefont {Q.}~\bibnamefont {Bao}}, \bibinfo {author} {\bibfnamefont
  {C.-W.}\ \bibnamefont {Qiu}}, \ and\ \bibinfo {author} {\bibfnamefont
  {A.}~\bibnamefont {Alù}},\ }\href {\doibase 10.1038/s41586-020-2359-9}
  {\bibfield  {journal} {\bibinfo  {journal} {Nature}\ }\textbf {\bibinfo
  {volume} {582}},\ \bibinfo {pages} {209} (\bibinfo {year}
  {2020})}\BibitemShut {NoStop}%
\bibitem [{\citenamefont {Chen}\ \emph {et~al.}(2020)\citenamefont {Chen},
  \citenamefont {Lin}, \citenamefont {Dinh}, \citenamefont {Zheng},
  \citenamefont {Shen}, \citenamefont {Ma}, \citenamefont {Chen}, \citenamefont
  {Jarillo-Herrero},\ and\ \citenamefont {Dai}}]{Chen2020}%
  \BibitemOpen
  \bibfield  {author} {\bibinfo {author} {\bibfnamefont {M.}~\bibnamefont
  {Chen}}, \bibinfo {author} {\bibfnamefont {X.}~\bibnamefont {Lin}}, \bibinfo
  {author} {\bibfnamefont {T.~H.}\ \bibnamefont {Dinh}}, \bibinfo {author}
  {\bibfnamefont {Z.}~\bibnamefont {Zheng}}, \bibinfo {author} {\bibfnamefont
  {J.}~\bibnamefont {Shen}}, \bibinfo {author} {\bibfnamefont {Q.}~\bibnamefont
  {Ma}}, \bibinfo {author} {\bibfnamefont {H.}~\bibnamefont {Chen}}, \bibinfo
  {author} {\bibfnamefont {P.}~\bibnamefont {Jarillo-Herrero}}, \ and\ \bibinfo
  {author} {\bibfnamefont {S.}~\bibnamefont {Dai}},\ }\href {\doibase
  10.1038/s41563-020-0732-6} {\bibfield  {journal} {\bibinfo  {journal} {Nature
  Materials}\ }\textbf {\bibinfo {volume} {19}},\ \bibinfo {pages} {1307}
  (\bibinfo {year} {2020})}\BibitemShut {NoStop}%
\bibitem [{\citenamefont {Stern}\ \emph {et~al.}(2021)\citenamefont {Stern},
  \citenamefont {Waschitz}, \citenamefont {Cao}, \citenamefont {Nevo},
  \citenamefont {Watanabe}, \citenamefont {Taniguchi}, \citenamefont {Sela},
  \citenamefont {Urbakh}, \citenamefont {Hod},\ and\ \citenamefont
  {Shalom}}]{Vizner2021}%
  \BibitemOpen
  \bibfield  {author} {\bibinfo {author} {\bibfnamefont {M.~V.}\ \bibnamefont
  {Stern}}, \bibinfo {author} {\bibfnamefont {Y.}~\bibnamefont {Waschitz}},
  \bibinfo {author} {\bibfnamefont {W.}~\bibnamefont {Cao}}, \bibinfo {author}
  {\bibfnamefont {I.}~\bibnamefont {Nevo}}, \bibinfo {author} {\bibfnamefont
  {K.}~\bibnamefont {Watanabe}}, \bibinfo {author} {\bibfnamefont
  {T.}~\bibnamefont {Taniguchi}}, \bibinfo {author} {\bibfnamefont
  {E.}~\bibnamefont {Sela}}, \bibinfo {author} {\bibfnamefont {M.}~\bibnamefont
  {Urbakh}}, \bibinfo {author} {\bibfnamefont {O.}~\bibnamefont {Hod}}, \ and\
  \bibinfo {author} {\bibfnamefont {M.~B.}\ \bibnamefont {Shalom}},\ }\href
  {\doibase 10.1126/science.abe8177} {\bibfield  {journal} {\bibinfo  {journal}
  {Science}\ }\textbf {\bibinfo {volume} {372}},\ \bibinfo {pages} {1462}
  (\bibinfo {year} {2021})}\BibitemShut {NoStop}%
\bibitem [{\citenamefont {Yasuda}\ \emph {et~al.}(2021)\citenamefont {Yasuda},
  \citenamefont {Wang}, \citenamefont {Watanabe}, \citenamefont {Taniguchi},\
  and\ \citenamefont {Jarillo-Herrero}}]{Yasuda2021}%
  \BibitemOpen
  \bibfield  {author} {\bibinfo {author} {\bibfnamefont {K.}~\bibnamefont
  {Yasuda}}, \bibinfo {author} {\bibfnamefont {X.}~\bibnamefont {Wang}},
  \bibinfo {author} {\bibfnamefont {K.}~\bibnamefont {Watanabe}}, \bibinfo
  {author} {\bibfnamefont {T.}~\bibnamefont {Taniguchi}}, \ and\ \bibinfo
  {author} {\bibfnamefont {P.}~\bibnamefont {Jarillo-Herrero}},\ }\href
  {\doibase 10.1126/science.abd3230} {\bibfield  {journal} {\bibinfo  {journal}
  {Science}\ }\textbf {\bibinfo {volume} {372}},\ \bibinfo {pages} {1458}
  (\bibinfo {year} {2021})}\BibitemShut {NoStop}%
\bibitem [{\citenamefont {Hao}\ \emph {et~al.}(2021)\citenamefont {Hao},
  \citenamefont {Zimmerman}, \citenamefont {Ledwith}, \citenamefont {Khalaf},
  \citenamefont {Najafabadi}, \citenamefont {Watanabe}, \citenamefont
  {Taniguchi}, \citenamefont {Vishwanath},\ and\ \citenamefont
  {Kim}}]{Hao2021}%
  \BibitemOpen
  \bibfield  {author} {\bibinfo {author} {\bibfnamefont {Z.}~\bibnamefont
  {Hao}}, \bibinfo {author} {\bibfnamefont {A.~M.}\ \bibnamefont {Zimmerman}},
  \bibinfo {author} {\bibfnamefont {P.}~\bibnamefont {Ledwith}}, \bibinfo
  {author} {\bibfnamefont {E.}~\bibnamefont {Khalaf}}, \bibinfo {author}
  {\bibfnamefont {D.~H.}\ \bibnamefont {Najafabadi}}, \bibinfo {author}
  {\bibfnamefont {K.}~\bibnamefont {Watanabe}}, \bibinfo {author}
  {\bibfnamefont {T.}~\bibnamefont {Taniguchi}}, \bibinfo {author}
  {\bibfnamefont {A.}~\bibnamefont {Vishwanath}}, \ and\ \bibinfo {author}
  {\bibfnamefont {P.}~\bibnamefont {Kim}},\ }\href {\doibase
  10.1126/science.abg0399} {\bibfield  {journal} {\bibinfo  {journal}
  {Science}\ }\textbf {\bibinfo {volume} {371}},\ \bibinfo {pages} {1133}
  (\bibinfo {year} {2021})}\BibitemShut {NoStop}%
\bibitem [{\citenamefont {Quintanilla}\ and\ \citenamefont
  {Hooley}(2009)}]{Meda2009}%
  \BibitemOpen
  \bibfield  {author} {\bibinfo {author} {\bibfnamefont {J.}~\bibnamefont
  {Quintanilla}}\ and\ \bibinfo {author} {\bibfnamefont {C.}~\bibnamefont
  {Hooley}},\ }\href@noop {} {\bibfield  {journal} {\bibinfo  {journal}
  {Physics World}\ }\textbf {\bibinfo {volume} {22}},\ \bibinfo {pages} {32}
  (\bibinfo {year} {2009})}\BibitemShut {NoStop}%
\bibitem [{\citenamefont {Jiang}\ and\ \citenamefont
  {Devereaux}(2019)}]{jiang2019superconductivity}%
  \BibitemOpen
  \bibfield  {author} {\bibinfo {author} {\bibfnamefont {H.-C.}\ \bibnamefont
  {Jiang}}\ and\ \bibinfo {author} {\bibfnamefont {T.~P.}\ \bibnamefont
  {Devereaux}},\ }\href {\doibase 10.1126/science.aal5304} {\bibfield
  {journal} {\bibinfo  {journal} {Science}\ }\textbf {\bibinfo {volume}
  {365}},\ \bibinfo {pages} {1424} (\bibinfo {year} {2019})}\BibitemShut
  {NoStop}%
\bibitem [{\citenamefont {Tarruell}\ and\ \citenamefont
  {Sanchez-Palencia}(2018)}]{Tarruell2018a}%
  \BibitemOpen
  \bibfield  {author} {\bibinfo {author} {\bibfnamefont {L.}~\bibnamefont
  {Tarruell}}\ and\ \bibinfo {author} {\bibfnamefont {L.}~\bibnamefont
  {Sanchez-Palencia}},\ }\href {\doibase 10.1016/J.CRHY.2018.10.013} {\bibfield
   {journal} {\bibinfo  {journal} {Comptes Rendus Physique}\ }\textbf {\bibinfo
  {volume} {19}},\ \bibinfo {pages} {365} (\bibinfo {year} {2018})},\ \Eprint
  {http://arxiv.org/abs/1809.00571} {1809.00571} \BibitemShut {NoStop}%
\bibitem [{\citenamefont {Gross}\ and\ \citenamefont {Bloch}(2017)}]{Gross}%
  \BibitemOpen
  \bibfield  {author} {\bibinfo {author} {\bibfnamefont {C.}~\bibnamefont
  {Gross}}\ and\ \bibinfo {author} {\bibfnamefont {I.}~\bibnamefont {Bloch}},\
  }\href {\doibase 10.1126/science.aal3837} {\bibfield  {journal} {\bibinfo
  {journal} {Science}\ }\textbf {\bibinfo {volume} {357}},\ \bibinfo {pages}
  {995} (\bibinfo {year} {2017})}\BibitemShut {NoStop}%
\bibitem [{\citenamefont {Wu}\ \emph {et~al.}(2018)\citenamefont {Wu},
  \citenamefont {Lovorn}, \citenamefont {Tutuc},\ and\ \citenamefont
  {Macdonald}}]{Wu2018a}%
  \BibitemOpen
  \bibfield  {author} {\bibinfo {author} {\bibfnamefont {F.}~\bibnamefont
  {Wu}}, \bibinfo {author} {\bibfnamefont {T.}~\bibnamefont {Lovorn}}, \bibinfo
  {author} {\bibfnamefont {E.}~\bibnamefont {Tutuc}}, \ and\ \bibinfo {author}
  {\bibfnamefont {A.~H.}\ \bibnamefont {Macdonald}},\ }\href {\doibase
  10.1103/PhysRevLett.121.026402} {\bibfield  {journal} {\bibinfo  {journal}
  {Physical Review Letters}\ }\textbf {\bibinfo {volume} {121}},\ \bibinfo
  {pages} {26402} (\bibinfo {year} {2018})}\BibitemShut {NoStop}%
\bibitem [{\citenamefont {Pan}\ \emph {et~al.}(2020)\citenamefont {Pan},
  \citenamefont {Wu},\ and\ \citenamefont {{Das Sarma}}}]{Pan2020a}%
  \BibitemOpen
  \bibfield  {author} {\bibinfo {author} {\bibfnamefont {H.}~\bibnamefont
  {Pan}}, \bibinfo {author} {\bibfnamefont {F.}~\bibnamefont {Wu}}, \ and\
  \bibinfo {author} {\bibfnamefont {S.}~\bibnamefont {{Das Sarma}}},\ }\href
  {\doibase 10.1103/PhysRevB.102.201104} {\bibfield  {journal} {\bibinfo
  {journal} {Physical Review B}\ }\textbf {\bibinfo {volume} {102}},\ \bibinfo
  {pages} {201104} (\bibinfo {year} {2020})},\ \Eprint
  {http://arxiv.org/abs/2008.08998} {2008.08998} \BibitemShut {NoStop}%
\bibitem [{\citenamefont {Georges}\ \emph {et~al.}(2000)\citenamefont
  {Georges}, \citenamefont {Giamarchi},\ and\ \citenamefont
  {Sandler}}]{Georges2000}%
  \BibitemOpen
  \bibfield  {author} {\bibinfo {author} {\bibfnamefont {A.}~\bibnamefont
  {Georges}}, \bibinfo {author} {\bibfnamefont {T.}~\bibnamefont {Giamarchi}},
  \ and\ \bibinfo {author} {\bibfnamefont {N.}~\bibnamefont {Sandler}},\ }\href
  {\doibase 10.1103/PhysRevB.61.16393} {\bibfield  {journal} {\bibinfo
  {journal} {Physical Review B}\ }\textbf {\bibinfo {volume} {61}},\ \bibinfo
  {pages} {16393} (\bibinfo {year} {2000})}\BibitemShut {NoStop}%
\bibitem [{\citenamefont {Lau}\ \emph {et~al.}(2022)\citenamefont {Lau},
  \citenamefont {Bockrath}, \citenamefont {Mak},\ and\ \citenamefont
  {Zhang}}]{lau2022reproducibility}%
  \BibitemOpen
  \bibfield  {author} {\bibinfo {author} {\bibfnamefont {C.~N.}\ \bibnamefont
  {Lau}}, \bibinfo {author} {\bibfnamefont {M.~W.}\ \bibnamefont {Bockrath}},
  \bibinfo {author} {\bibfnamefont {K.~F.}\ \bibnamefont {Mak}}, \ and\
  \bibinfo {author} {\bibfnamefont {F.}~\bibnamefont {Zhang}},\ }\href
  {\doibase 10.1038/s41586-021-04173-z} {\bibfield  {journal} {\bibinfo
  {journal} {Nature}\ }\textbf {\bibinfo {volume} {602}},\ \bibinfo {pages}
  {41} (\bibinfo {year} {2022})}\BibitemShut {NoStop}%
\bibitem [{\citenamefont {Kuwabara}\ \emph {et~al.}(1990)\citenamefont
  {Kuwabara}, \citenamefont {Clarke},\ and\ \citenamefont
  {Smith}}]{kuwabara1990anomalous}%
  \BibitemOpen
  \bibfield  {author} {\bibinfo {author} {\bibfnamefont {M.}~\bibnamefont
  {Kuwabara}}, \bibinfo {author} {\bibfnamefont {D.~R.}\ \bibnamefont
  {Clarke}}, \ and\ \bibinfo {author} {\bibfnamefont {D.}~\bibnamefont
  {Smith}},\ }\href {\doibase 10.1063/1.102906} {\bibfield  {journal} {\bibinfo
   {journal} {Applied physics letters}\ }\textbf {\bibinfo {volume} {56}},\
  \bibinfo {pages} {2396} (\bibinfo {year} {1990})}\BibitemShut {NoStop}%
\bibitem [{\citenamefont {Ru}\ \emph {et~al.}(2020)\citenamefont {Ru},
  \citenamefont {Qi}, \citenamefont {Tang}, \citenamefont {Wei},\ and\
  \citenamefont {Xue}}]{Ru2020}%
  \BibitemOpen
  \bibfield  {author} {\bibinfo {author} {\bibfnamefont {G.}~\bibnamefont
  {Ru}}, \bibinfo {author} {\bibfnamefont {W.}~\bibnamefont {Qi}}, \bibinfo
  {author} {\bibfnamefont {K.}~\bibnamefont {Tang}}, \bibinfo {author}
  {\bibfnamefont {Y.}~\bibnamefont {Wei}}, \ and\ \bibinfo {author}
  {\bibfnamefont {T.}~\bibnamefont {Xue}},\ }\href {\doibase
  10.1016/j.triboint.2020.106483} {\bibfield  {journal} {\bibinfo  {journal}
  {Tribology International}\ }\textbf {\bibinfo {volume} {151}} (\bibinfo
  {year} {2020}),\ 10.1016/j.triboint.2020.106483}\BibitemShut {NoStop}%
\bibitem [{\citenamefont {Song}\ \emph {et~al.}(2018)\citenamefont {Song},
  \citenamefont {Mandelli}, \citenamefont {Hod}, \citenamefont {Urbakh},
  \citenamefont {Ma},\ and\ \citenamefont {Zheng}}]{Song2018a}%
  \BibitemOpen
  \bibfield  {author} {\bibinfo {author} {\bibfnamefont {Y.}~\bibnamefont
  {Song}}, \bibinfo {author} {\bibfnamefont {D.}~\bibnamefont {Mandelli}},
  \bibinfo {author} {\bibfnamefont {O.}~\bibnamefont {Hod}}, \bibinfo {author}
  {\bibfnamefont {M.}~\bibnamefont {Urbakh}}, \bibinfo {author} {\bibfnamefont
  {M.}~\bibnamefont {Ma}}, \ and\ \bibinfo {author} {\bibfnamefont
  {Q.}~\bibnamefont {Zheng}},\ }\href {\doibase 10.1038/s41563-018-0144-z}
  {\bibfield  {journal} {\bibinfo  {journal} {Nature Materials}\ }\textbf
  {\bibinfo {volume} {17}},\ \bibinfo {pages} {894} (\bibinfo {year}
  {2018})}\BibitemShut {NoStop}%
\bibitem [{\citenamefont {Ribeiro-Palau}\ \emph {et~al.}(2018)\citenamefont
  {Ribeiro-Palau}, \citenamefont {Zhang}, \citenamefont {Watanabe},
  \citenamefont {Taniguchi}, \citenamefont {Hone},\ and\ \citenamefont
  {Dean}}]{RN35}%
  \BibitemOpen
  \bibfield  {author} {\bibinfo {author} {\bibfnamefont {R.}~\bibnamefont
  {Ribeiro-Palau}}, \bibinfo {author} {\bibfnamefont {C.}~\bibnamefont
  {Zhang}}, \bibinfo {author} {\bibfnamefont {K.}~\bibnamefont {Watanabe}},
  \bibinfo {author} {\bibfnamefont {T.}~\bibnamefont {Taniguchi}}, \bibinfo
  {author} {\bibfnamefont {J.}~\bibnamefont {Hone}}, \ and\ \bibinfo {author}
  {\bibfnamefont {C.~R.}\ \bibnamefont {Dean}},\ }\href {\doibase
  10.1126/science.aat6981} {\bibfield  {journal} {\bibinfo  {journal}
  {Science}\ }\textbf {\bibinfo {volume} {361}},\ \bibinfo {pages} {690}
  (\bibinfo {year} {2018})},\ \Eprint {http://arxiv.org/abs/1804.02038}
  {1804.02038} \BibitemShut {NoStop}%
\bibitem [{\citenamefont {Bi}\ \emph {et~al.}(2019)\citenamefont {Bi},
  \citenamefont {Yuan},\ and\ \citenamefont {Fu}}]{Bi2019}%
  \BibitemOpen
  \bibfield  {author} {\bibinfo {author} {\bibfnamefont {Z.}~\bibnamefont
  {Bi}}, \bibinfo {author} {\bibfnamefont {N.~F.}\ \bibnamefont {Yuan}}, \ and\
  \bibinfo {author} {\bibfnamefont {L.}~\bibnamefont {Fu}},\ }\href {\doibase
  10.1103/PhysRevB.100.035448} {\bibfield  {journal} {\bibinfo  {journal}
  {Physical Review B}\ }\textbf {\bibinfo {volume} {100}},\ \bibinfo {pages}
  {1} (\bibinfo {year} {2019})}\BibitemShut {NoStop}%
\bibitem [{\citenamefont {Halbertal}\ \emph {et~al.}(2021)\citenamefont
  {Halbertal}, \citenamefont {Finney}, \citenamefont {Sunku}, \citenamefont
  {Kerelsky}, \citenamefont {Rubio-Verd{\'u}}, \citenamefont {Shabani},
  \citenamefont {Xian}, \citenamefont {Carr}, \citenamefont {Chen},
  \citenamefont {Zhang} \emph {et~al.}}]{halbertal2021moire}%
  \BibitemOpen
  \bibfield  {author} {\bibinfo {author} {\bibfnamefont {D.}~\bibnamefont
  {Halbertal}}, \bibinfo {author} {\bibfnamefont {N.~R.}\ \bibnamefont
  {Finney}}, \bibinfo {author} {\bibfnamefont {S.~S.}\ \bibnamefont {Sunku}},
  \bibinfo {author} {\bibfnamefont {A.}~\bibnamefont {Kerelsky}}, \bibinfo
  {author} {\bibfnamefont {C.}~\bibnamefont {Rubio-Verd{\'u}}}, \bibinfo
  {author} {\bibfnamefont {S.}~\bibnamefont {Shabani}}, \bibinfo {author}
  {\bibfnamefont {L.}~\bibnamefont {Xian}}, \bibinfo {author} {\bibfnamefont
  {S.}~\bibnamefont {Carr}}, \bibinfo {author} {\bibfnamefont {S.}~\bibnamefont
  {Chen}}, \bibinfo {author} {\bibfnamefont {C.}~\bibnamefont {Zhang}},  \emph
  {et~al.},\ }\href {\doibase 10.1038/s41467-020-20428-1} {\bibfield  {journal}
  {\bibinfo  {journal} {Nature communications}\ }\textbf {\bibinfo {volume}
  {12}},\ \bibinfo {pages} {1} (\bibinfo {year} {2021})}\BibitemShut {NoStop}%
\bibitem [{\citenamefont {Cosma}\ \emph {et~al.}(2014)\citenamefont {Cosma},
  \citenamefont {Wallbank}, \citenamefont {Cheianov},\ and\ \citenamefont
  {Fal'Ko}}]{Cosma2014}%
  \BibitemOpen
  \bibfield  {author} {\bibinfo {author} {\bibfnamefont {D.~A.}\ \bibnamefont
  {Cosma}}, \bibinfo {author} {\bibfnamefont {J.~R.}\ \bibnamefont {Wallbank}},
  \bibinfo {author} {\bibfnamefont {V.}~\bibnamefont {Cheianov}}, \ and\
  \bibinfo {author} {\bibfnamefont {V.~I.}\ \bibnamefont {Fal'Ko}},\ }\href
  {\doibase 10.1039/c4fd00146j} {\bibfield  {journal} {\bibinfo  {journal}
  {Faraday Discussions}\ }\textbf {\bibinfo {volume} {173}},\ \bibinfo {pages}
  {137} (\bibinfo {year} {2014})},\ \Eprint {http://arxiv.org/abs/1408.1624}
  {1408.1624} \BibitemShut {NoStop}%
\bibitem [{\citenamefont {Tong}\ \emph {et~al.}(2017)\citenamefont {Tong},
  \citenamefont {Yu}, \citenamefont {Zhu}, \citenamefont {Wang}, \citenamefont
  {Xu},\ and\ \citenamefont {Yao}}]{Tong2017a}%
  \BibitemOpen
  \bibfield  {author} {\bibinfo {author} {\bibfnamefont {Q.}~\bibnamefont
  {Tong}}, \bibinfo {author} {\bibfnamefont {H.}~\bibnamefont {Yu}}, \bibinfo
  {author} {\bibfnamefont {Q.}~\bibnamefont {Zhu}}, \bibinfo {author}
  {\bibfnamefont {Y.}~\bibnamefont {Wang}}, \bibinfo {author} {\bibfnamefont
  {X.}~\bibnamefont {Xu}}, \ and\ \bibinfo {author} {\bibfnamefont
  {W.}~\bibnamefont {Yao}},\ }\href {\doibase 10.1038/nphys3968} {\bibfield
  {journal} {\bibinfo  {journal} {Nature Physics}\ }\textbf {\bibinfo {volume}
  {13}},\ \bibinfo {pages} {356} (\bibinfo {year} {2017})},\ \Eprint
  {http://arxiv.org/abs/1608.00115} {1608.00115} \BibitemShut {NoStop}%
\bibitem [{\citenamefont {Bertolazzi}\ \emph {et~al.}(2011)\citenamefont
  {Bertolazzi}, \citenamefont {Brivio},\ and\ \citenamefont
  {Kis}}]{Bertolazzi2011}%
  \BibitemOpen
  \bibfield  {author} {\bibinfo {author} {\bibfnamefont {S.}~\bibnamefont
  {Bertolazzi}}, \bibinfo {author} {\bibfnamefont {J.}~\bibnamefont {Brivio}},
  \ and\ \bibinfo {author} {\bibfnamefont {A.}~\bibnamefont {Kis}},\ }\href
  {\doibase 10.1021/nn203879f} {\bibfield  {journal} {\bibinfo  {journal} {ACS
  Nano}\ }\textbf {\bibinfo {volume} {5}},\ \bibinfo {pages} {9703} (\bibinfo
  {year} {2011})}\BibitemShut {NoStop}%
\bibitem [{\citenamefont {Bai}\ \emph {et~al.}(2020)\citenamefont {Bai},
  \citenamefont {Zhou}, \citenamefont {Wang}, \citenamefont {Wu}, \citenamefont
  {McGilly}, \citenamefont {Halbertal}, \citenamefont {Lo}, \citenamefont
  {Liu}, \citenamefont {Ardelean}, \citenamefont {Rivera}, \citenamefont
  {Finney}, \citenamefont {Yang}, \citenamefont {Basov}, \citenamefont {Yao},
  \citenamefont {Xu}, \citenamefont {Hone}, \citenamefont {Pasupathy},\ and\
  \citenamefont {Zhu}}]{Bai2019}%
  \BibitemOpen
  \bibfield  {author} {\bibinfo {author} {\bibfnamefont {Y.}~\bibnamefont
  {Bai}}, \bibinfo {author} {\bibfnamefont {L.}~\bibnamefont {Zhou}}, \bibinfo
  {author} {\bibfnamefont {J.}~\bibnamefont {Wang}}, \bibinfo {author}
  {\bibfnamefont {W.}~\bibnamefont {Wu}}, \bibinfo {author} {\bibfnamefont
  {L.~J.}\ \bibnamefont {McGilly}}, \bibinfo {author} {\bibfnamefont
  {D.}~\bibnamefont {Halbertal}}, \bibinfo {author} {\bibfnamefont {C.~F.~B.}\
  \bibnamefont {Lo}}, \bibinfo {author} {\bibfnamefont {F.}~\bibnamefont
  {Liu}}, \bibinfo {author} {\bibfnamefont {J.}~\bibnamefont {Ardelean}},
  \bibinfo {author} {\bibfnamefont {P.}~\bibnamefont {Rivera}}, \bibinfo
  {author} {\bibfnamefont {N.~R.}\ \bibnamefont {Finney}}, \bibinfo {author}
  {\bibfnamefont {X.~C.}\ \bibnamefont {Yang}}, \bibinfo {author}
  {\bibfnamefont {D.~N.}\ \bibnamefont {Basov}}, \bibinfo {author}
  {\bibfnamefont {W.}~\bibnamefont {Yao}}, \bibinfo {author} {\bibfnamefont
  {X.}~\bibnamefont {Xu}}, \bibinfo {author} {\bibfnamefont {J.}~\bibnamefont
  {Hone}}, \bibinfo {author} {\bibfnamefont {A.~N.}\ \bibnamefont {Pasupathy}},
  \ and\ \bibinfo {author} {\bibfnamefont {X.~Y.}\ \bibnamefont {Zhu}},\ }\href
  {\doibase 10.1038/s41563-020-0730-8} {\bibfield  {journal} {\bibinfo
  {journal} {Nature Materials}\ } (\bibinfo {year} {2020}),\
  10.1038/s41563-020-0730-8}\BibitemShut {NoStop}%
\bibitem [{\citenamefont {Yao}\ \emph {et~al.}(2021)\citenamefont {Yao},
  \citenamefont {Finney}, \citenamefont {Zhang}, \citenamefont {Moore},
  \citenamefont {Xian}, \citenamefont {Tancogne-Dejean}, \citenamefont {Liu},
  \citenamefont {Ardelean}, \citenamefont {Xu}, \citenamefont {Halbertal},
  \citenamefont {Watanabe}, \citenamefont {Taniguchi}, \citenamefont {Ochoa},
  \citenamefont {Asenjo-Garcia}, \citenamefont {Zhu}, \citenamefont {Basov},
  \citenamefont {Rubio}, \citenamefont {Dean}, \citenamefont {Hone},\ and\
  \citenamefont {Schuck}}]{Yao2021}%
  \BibitemOpen
  \bibfield  {author} {\bibinfo {author} {\bibfnamefont {K.}~\bibnamefont
  {Yao}}, \bibinfo {author} {\bibfnamefont {N.~R.}\ \bibnamefont {Finney}},
  \bibinfo {author} {\bibfnamefont {J.}~\bibnamefont {Zhang}}, \bibinfo
  {author} {\bibfnamefont {S.~L.}\ \bibnamefont {Moore}}, \bibinfo {author}
  {\bibfnamefont {L.}~\bibnamefont {Xian}}, \bibinfo {author} {\bibfnamefont
  {N.}~\bibnamefont {Tancogne-Dejean}}, \bibinfo {author} {\bibfnamefont
  {F.}~\bibnamefont {Liu}}, \bibinfo {author} {\bibfnamefont {J.}~\bibnamefont
  {Ardelean}}, \bibinfo {author} {\bibfnamefont {X.}~\bibnamefont {Xu}},
  \bibinfo {author} {\bibfnamefont {D.}~\bibnamefont {Halbertal}}, \bibinfo
  {author} {\bibfnamefont {K.}~\bibnamefont {Watanabe}}, \bibinfo {author}
  {\bibfnamefont {T.}~\bibnamefont {Taniguchi}}, \bibinfo {author}
  {\bibfnamefont {H.}~\bibnamefont {Ochoa}}, \bibinfo {author} {\bibfnamefont
  {A.}~\bibnamefont {Asenjo-Garcia}}, \bibinfo {author} {\bibfnamefont
  {X.}~\bibnamefont {Zhu}}, \bibinfo {author} {\bibfnamefont {D.~N.}\
  \bibnamefont {Basov}}, \bibinfo {author} {\bibfnamefont {A.}~\bibnamefont
  {Rubio}}, \bibinfo {author} {\bibfnamefont {C.~R.}\ \bibnamefont {Dean}},
  \bibinfo {author} {\bibfnamefont {J.}~\bibnamefont {Hone}}, \ and\ \bibinfo
  {author} {\bibfnamefont {P.~J.}\ \bibnamefont {Schuck}},\ }\href {\doibase
  10.1126/sciadv.abe8691} {\bibfield  {journal} {\bibinfo  {journal} {Science
  Advances}\ }\textbf {\bibinfo {volume} {7}},\ \bibinfo {pages} {1} (\bibinfo
  {year} {2021})}\BibitemShut {NoStop}%
\bibitem [{\citenamefont {Hu}\ \emph {et~al.}(2022)\citenamefont {Hu},
  \citenamefont {Wu}, \citenamefont {Huang}, \citenamefont {Dong},
  \citenamefont {Chen}, \citenamefont {Zhang}, \citenamefont {Lyu},
  \citenamefont {Ma}, \citenamefont {Watanabe}, \citenamefont {Taniguchi},
  \citenamefont {Xie}, \citenamefont {Li}, \citenamefont {Liang},\ and\
  \citenamefont {Shi}}]{Hu2022}%
  \BibitemOpen
  \bibfield  {author} {\bibinfo {author} {\bibfnamefont {C.}~\bibnamefont
  {Hu}}, \bibinfo {author} {\bibfnamefont {T.}~\bibnamefont {Wu}}, \bibinfo
  {author} {\bibfnamefont {X.}~\bibnamefont {Huang}}, \bibinfo {author}
  {\bibfnamefont {Y.}~\bibnamefont {Dong}}, \bibinfo {author} {\bibfnamefont
  {J.}~\bibnamefont {Chen}}, \bibinfo {author} {\bibfnamefont {Z.}~\bibnamefont
  {Zhang}}, \bibinfo {author} {\bibfnamefont {B.}~\bibnamefont {Lyu}}, \bibinfo
  {author} {\bibfnamefont {S.}~\bibnamefont {Ma}}, \bibinfo {author}
  {\bibfnamefont {K.}~\bibnamefont {Watanabe}}, \bibinfo {author}
  {\bibfnamefont {T.}~\bibnamefont {Taniguchi}}, \bibinfo {author}
  {\bibfnamefont {G.}~\bibnamefont {Xie}}, \bibinfo {author} {\bibfnamefont
  {X.}~\bibnamefont {Li}}, \bibinfo {author} {\bibfnamefont {Q.}~\bibnamefont
  {Liang}}, \ and\ \bibinfo {author} {\bibfnamefont {Z.}~\bibnamefont {Shi}},\
  }\href {\doibase 10.1038/s41598-021-04030-z} {\bibfield  {journal} {\bibinfo
  {journal} {Scientific Reports}\ }\textbf {\bibinfo {volume} {12}},\ \bibinfo
  {pages} {1} (\bibinfo {year} {2022})}\BibitemShut {NoStop}%
\bibitem [{\citenamefont {Finney}\ \emph {et~al.}(2019)\citenamefont {Finney},
  \citenamefont {Yankowitz}, \citenamefont {Muraleetharan}, \citenamefont
  {Watanabe}, \citenamefont {Taniguchi}, \citenamefont {Dean},\ and\
  \citenamefont {Hone}}]{RN33}%
  \BibitemOpen
  \bibfield  {author} {\bibinfo {author} {\bibfnamefont {N.~R.}\ \bibnamefont
  {Finney}}, \bibinfo {author} {\bibfnamefont {M.}~\bibnamefont {Yankowitz}},
  \bibinfo {author} {\bibfnamefont {L.}~\bibnamefont {Muraleetharan}}, \bibinfo
  {author} {\bibfnamefont {K.}~\bibnamefont {Watanabe}}, \bibinfo {author}
  {\bibfnamefont {T.}~\bibnamefont {Taniguchi}}, \bibinfo {author}
  {\bibfnamefont {C.~R.}\ \bibnamefont {Dean}}, \ and\ \bibinfo {author}
  {\bibfnamefont {J.}~\bibnamefont {Hone}},\ }\href {\doibase
  10.1038/s41565-019-0547-2} {\bibfield  {journal} {\bibinfo  {journal} {Nature
  Nanotechnology}\ }\textbf {\bibinfo {volume} {14}},\ \bibinfo {pages} {1029}
  (\bibinfo {year} {2019})},\ \Eprint {http://arxiv.org/abs/1903.11191}
  {1903.11191} \BibitemShut {NoStop}%
\bibitem [{\citenamefont {Kang}\ \emph {et~al.}(2013)\citenamefont {Kang},
  \citenamefont {Tongay}, \citenamefont {Zhou}, \citenamefont {Li},\ and\
  \citenamefont {Wu}}]{Kang2013a}%
  \BibitemOpen
  \bibfield  {author} {\bibinfo {author} {\bibfnamefont {J.}~\bibnamefont
  {Kang}}, \bibinfo {author} {\bibfnamefont {S.}~\bibnamefont {Tongay}},
  \bibinfo {author} {\bibfnamefont {J.}~\bibnamefont {Zhou}}, \bibinfo {author}
  {\bibfnamefont {J.}~\bibnamefont {Li}}, \ and\ \bibinfo {author}
  {\bibfnamefont {J.}~\bibnamefont {Wu}},\ }\href {\doibase 10.1063/1.4774090}
  {\bibfield  {journal} {\bibinfo  {journal} {Applied Physics Letters}\
  }\textbf {\bibinfo {volume} {102}} (\bibinfo {year} {2013}),\
  10.1063/1.4774090}\BibitemShut {NoStop}%
\bibitem [{\citenamefont {Project}(2020)}]{osti_1192989}%
  \BibitemOpen
  \bibfield  {author} {\bibinfo {author} {\bibfnamefont {T.~M.}\ \bibnamefont
  {Project}},\ }\href {\doibase 10.17188/1192989} {\  (\bibinfo {year}
  {2020}),\ 10.17188/1192989}\BibitemShut {NoStop}%
\bibitem [{\citenamefont {Schutte}\ \emph {et~al.}(1987)\citenamefont
  {Schutte}, \citenamefont {{De Boer}},\ and\ \citenamefont
  {Jellinek}}]{SCHUTTE1987207}%
  \BibitemOpen
  \bibfield  {author} {\bibinfo {author} {\bibfnamefont {W.}~\bibnamefont
  {Schutte}}, \bibinfo {author} {\bibfnamefont {J.}~\bibnamefont {{De Boer}}},
  \ and\ \bibinfo {author} {\bibfnamefont {F.}~\bibnamefont {Jellinek}},\
  }\href {\doibase https://doi.org/10.1016/0022-4596(87)90057-0} {\bibfield
  {journal} {\bibinfo  {journal} {Journal of Solid State Chemistry}\ }\textbf
  {\bibinfo {volume} {70}},\ \bibinfo {pages} {207} (\bibinfo {year}
  {1987})}\BibitemShut {NoStop}%
\bibitem [{\citenamefont {Yankowitz}\ \emph {et~al.}(2012)\citenamefont
  {Yankowitz}, \citenamefont {Xue}, \citenamefont {Cormode}, \citenamefont
  {Sanchez-Yamagishi}, \citenamefont {Watanabe}, \citenamefont {Taniguchi},
  \citenamefont {Jarillo-Herrero}, \citenamefont {Jacquod},\ and\ \citenamefont
  {Leroy}}]{Yankowitz2012}%
  \BibitemOpen
  \bibfield  {author} {\bibinfo {author} {\bibfnamefont {M.}~\bibnamefont
  {Yankowitz}}, \bibinfo {author} {\bibfnamefont {J.}~\bibnamefont {Xue}},
  \bibinfo {author} {\bibfnamefont {D.}~\bibnamefont {Cormode}}, \bibinfo
  {author} {\bibfnamefont {J.~D.}\ \bibnamefont {Sanchez-Yamagishi}}, \bibinfo
  {author} {\bibfnamefont {K.}~\bibnamefont {Watanabe}}, \bibinfo {author}
  {\bibfnamefont {T.}~\bibnamefont {Taniguchi}}, \bibinfo {author}
  {\bibfnamefont {P.}~\bibnamefont {Jarillo-Herrero}}, \bibinfo {author}
  {\bibfnamefont {P.}~\bibnamefont {Jacquod}}, \ and\ \bibinfo {author}
  {\bibfnamefont {B.~J.}\ \bibnamefont {Leroy}},\ }\href {\doibase
  10.1038/nphys2272} {\bibfield  {journal} {\bibinfo  {journal} {Nature
  Physics}\ }\textbf {\bibinfo {volume} {8}},\ \bibinfo {pages} {382} (\bibinfo
  {year} {2012})},\ \Eprint {http://arxiv.org/abs/1202.2870} {1202.2870}
  \BibitemShut {NoStop}%
\bibitem [{ESM()}]{ESM}%
  \BibitemOpen
  \href@noop {} {\bibinfo  {journal}
  {https://github.com/QuantumPhotonicsLab/Strained-Moire-Visualization}\
  }\BibitemShut {NoStop}%
\bibitem [{\citenamefont {Iff}\ \emph {et~al.}(2019)\citenamefont {Iff},
  \citenamefont {Tedeschi}, \citenamefont {Mart{\'{i}}n-S{\'{a}}nchez},
  \citenamefont {Mocza{\l}a-Dusanowska}, \citenamefont {Tongay}, \citenamefont
  {Yumigeta}, \citenamefont {Taboada-Guti{\'{e}}rrez}, \citenamefont
  {Savaresi}, \citenamefont {Rastelli}, \citenamefont {Alonso-Gonz{\'{a}}lez},
  \citenamefont {H{\"{o}}fling}, \citenamefont {Trotta},\ and\ \citenamefont
  {Schneider}}]{Iff2019}%
  \BibitemOpen
\bibfield  {journal} {  }\bibfield  {author} {\bibinfo {author} {\bibfnamefont
  {O.}~\bibnamefont {Iff}}, \bibinfo {author} {\bibfnamefont {D.}~\bibnamefont
  {Tedeschi}}, \bibinfo {author} {\bibfnamefont {J.}~\bibnamefont
  {Mart{\'{i}}n-S{\'{a}}nchez}}, \bibinfo {author} {\bibfnamefont
  {M.}~\bibnamefont {Mocza{\l}a-Dusanowska}}, \bibinfo {author} {\bibfnamefont
  {S.}~\bibnamefont {Tongay}}, \bibinfo {author} {\bibfnamefont
  {K.}~\bibnamefont {Yumigeta}}, \bibinfo {author} {\bibfnamefont
  {J.}~\bibnamefont {Taboada-Guti{\'{e}}rrez}}, \bibinfo {author}
  {\bibfnamefont {M.}~\bibnamefont {Savaresi}}, \bibinfo {author}
  {\bibfnamefont {A.}~\bibnamefont {Rastelli}}, \bibinfo {author}
  {\bibfnamefont {P.}~\bibnamefont {Alonso-Gonz{\'{a}}lez}}, \bibinfo {author}
  {\bibfnamefont {S.}~\bibnamefont {H{\"{o}}fling}}, \bibinfo {author}
  {\bibfnamefont {R.}~\bibnamefont {Trotta}}, \ and\ \bibinfo {author}
  {\bibfnamefont {C.}~\bibnamefont {Schneider}},\ }\href {\doibase
  10.1021/acs.nanolett.9b02221} {\bibfield  {journal} {\bibinfo  {journal}
  {Nano Letters}\ }\textbf {\bibinfo {volume} {19}},\ \bibinfo {pages} {6931}
  (\bibinfo {year} {2019})},\ \Eprint {http://arxiv.org/abs/1909.01784}
  {1909.01784} \BibitemShut {NoStop}%
\bibitem [{\citenamefont {Wang}\ \emph
  {et~al.}(2020{\natexlab{b}})\citenamefont {Wang}, \citenamefont {Makk},
  \citenamefont {Zihlmann}, \citenamefont {Baumgartner}, \citenamefont
  {Indolese}, \citenamefont {Watanabe}, \citenamefont {Taniguchi},\ and\
  \citenamefont {Sch{\"{o}}nenberger}}]{Wang2020}%
  \BibitemOpen
  \bibfield  {author} {\bibinfo {author} {\bibfnamefont {L.}~\bibnamefont
  {Wang}}, \bibinfo {author} {\bibfnamefont {P.}~\bibnamefont {Makk}}, \bibinfo
  {author} {\bibfnamefont {S.}~\bibnamefont {Zihlmann}}, \bibinfo {author}
  {\bibfnamefont {A.}~\bibnamefont {Baumgartner}}, \bibinfo {author}
  {\bibfnamefont {D.~I.}\ \bibnamefont {Indolese}}, \bibinfo {author}
  {\bibfnamefont {K.}~\bibnamefont {Watanabe}}, \bibinfo {author}
  {\bibfnamefont {T.}~\bibnamefont {Taniguchi}}, \ and\ \bibinfo {author}
  {\bibfnamefont {C.}~\bibnamefont {Sch{\"{o}}nenberger}},\ }\href {\doibase
  10.1103/physrevlett.124.157701} {\bibfield  {journal} {\bibinfo  {journal}
  {Physical Review Letters}\ }\textbf {\bibinfo {volume} {124}},\ \bibinfo
  {pages} {1} (\bibinfo {year} {2020}{\natexlab{b}})}\BibitemShut {NoStop}%
\bibitem [{\citenamefont {Hicks}\ \emph {et~al.}(2014)\citenamefont {Hicks},
  \citenamefont {Barber}, \citenamefont {Edkins}, \citenamefont {Brodsky},\
  and\ \citenamefont {Mackenzie}}]{Hicks2014}%
  \BibitemOpen
  \bibfield  {author} {\bibinfo {author} {\bibfnamefont {C.~W.}\ \bibnamefont
  {Hicks}}, \bibinfo {author} {\bibfnamefont {M.~E.}\ \bibnamefont {Barber}},
  \bibinfo {author} {\bibfnamefont {S.~D.}\ \bibnamefont {Edkins}}, \bibinfo
  {author} {\bibfnamefont {D.~O.}\ \bibnamefont {Brodsky}}, \ and\ \bibinfo
  {author} {\bibfnamefont {A.~P.}\ \bibnamefont {Mackenzie}},\ }\href {\doibase
  10.1063/1.4881611} {\bibfield  {journal} {\bibinfo  {journal} {Review of
  Scientific Instruments}\ }\textbf {\bibinfo {volume} {85}} (\bibinfo {year}
  {2014}),\ 10.1063/1.4881611},\ \Eprint {http://arxiv.org/abs/1403.4368}
  {1403.4368} \BibitemShut {NoStop}%
\bibitem [{\citenamefont {Zhang}\ \emph {et~al.}(2022)\citenamefont {Zhang},
  \citenamefont {Wang}, \citenamefont {Hu}, \citenamefont {Wan}, \citenamefont
  {Zheliuk}, \citenamefont {Liang}, \citenamefont {Peng}, \citenamefont
  {Zeng},\ and\ \citenamefont {Ye}}]{Zhang2022}%
  \BibitemOpen
  \bibfield  {author} {\bibinfo {author} {\bibfnamefont {L.}~\bibnamefont
  {Zhang}}, \bibinfo {author} {\bibfnamefont {Y.}~\bibnamefont {Wang}},
  \bibinfo {author} {\bibfnamefont {R.}~\bibnamefont {Hu}}, \bibinfo {author}
  {\bibfnamefont {P.}~\bibnamefont {Wan}}, \bibinfo {author} {\bibfnamefont
  {O.}~\bibnamefont {Zheliuk}}, \bibinfo {author} {\bibfnamefont
  {M.}~\bibnamefont {Liang}}, \bibinfo {author} {\bibfnamefont
  {X.}~\bibnamefont {Peng}}, \bibinfo {author} {\bibfnamefont {Y.-J.}\
  \bibnamefont {Zeng}}, \ and\ \bibinfo {author} {\bibfnamefont
  {J.}~\bibnamefont {Ye}},\ }\href {\doibase 10.1021/acs.nanolett.1c04400}
  {\bibfield  {journal} {\bibinfo  {journal} {Nano Letters}\ }\textbf {\bibinfo
  {volume} {22}},\ \bibinfo {pages} {3204} (\bibinfo {year}
  {2022})}\BibitemShut {NoStop}%
\bibitem [{\citenamefont {Wang}\ \emph
  {et~al.}(2019{\natexlab{a}})\citenamefont {Wang}, \citenamefont {Zihlmann},
  \citenamefont {Baumgartner}, \citenamefont {Overbeck}, \citenamefont
  {Watanabe}, \citenamefont {Taniguchi}, \citenamefont {Makk},\ and\
  \citenamefont {Sch{\"{o}}nenberger}}]{RN125}%
  \BibitemOpen
  \bibfield  {author} {\bibinfo {author} {\bibfnamefont {L.}~\bibnamefont
  {Wang}}, \bibinfo {author} {\bibfnamefont {S.}~\bibnamefont {Zihlmann}},
  \bibinfo {author} {\bibfnamefont {A.}~\bibnamefont {Baumgartner}}, \bibinfo
  {author} {\bibfnamefont {J.}~\bibnamefont {Overbeck}}, \bibinfo {author}
  {\bibfnamefont {K.}~\bibnamefont {Watanabe}}, \bibinfo {author}
  {\bibfnamefont {T.}~\bibnamefont {Taniguchi}}, \bibinfo {author}
  {\bibfnamefont {P.}~\bibnamefont {Makk}}, \ and\ \bibinfo {author}
  {\bibfnamefont {C.}~\bibnamefont {Sch{\"{o}}nenberger}},\ }\href {\doibase
  10.1021/acs.nanolett.9b01491} {\bibfield  {journal} {\bibinfo  {journal}
  {Nano Letters}\ }\textbf {\bibinfo {volume} {19}},\ \bibinfo {pages} {4097}
  (\bibinfo {year} {2019}{\natexlab{a}})},\ \Eprint
  {http://arxiv.org/abs/1904.06737} {1904.06737} \BibitemShut {NoStop}%
\bibitem [{\citenamefont {Liu}\ \emph {et~al.}(2018)\citenamefont {Liu},
  \citenamefont {Guo}, \citenamefont {Zhu}, \citenamefont {Liao}, \citenamefont
  {Lee}, \citenamefont {Ding}, \citenamefont {Shakir}, \citenamefont {Gambin},
  \citenamefont {Huang},\ and\ \citenamefont {Duan}}]{Liu2018}%
  \BibitemOpen
  \bibfield  {author} {\bibinfo {author} {\bibfnamefont {Y.}~\bibnamefont
  {Liu}}, \bibinfo {author} {\bibfnamefont {J.}~\bibnamefont {Guo}}, \bibinfo
  {author} {\bibfnamefont {E.}~\bibnamefont {Zhu}}, \bibinfo {author}
  {\bibfnamefont {L.}~\bibnamefont {Liao}}, \bibinfo {author} {\bibfnamefont
  {S.~J.}\ \bibnamefont {Lee}}, \bibinfo {author} {\bibfnamefont
  {M.}~\bibnamefont {Ding}}, \bibinfo {author} {\bibfnamefont {I.}~\bibnamefont
  {Shakir}}, \bibinfo {author} {\bibfnamefont {V.}~\bibnamefont {Gambin}},
  \bibinfo {author} {\bibfnamefont {Y.}~\bibnamefont {Huang}}, \ and\ \bibinfo
  {author} {\bibfnamefont {X.}~\bibnamefont {Duan}},\ }\href {\doibase
  10.1038/s41586-018-0129-8} {\bibfield  {journal} {\bibinfo  {journal}
  {Nature}\ }\textbf {\bibinfo {volume} {557}},\ \bibinfo {pages} {696}
  (\bibinfo {year} {2018})}\BibitemShut {NoStop}%
\bibitem [{\citenamefont {Huder}\ \emph {et~al.}(2018)\citenamefont {Huder},
  \citenamefont {Artaud}, \citenamefont {Quang}, \citenamefont
  {de~Laissardière}, \citenamefont {Jansen}, \citenamefont {Lapertot},
  \citenamefont {Chapelier},\ and\ \citenamefont {Renard}}]{Huder2018}%
  \BibitemOpen
  \bibfield  {author} {\bibinfo {author} {\bibfnamefont {L.}~\bibnamefont
  {Huder}}, \bibinfo {author} {\bibfnamefont {A.}~\bibnamefont {Artaud}},
  \bibinfo {author} {\bibfnamefont {T.~L.}\ \bibnamefont {Quang}}, \bibinfo
  {author} {\bibfnamefont {G.~T.}\ \bibnamefont {de~Laissardière}}, \bibinfo
  {author} {\bibfnamefont {A.~G.}\ \bibnamefont {Jansen}}, \bibinfo {author}
  {\bibfnamefont {G.}~\bibnamefont {Lapertot}}, \bibinfo {author}
  {\bibfnamefont {C.}~\bibnamefont {Chapelier}}, \ and\ \bibinfo {author}
  {\bibfnamefont {V.~T.}\ \bibnamefont {Renard}},\ }\href {\doibase
  10.1103/PhysRevLett.120.156405} {\bibfield  {journal} {\bibinfo  {journal}
  {Physical Review Letters}\ }\textbf {\bibinfo {volume} {120}},\ \bibinfo
  {pages} {156405} (\bibinfo {year} {2018})}\BibitemShut {NoStop}%
\bibitem [{\citenamefont {Guan}\ \emph {et~al.}(2015)\citenamefont {Guan},
  \citenamefont {Kumaravadivel}, \citenamefont {Averin},\ and\ \citenamefont
  {Du}}]{Guan2015}%
  \BibitemOpen
  \bibfield  {author} {\bibinfo {author} {\bibfnamefont {F.}~\bibnamefont
  {Guan}}, \bibinfo {author} {\bibfnamefont {P.}~\bibnamefont {Kumaravadivel}},
  \bibinfo {author} {\bibfnamefont {D.~V.}\ \bibnamefont {Averin}}, \ and\
  \bibinfo {author} {\bibfnamefont {X.}~\bibnamefont {Du}},\ }\href {\doibase
  10.1063/1.4935239} {\bibfield  {journal} {\bibinfo  {journal} {Applied
  Physics Letters}\ }\textbf {\bibinfo {volume} {107}} (\bibinfo {year}
  {2015}),\ 10.1063/1.4935239},\ \Eprint {http://arxiv.org/abs/1506.01643}
  {1506.01643} \BibitemShut {NoStop}%
\bibitem [{\citenamefont {McRae}\ \emph {et~al.}(2019)\citenamefont {McRae},
  \citenamefont {Wei},\ and\ \citenamefont {Champagne}}]{McRae2019}%
  \BibitemOpen
  \bibfield  {author} {\bibinfo {author} {\bibfnamefont {A.~C.}\ \bibnamefont
  {McRae}}, \bibinfo {author} {\bibfnamefont {G.}~\bibnamefont {Wei}}, \ and\
  \bibinfo {author} {\bibfnamefont {A.~R.}\ \bibnamefont {Champagne}},\ }\href
  {\doibase 10.1103/PhysRevApplied.11.054019} {\bibfield  {journal} {\bibinfo
  {journal} {Physical Review Applied}\ }\textbf {\bibinfo {volume} {11}},\
  \bibinfo {pages} {1} (\bibinfo {year} {2019})},\ \Eprint
  {http://arxiv.org/abs/1809.09679} {1809.09679} \BibitemShut {NoStop}%
\bibitem [{\citenamefont {Wang}\ \emph
  {et~al.}(2020{\natexlab{c}})\citenamefont {Wang}, \citenamefont {Makk},
  \citenamefont {Zihlmann}, \citenamefont {Baumgartner}, \citenamefont
  {Indolese}, \citenamefont {Watanabe}, \citenamefont {Taniguchi},\ and\
  \citenamefont {Sch{\"{o}}nenberger}}]{Wang2020f}%
  \BibitemOpen
  \bibfield  {author} {\bibinfo {author} {\bibfnamefont {L.}~\bibnamefont
  {Wang}}, \bibinfo {author} {\bibfnamefont {P.}~\bibnamefont {Makk}}, \bibinfo
  {author} {\bibfnamefont {S.}~\bibnamefont {Zihlmann}}, \bibinfo {author}
  {\bibfnamefont {A.}~\bibnamefont {Baumgartner}}, \bibinfo {author}
  {\bibfnamefont {D.~I.}\ \bibnamefont {Indolese}}, \bibinfo {author}
  {\bibfnamefont {K.}~\bibnamefont {Watanabe}}, \bibinfo {author}
  {\bibfnamefont {T.}~\bibnamefont {Taniguchi}}, \ and\ \bibinfo {author}
  {\bibfnamefont {C.}~\bibnamefont {Sch{\"{o}}nenberger}},\ }\href {\doibase
  10.1103/PhysRevLett.124.157701} {\bibfield  {journal} {\bibinfo  {journal}
  {Physical Review Letters}\ }\textbf {\bibinfo {volume} {124}},\ \bibinfo
  {pages} {1} (\bibinfo {year} {2020}{\natexlab{c}})},\ \Eprint
  {http://arxiv.org/abs/1909.13484} {1909.13484} \BibitemShut {NoStop}%
\bibitem [{\citenamefont {Guan}\ and\ \citenamefont {Du}(2017)}]{Guan2017}%
  \BibitemOpen
  \bibfield  {author} {\bibinfo {author} {\bibfnamefont {F.}~\bibnamefont
  {Guan}}\ and\ \bibinfo {author} {\bibfnamefont {X.}~\bibnamefont {Du}},\
  }\href {\doibase 10.1021/acs.nanolett.7b03618} {\bibfield  {journal}
  {\bibinfo  {journal} {Nano Letters}\ }\textbf {\bibinfo {volume} {17}},\
  \bibinfo {pages} {7009} (\bibinfo {year} {2017})}\BibitemShut {NoStop}%
\bibitem [{\citenamefont {Frisenda}\ \emph {et~al.}(2017)\citenamefont
  {Frisenda}, \citenamefont {Dr{\"{u}}ppel}, \citenamefont {Schmidt},
  \citenamefont {{Michaelis de Vasconcellos}}, \citenamefont {{Perez de Lara}},
  \citenamefont {Bratschitsch}, \citenamefont {Rohlfing},\ and\ \citenamefont
  {Castellanos-Gomez}}]{RN83}%
  \BibitemOpen
  \bibfield  {author} {\bibinfo {author} {\bibfnamefont {R.}~\bibnamefont
  {Frisenda}}, \bibinfo {author} {\bibfnamefont {M.}~\bibnamefont
  {Dr{\"{u}}ppel}}, \bibinfo {author} {\bibfnamefont {R.}~\bibnamefont
  {Schmidt}}, \bibinfo {author} {\bibfnamefont {S.}~\bibnamefont {{Michaelis de
  Vasconcellos}}}, \bibinfo {author} {\bibfnamefont {D.}~\bibnamefont {{Perez
  de Lara}}}, \bibinfo {author} {\bibfnamefont {R.}~\bibnamefont
  {Bratschitsch}}, \bibinfo {author} {\bibfnamefont {M.}~\bibnamefont
  {Rohlfing}}, \ and\ \bibinfo {author} {\bibfnamefont {A.}~\bibnamefont
  {Castellanos-Gomez}},\ }\href {\doibase 10.1038/s41699-017-0013-7} {\bibfield
   {journal} {\bibinfo  {journal} {npj 2D Materials and Applications}\ }\textbf
  {\bibinfo {volume} {1}},\ \bibinfo {pages} {10} (\bibinfo {year} {2017})},\
  \Eprint {http://arxiv.org/abs/1703.02831} {1703.02831} \BibitemShut {NoStop}%
\bibitem [{\citenamefont {Aslan}\ \emph {et~al.}(2018)\citenamefont {Aslan},
  \citenamefont {Deng},\ and\ \citenamefont {Heinz}}]{Aslan2018}%
  \BibitemOpen
  \bibfield  {author} {\bibinfo {author} {\bibfnamefont {O.~B.}\ \bibnamefont
  {Aslan}}, \bibinfo {author} {\bibfnamefont {M.}~\bibnamefont {Deng}}, \ and\
  \bibinfo {author} {\bibfnamefont {T.~F.}\ \bibnamefont {Heinz}},\ }\href
  {\doibase 10.1103/PhysRevB.98.115308} {\bibfield  {journal} {\bibinfo
  {journal} {Physical Review B}\ }\textbf {\bibinfo {volume} {98}},\ \bibinfo
  {pages} {115308} (\bibinfo {year} {2018})}\BibitemShut {NoStop}%
\bibitem [{\citenamefont {Hui}\ \emph {et~al.}(2013)\citenamefont {Hui},
  \citenamefont {Liu}, \citenamefont {Jie}, \citenamefont {Chan}, \citenamefont
  {Hao}, \citenamefont {Hsu}, \citenamefont {Li}, \citenamefont {Guo},\ and\
  \citenamefont {Lau}}]{Hui2013}%
  \BibitemOpen
  \bibfield  {author} {\bibinfo {author} {\bibfnamefont {Y.~Y.}\ \bibnamefont
  {Hui}}, \bibinfo {author} {\bibfnamefont {X.}~\bibnamefont {Liu}}, \bibinfo
  {author} {\bibfnamefont {W.}~\bibnamefont {Jie}}, \bibinfo {author}
  {\bibfnamefont {N.~Y.}\ \bibnamefont {Chan}}, \bibinfo {author}
  {\bibfnamefont {J.}~\bibnamefont {Hao}}, \bibinfo {author} {\bibfnamefont
  {Y.~T.}\ \bibnamefont {Hsu}}, \bibinfo {author} {\bibfnamefont {L.~J.}\
  \bibnamefont {Li}}, \bibinfo {author} {\bibfnamefont {W.}~\bibnamefont
  {Guo}}, \ and\ \bibinfo {author} {\bibfnamefont {S.~P.}\ \bibnamefont
  {Lau}},\ }\href {\doibase 10.1021/nn4024834} {\bibfield  {journal} {\bibinfo
  {journal} {ACS Nano}\ }\textbf {\bibinfo {volume} {7}},\ \bibinfo {pages}
  {7126} (\bibinfo {year} {2013})}\BibitemShut {NoStop}%
\bibitem [{\citenamefont {Cenker}\ \emph {et~al.}(2022)\citenamefont {Cenker},
  \citenamefont {Sivakumar}, \citenamefont {Xie}, \citenamefont {Miller},
  \citenamefont {Thijssen}, \citenamefont {Liu}, \citenamefont {Dismukes},
  \citenamefont {Fonseca}, \citenamefont {Anderson}, \citenamefont {Zhu},
  \citenamefont {Roy}, \citenamefont {Xiao}, \citenamefont {Chu}, \citenamefont
  {Cao},\ and\ \citenamefont {Xu}}]{Cenker2022}%
  \BibitemOpen
  \bibfield  {author} {\bibinfo {author} {\bibfnamefont {J.}~\bibnamefont
  {Cenker}}, \bibinfo {author} {\bibfnamefont {S.}~\bibnamefont {Sivakumar}},
  \bibinfo {author} {\bibfnamefont {K.}~\bibnamefont {Xie}}, \bibinfo {author}
  {\bibfnamefont {A.}~\bibnamefont {Miller}}, \bibinfo {author} {\bibfnamefont
  {P.}~\bibnamefont {Thijssen}}, \bibinfo {author} {\bibfnamefont
  {Z.}~\bibnamefont {Liu}}, \bibinfo {author} {\bibfnamefont {A.}~\bibnamefont
  {Dismukes}}, \bibinfo {author} {\bibfnamefont {J.}~\bibnamefont {Fonseca}},
  \bibinfo {author} {\bibfnamefont {E.}~\bibnamefont {Anderson}}, \bibinfo
  {author} {\bibfnamefont {X.}~\bibnamefont {Zhu}}, \bibinfo {author}
  {\bibfnamefont {X.}~\bibnamefont {Roy}}, \bibinfo {author} {\bibfnamefont
  {D.}~\bibnamefont {Xiao}}, \bibinfo {author} {\bibfnamefont {J.~H.}\
  \bibnamefont {Chu}}, \bibinfo {author} {\bibfnamefont {T.}~\bibnamefont
  {Cao}}, \ and\ \bibinfo {author} {\bibfnamefont {X.}~\bibnamefont {Xu}},\
  }\href {\doibase 10.1038/s41565-021-01052-6} {\bibfield  {journal} {\bibinfo
  {journal} {Nature Nanotechnology}\ }\textbf {\bibinfo {volume} {17}},\
  \bibinfo {pages} {256} (\bibinfo {year} {2022})}\BibitemShut {NoStop}%
\bibitem [{\citenamefont {Kuklewicz}\ \emph {et~al.}(2012)\citenamefont
  {Kuklewicz}, \citenamefont {Malein}, \citenamefont {Petroff},\ and\
  \citenamefont {Gerardot}}]{Kuklewicz2012}%
  \BibitemOpen
  \bibfield  {author} {\bibinfo {author} {\bibfnamefont {C.~E.}\ \bibnamefont
  {Kuklewicz}}, \bibinfo {author} {\bibfnamefont {R.~N.~E.}\ \bibnamefont
  {Malein}}, \bibinfo {author} {\bibfnamefont {P.~M.}\ \bibnamefont {Petroff}},
  \ and\ \bibinfo {author} {\bibfnamefont {B.~D.}\ \bibnamefont {Gerardot}},\
  }\href {\doibase 10.1021/nl301621u} {\bibfield  {journal} {\bibinfo
  {journal} {Nano Letters}\ }\textbf {\bibinfo {volume} {12}},\ \bibinfo
  {pages} {3761} (\bibinfo {year} {2012})}\BibitemShut {NoStop}%
\bibitem [{\citenamefont {Seidl}\ \emph {et~al.}(2006)\citenamefont {Seidl},
  \citenamefont {Kroner}, \citenamefont {Högele}, \citenamefont {Karrai},
  \citenamefont {Warburton}, \citenamefont {Badolato},\ and\ \citenamefont
  {Petroff}}]{Seidl2006}%
  \BibitemOpen
  \bibfield  {author} {\bibinfo {author} {\bibfnamefont {S.}~\bibnamefont
  {Seidl}}, \bibinfo {author} {\bibfnamefont {M.}~\bibnamefont {Kroner}},
  \bibinfo {author} {\bibfnamefont {A.}~\bibnamefont {Högele}}, \bibinfo
  {author} {\bibfnamefont {K.}~\bibnamefont {Karrai}}, \bibinfo {author}
  {\bibfnamefont {R.~J.}\ \bibnamefont {Warburton}}, \bibinfo {author}
  {\bibfnamefont {A.}~\bibnamefont {Badolato}}, \ and\ \bibinfo {author}
  {\bibfnamefont {P.~M.}\ \bibnamefont {Petroff}},\ }\href {\doibase
  10.1063/1.2204843} {\bibfield  {journal} {\bibinfo  {journal} {Applied
  Physics Letters}\ }\textbf {\bibinfo {volume} {88}},\ \bibinfo {pages}
  {203113} (\bibinfo {year} {2006})}\BibitemShut {NoStop}%
\bibitem [{\citenamefont {Woods}\ \emph {et~al.}(2014)\citenamefont {Woods},
  \citenamefont {Britnell}, \citenamefont {Eckmann}, \citenamefont {Ma},
  \citenamefont {Lu}, \citenamefont {Guo}, \citenamefont {Lin}, \citenamefont
  {Yu}, \citenamefont {Cao}, \citenamefont {Gorbachev}, \citenamefont
  {Kretinin}, \citenamefont {Park}, \citenamefont {Ponomarenko}, \citenamefont
  {Katsnelson}, \citenamefont {Gornostyrev}, \citenamefont {Watanabe},
  \citenamefont {Taniguchi}, \citenamefont {Casiraghi}, \citenamefont {Gao},
  \citenamefont {Geim},\ and\ \citenamefont {Novoselov}}]{Woods2014}%
  \BibitemOpen
  \bibfield  {author} {\bibinfo {author} {\bibfnamefont {C.~R.}\ \bibnamefont
  {Woods}}, \bibinfo {author} {\bibfnamefont {L.}~\bibnamefont {Britnell}},
  \bibinfo {author} {\bibfnamefont {A.}~\bibnamefont {Eckmann}}, \bibinfo
  {author} {\bibfnamefont {R.~S.}\ \bibnamefont {Ma}}, \bibinfo {author}
  {\bibfnamefont {J.~C.}\ \bibnamefont {Lu}}, \bibinfo {author} {\bibfnamefont
  {H.~M.}\ \bibnamefont {Guo}}, \bibinfo {author} {\bibfnamefont
  {X.}~\bibnamefont {Lin}}, \bibinfo {author} {\bibfnamefont {G.~L.}\
  \bibnamefont {Yu}}, \bibinfo {author} {\bibfnamefont {Y.}~\bibnamefont
  {Cao}}, \bibinfo {author} {\bibfnamefont {R.~V.}\ \bibnamefont {Gorbachev}},
  \bibinfo {author} {\bibfnamefont {A.~V.}\ \bibnamefont {Kretinin}}, \bibinfo
  {author} {\bibfnamefont {J.}~\bibnamefont {Park}}, \bibinfo {author}
  {\bibfnamefont {L.~A.}\ \bibnamefont {Ponomarenko}}, \bibinfo {author}
  {\bibfnamefont {M.~I.}\ \bibnamefont {Katsnelson}}, \bibinfo {author}
  {\bibfnamefont {Y.~N.}\ \bibnamefont {Gornostyrev}}, \bibinfo {author}
  {\bibfnamefont {K.}~\bibnamefont {Watanabe}}, \bibinfo {author}
  {\bibfnamefont {T.}~\bibnamefont {Taniguchi}}, \bibinfo {author}
  {\bibfnamefont {C.}~\bibnamefont {Casiraghi}}, \bibinfo {author}
  {\bibfnamefont {H.~J.}\ \bibnamefont {Gao}}, \bibinfo {author} {\bibfnamefont
  {A.~K.}\ \bibnamefont {Geim}}, \ and\ \bibinfo {author} {\bibfnamefont
  {K.~S.}\ \bibnamefont {Novoselov}},\ }\href {\doibase 10.1038/nphys2954}
  {\bibfield  {journal} {\bibinfo  {journal} {Nature Physics}\ }\textbf
  {\bibinfo {volume} {10}},\ \bibinfo {pages} {451} (\bibinfo {year} {2014})},\
  \Eprint {http://arxiv.org/abs/1401.2637} {1401.2637} \BibitemShut {NoStop}%
\bibitem [{\citenamefont {Weston}\ \emph {et~al.}(2020)\citenamefont {Weston},
  \citenamefont {Zou}, \citenamefont {Enaldiev}, \citenamefont {Summerfield},
  \citenamefont {Clark}, \citenamefont {Z{\'{o}}lyomi}, \citenamefont {Graham},
  \citenamefont {Yelgel}, \citenamefont {Magorrian}, \citenamefont {Zhou},
  \citenamefont {Zultak}, \citenamefont {Hopkinson}, \citenamefont {Barinov},
  \citenamefont {Bointon}, \citenamefont {Kretinin}, \citenamefont {Wilson},
  \citenamefont {Beton}, \citenamefont {Fal'ko}, \citenamefont {Haigh},\ and\
  \citenamefont {Gorbachev}}]{RN88}%
  \BibitemOpen
  \bibfield  {author} {\bibinfo {author} {\bibfnamefont {A.}~\bibnamefont
  {Weston}}, \bibinfo {author} {\bibfnamefont {Y.}~\bibnamefont {Zou}},
  \bibinfo {author} {\bibfnamefont {V.}~\bibnamefont {Enaldiev}}, \bibinfo
  {author} {\bibfnamefont {A.}~\bibnamefont {Summerfield}}, \bibinfo {author}
  {\bibfnamefont {N.}~\bibnamefont {Clark}}, \bibinfo {author} {\bibfnamefont
  {V.}~\bibnamefont {Z{\'{o}}lyomi}}, \bibinfo {author} {\bibfnamefont
  {A.}~\bibnamefont {Graham}}, \bibinfo {author} {\bibfnamefont
  {C.}~\bibnamefont {Yelgel}}, \bibinfo {author} {\bibfnamefont
  {S.}~\bibnamefont {Magorrian}}, \bibinfo {author} {\bibfnamefont
  {M.}~\bibnamefont {Zhou}}, \bibinfo {author} {\bibfnamefont {J.}~\bibnamefont
  {Zultak}}, \bibinfo {author} {\bibfnamefont {D.}~\bibnamefont {Hopkinson}},
  \bibinfo {author} {\bibfnamefont {A.}~\bibnamefont {Barinov}}, \bibinfo
  {author} {\bibfnamefont {T.~H.}\ \bibnamefont {Bointon}}, \bibinfo {author}
  {\bibfnamefont {A.}~\bibnamefont {Kretinin}}, \bibinfo {author}
  {\bibfnamefont {N.~R.}\ \bibnamefont {Wilson}}, \bibinfo {author}
  {\bibfnamefont {P.~H.}\ \bibnamefont {Beton}}, \bibinfo {author}
  {\bibfnamefont {V.~I.}\ \bibnamefont {Fal'ko}}, \bibinfo {author}
  {\bibfnamefont {S.~J.}\ \bibnamefont {Haigh}}, \ and\ \bibinfo {author}
  {\bibfnamefont {R.}~\bibnamefont {Gorbachev}},\ }\href {\doibase
  10.1038/s41565-020-0682-9} {\bibfield  {journal} {\bibinfo  {journal} {Nature
  Nanotechnology}\ }\textbf {\bibinfo {volume} {15}} (\bibinfo {year} {2020}),\
  10.1038/s41565-020-0682-9},\ \Eprint {http://arxiv.org/abs/1911.12664}
  {1911.12664} \BibitemShut {NoStop}%
\bibitem [{\citenamefont {Lin}\ \emph {et~al.}(2021)\citenamefont {Lin},
  \citenamefont {Holler}, \citenamefont {Bauer}, \citenamefont {Parzefall},
  \citenamefont {Scheuck}, \citenamefont {Peng}, \citenamefont {Korn},
  \citenamefont {Bange}, \citenamefont {Lupton},\ and\ \citenamefont
  {Sch{\"u}ller}}]{lin2021large}%
  \BibitemOpen
  \bibfield  {author} {\bibinfo {author} {\bibfnamefont {K.-Q.}\ \bibnamefont
  {Lin}}, \bibinfo {author} {\bibfnamefont {J.}~\bibnamefont {Holler}},
  \bibinfo {author} {\bibfnamefont {J.~M.}\ \bibnamefont {Bauer}}, \bibinfo
  {author} {\bibfnamefont {P.}~\bibnamefont {Parzefall}}, \bibinfo {author}
  {\bibfnamefont {M.}~\bibnamefont {Scheuck}}, \bibinfo {author} {\bibfnamefont
  {B.}~\bibnamefont {Peng}}, \bibinfo {author} {\bibfnamefont {T.}~\bibnamefont
  {Korn}}, \bibinfo {author} {\bibfnamefont {S.}~\bibnamefont {Bange}},
  \bibinfo {author} {\bibfnamefont {J.~M.}\ \bibnamefont {Lupton}}, \ and\
  \bibinfo {author} {\bibfnamefont {C.}~\bibnamefont {Sch{\"u}ller}},\
  }\href@noop {} {\bibfield  {journal} {\bibinfo  {journal} {Advanced
  Materials}\ }\textbf {\bibinfo {volume} {33}},\ \bibinfo {pages} {2008333}
  (\bibinfo {year} {2021})}\BibitemShut {NoStop}%
\bibitem [{\citenamefont {Andersen}\ \emph {et~al.}(2021)\citenamefont
  {Andersen}, \citenamefont {Scuri}, \citenamefont {Sushko}, \citenamefont {{De
  Greve}}, \citenamefont {Sung}, \citenamefont {Zhou}, \citenamefont {Wild},
  \citenamefont {Gelly}, \citenamefont {Heo}, \citenamefont
  {B{\'{e}}rub{\'{e}}}, \citenamefont {Joe}, \citenamefont {Jauregui},
  \citenamefont {Watanabe}, \citenamefont {Taniguchi}, \citenamefont {Kim},
  \citenamefont {Park},\ and\ \citenamefont {Lukin}}]{Andersen2021}%
  \BibitemOpen
  \bibfield  {author} {\bibinfo {author} {\bibfnamefont {T.~I.}\ \bibnamefont
  {Andersen}}, \bibinfo {author} {\bibfnamefont {G.}~\bibnamefont {Scuri}},
  \bibinfo {author} {\bibfnamefont {A.}~\bibnamefont {Sushko}}, \bibinfo
  {author} {\bibfnamefont {K.}~\bibnamefont {{De Greve}}}, \bibinfo {author}
  {\bibfnamefont {J.}~\bibnamefont {Sung}}, \bibinfo {author} {\bibfnamefont
  {Y.}~\bibnamefont {Zhou}}, \bibinfo {author} {\bibfnamefont {D.~S.}\
  \bibnamefont {Wild}}, \bibinfo {author} {\bibfnamefont {R.~J.}\ \bibnamefont
  {Gelly}}, \bibinfo {author} {\bibfnamefont {H.}~\bibnamefont {Heo}}, \bibinfo
  {author} {\bibfnamefont {D.}~\bibnamefont {B{\'{e}}rub{\'{e}}}}, \bibinfo
  {author} {\bibfnamefont {A.~Y.}\ \bibnamefont {Joe}}, \bibinfo {author}
  {\bibfnamefont {L.~A.}\ \bibnamefont {Jauregui}}, \bibinfo {author}
  {\bibfnamefont {K.}~\bibnamefont {Watanabe}}, \bibinfo {author}
  {\bibfnamefont {T.}~\bibnamefont {Taniguchi}}, \bibinfo {author}
  {\bibfnamefont {P.}~\bibnamefont {Kim}}, \bibinfo {author} {\bibfnamefont
  {H.}~\bibnamefont {Park}}, \ and\ \bibinfo {author} {\bibfnamefont {M.~D.}\
  \bibnamefont {Lukin}},\ }\href {\doibase 10.1038/s41563-020-00873-5}
  {\bibfield  {journal} {\bibinfo  {journal} {Nature Materials}\ } (\bibinfo
  {year} {2021}),\ 10.1038/s41563-020-00873-5}\BibitemShut {NoStop}%
\bibitem [{\citenamefont {Enaldiev}\ \emph {et~al.}(2020)\citenamefont
  {Enaldiev}, \citenamefont {Z{\'{o}}lyomi}, \citenamefont {Yelgel},
  \citenamefont {Magorrian},\ and\ \citenamefont {Fal'ko}}]{Enaldiev2020}%
  \BibitemOpen
  \bibfield  {author} {\bibinfo {author} {\bibfnamefont {V.~V.}\ \bibnamefont
  {Enaldiev}}, \bibinfo {author} {\bibfnamefont {V.}~\bibnamefont
  {Z{\'{o}}lyomi}}, \bibinfo {author} {\bibfnamefont {C.}~\bibnamefont
  {Yelgel}}, \bibinfo {author} {\bibfnamefont {S.~J.}\ \bibnamefont
  {Magorrian}}, \ and\ \bibinfo {author} {\bibfnamefont {V.~I.}\ \bibnamefont
  {Fal'ko}},\ }\href {\doibase 10.1103/PhysRevLett.124.206101} {\bibfield
  {journal} {\bibinfo  {journal} {Physical Review Letters}\ }\textbf {\bibinfo
  {volume} {124}},\ \bibinfo {pages} {1} (\bibinfo {year} {2020})},\ \Eprint
  {http://arxiv.org/abs/1911.12804} {1911.12804} \BibitemShut {NoStop}%
\bibitem [{\citenamefont {Wang}\ \emph
  {et~al.}(2019{\natexlab{b}})\citenamefont {Wang}, \citenamefont {Ouyang},
  \citenamefont {Cao}, \citenamefont {Ma},\ and\ \citenamefont
  {Zheng}}]{Wang2019b}%
  \BibitemOpen
  \bibfield  {author} {\bibinfo {author} {\bibfnamefont {K.}~\bibnamefont
  {Wang}}, \bibinfo {author} {\bibfnamefont {W.}~\bibnamefont {Ouyang}},
  \bibinfo {author} {\bibfnamefont {W.}~\bibnamefont {Cao}}, \bibinfo {author}
  {\bibfnamefont {M.}~\bibnamefont {Ma}}, \ and\ \bibinfo {author}
  {\bibfnamefont {Q.}~\bibnamefont {Zheng}},\ }\href {\doibase
  10.1039/c8nr07963c} {\bibfield  {journal} {\bibinfo  {journal} {Nanoscale}\
  }\textbf {\bibinfo {volume} {11}},\ \bibinfo {pages} {2186} (\bibinfo {year}
  {2019}{\natexlab{b}})}\BibitemShut {NoStop}%
\bibitem [{\citenamefont {Dadgar}\ \emph {et~al.}(2018)\citenamefont {Dadgar},
  \citenamefont {Scullion}, \citenamefont {Kang}, \citenamefont {Esposito},
  \citenamefont {Yang}, \citenamefont {Herman}, \citenamefont {Pimenta},
  \citenamefont {Santos},\ and\ \citenamefont {Pasupathy}}]{dadgar2018strain}%
  \BibitemOpen
  \bibfield  {author} {\bibinfo {author} {\bibfnamefont {A.}~\bibnamefont
  {Dadgar}}, \bibinfo {author} {\bibfnamefont {D.}~\bibnamefont {Scullion}},
  \bibinfo {author} {\bibfnamefont {K.}~\bibnamefont {Kang}}, \bibinfo {author}
  {\bibfnamefont {D.}~\bibnamefont {Esposito}}, \bibinfo {author}
  {\bibfnamefont {E.}~\bibnamefont {Yang}}, \bibinfo {author} {\bibfnamefont
  {I.}~\bibnamefont {Herman}}, \bibinfo {author} {\bibfnamefont
  {M.}~\bibnamefont {Pimenta}}, \bibinfo {author} {\bibfnamefont {E.-J.}\
  \bibnamefont {Santos}}, \ and\ \bibinfo {author} {\bibfnamefont
  {A.}~\bibnamefont {Pasupathy}},\ }\href@noop {} {\bibfield  {journal}
  {\bibinfo  {journal} {Chemistry of Materials}\ }\textbf {\bibinfo {volume}
  {30}},\ \bibinfo {pages} {5148} (\bibinfo {year} {2018})}\BibitemShut
  {NoStop}%
\bibitem [{\citenamefont {Kim}\ \emph {et~al.}(2019)\citenamefont {Kim},
  \citenamefont {Moon}, \citenamefont {Noh}, \citenamefont {Lee},\ and\
  \citenamefont {Kim}}]{RN24}%
  \BibitemOpen
  \bibfield  {author} {\bibinfo {author} {\bibfnamefont {H.}~\bibnamefont
  {Kim}}, \bibinfo {author} {\bibfnamefont {J.~S.}\ \bibnamefont {Moon}},
  \bibinfo {author} {\bibfnamefont {G.}~\bibnamefont {Noh}}, \bibinfo {author}
  {\bibfnamefont {J.}~\bibnamefont {Lee}}, \ and\ \bibinfo {author}
  {\bibfnamefont {J.~H.}\ \bibnamefont {Kim}},\ }\href {\doibase
  10.1021/acs.nanolett.9b03421} {\bibfield  {journal} {\bibinfo  {journal}
  {Nano Letters}\ }\textbf {\bibinfo {volume} {19}},\ \bibinfo {pages} {7534}
  (\bibinfo {year} {2019})}\BibitemShut {NoStop}%
\bibitem [{\citenamefont {Yankowitz}\ \emph {et~al.}(2019)\citenamefont
  {Yankowitz}, \citenamefont {Chen}, \citenamefont {Polshyn}, \citenamefont
  {Zhang}, \citenamefont {Watanabe}, \citenamefont {Taniguchi}, \citenamefont
  {Graf}, \citenamefont {Young},\ and\ \citenamefont {Dean}}]{Yankowitz2019}%
  \BibitemOpen
  \bibfield  {author} {\bibinfo {author} {\bibfnamefont {M.}~\bibnamefont
  {Yankowitz}}, \bibinfo {author} {\bibfnamefont {S.}~\bibnamefont {Chen}},
  \bibinfo {author} {\bibfnamefont {H.}~\bibnamefont {Polshyn}}, \bibinfo
  {author} {\bibfnamefont {Y.}~\bibnamefont {Zhang}}, \bibinfo {author}
  {\bibfnamefont {K.}~\bibnamefont {Watanabe}}, \bibinfo {author}
  {\bibfnamefont {T.}~\bibnamefont {Taniguchi}}, \bibinfo {author}
  {\bibfnamefont {D.}~\bibnamefont {Graf}}, \bibinfo {author} {\bibfnamefont
  {A.~F.}\ \bibnamefont {Young}}, \ and\ \bibinfo {author} {\bibfnamefont
  {C.~R.}\ \bibnamefont {Dean}},\ }\href {\doibase 10.1126/science.aav1910}
  {\bibfield  {journal} {\bibinfo  {journal} {Science}\ }\textbf {\bibinfo
  {volume} {363}},\ \bibinfo {pages} {1059} (\bibinfo {year}
  {2019})}\BibitemShut {NoStop}%
\bibitem [{\citenamefont {Schindler}\ \emph {et~al.}(2020)\citenamefont
  {Schindler}, \citenamefont {Noky}, \citenamefont {Schmidt}, \citenamefont
  {Felser}, \citenamefont {Wosnitza},\ and\ \citenamefont
  {Gooth}}]{Schindler2020}%
  \BibitemOpen
  \bibfield  {author} {\bibinfo {author} {\bibfnamefont {C.}~\bibnamefont
  {Schindler}}, \bibinfo {author} {\bibfnamefont {J.}~\bibnamefont {Noky}},
  \bibinfo {author} {\bibfnamefont {M.}~\bibnamefont {Schmidt}}, \bibinfo
  {author} {\bibfnamefont {C.}~\bibnamefont {Felser}}, \bibinfo {author}
  {\bibfnamefont {J.}~\bibnamefont {Wosnitza}}, \ and\ \bibinfo {author}
  {\bibfnamefont {J.}~\bibnamefont {Gooth}},\ }\href {\doibase
  10.1103/PhysRevB.102.035132} {\bibfield  {journal} {\bibinfo  {journal}
  {Physical Review B}\ }\textbf {\bibinfo {volume} {102}},\ \bibinfo {pages}
  {35132} (\bibinfo {year} {2020})}\BibitemShut {NoStop}%
\bibitem [{\citenamefont {Li}\ \emph {et~al.}(2017)\citenamefont {Li},
  \citenamefont {Wang}, \citenamefont {Gao}, \citenamefont {Chen},
  \citenamefont {Peng}, \citenamefont {Liu},\ and\ \citenamefont {Wei}}]{RN30}%
  \BibitemOpen
  \bibfield  {author} {\bibinfo {author} {\bibfnamefont {H.}~\bibnamefont
  {Li}}, \bibinfo {author} {\bibfnamefont {J.}~\bibnamefont {Wang}}, \bibinfo
  {author} {\bibfnamefont {S.}~\bibnamefont {Gao}}, \bibinfo {author}
  {\bibfnamefont {Q.}~\bibnamefont {Chen}}, \bibinfo {author} {\bibfnamefont
  {L.}~\bibnamefont {Peng}}, \bibinfo {author} {\bibfnamefont {K.}~\bibnamefont
  {Liu}}, \ and\ \bibinfo {author} {\bibfnamefont {X.}~\bibnamefont {Wei}},\
  }\href {\doibase 10.1002/adma.201701474} {\bibfield  {journal} {\bibinfo
  {journal} {Advanced Materials}\ }\textbf {\bibinfo {volume} {29}} (\bibinfo
  {year} {2017}),\ 10.1002/adma.201701474}\BibitemShut {NoStop}%
\bibitem [{\citenamefont {Kobayashi}\ \emph {et~al.}(2017)\citenamefont
  {Kobayashi}, \citenamefont {Taniguchi}, \citenamefont {Watanabe},
  \citenamefont {Maniwa},\ and\ \citenamefont {Miyata}}]{Kobayashi2017}%
  \BibitemOpen
  \bibfield  {author} {\bibinfo {author} {\bibfnamefont {Y.}~\bibnamefont
  {Kobayashi}}, \bibinfo {author} {\bibfnamefont {T.}~\bibnamefont
  {Taniguchi}}, \bibinfo {author} {\bibfnamefont {K.}~\bibnamefont {Watanabe}},
  \bibinfo {author} {\bibfnamefont {Y.}~\bibnamefont {Maniwa}}, \ and\ \bibinfo
  {author} {\bibfnamefont {Y.}~\bibnamefont {Miyata}},\ }\href {\doibase
  10.7567/APEX.10.045201} {\bibfield  {journal} {\bibinfo  {journal} {Applied
  Physics Express}\ }\textbf {\bibinfo {volume} {10}},\ \bibinfo {pages}
  {045201} (\bibinfo {year} {2017})}\BibitemShut {NoStop}%
\bibitem [{\citenamefont {Georgoulea}\ \emph {et~al.}(2022)\citenamefont
  {Georgoulea}, \citenamefont {Power},\ and\ \citenamefont
  {Caffrey}}]{Georgoulea2022}%
  \BibitemOpen
  \bibfield  {author} {\bibinfo {author} {\bibfnamefont {N.~C.}\ \bibnamefont
  {Georgoulea}}, \bibinfo {author} {\bibfnamefont {S.~R.}\ \bibnamefont
  {Power}}, \ and\ \bibinfo {author} {\bibfnamefont {N.~M.}\ \bibnamefont
  {Caffrey}},\ }\href@noop {} {\bibfield  {journal} {\bibinfo  {journal} {arXiv
  preprint arXiv:2206.01455}\ } (\bibinfo {year} {2022})}\BibitemShut {NoStop}%
\end{thebibliography}%

\end{document}